\documentclass[prd,aps,preprintnumbers,floats,floatfix,superscriptaddress,preprintnumbers,
showpacs,eqsecnum,nofootinbib,twocolumn]{revtex4-2}
\usepackage[utf8]{inputenc}
\usepackage{latexsym,array,theorem,mathrsfs,bm,float}
\usepackage{psfrag}
\usepackage{amsfonts,amsmath,amssymb,array,afterpage,
epsfig,color,graphicx,tabularx,here,multirow,hyperref}
\usepackage{tikz}
\usetikzlibrary{calc,decorations.pathmorphing}

\newcommand{\nn}{\nonumber \\}

\newcommand{\D}{{\rm d}}

\newcommand{\Disc}{{\rm Disc}}
\newcommand{\sumint}{ \sum \hspace{-1.3em} \int }

\newcommand{\Amp}{\mathcal{A}}
\newcommand{\ReM}{{\rm Re} M^2}

\newcommand{\Atwo}[1]{
\begin{tikzpicture}[baseline=-2]
\draw (0.6,0.15) -- (-0.6,0.15);
\draw (0.6,-0.15) -- (-0.6,-0.15);
\filldraw [fill=white] (0,0) circle [radius=4mm] ;
\node (O) at (0,0) {$#1$} ;
\end{tikzpicture}}

\newcommand{\Athree}[1]{
\begin{tikzpicture}[baseline=-2]
\draw (0.6,0.2) -- (-0.6,0.2);
\draw (0.6,0) -- (-0.6,0);
\draw (0.6,-0.2) -- (-0.6,-0.2);
\filldraw [fill=white] (0,0) circle [radius=4mm] ;
\node (O) at (0,0) {$#1$} ;
\end{tikzpicture}
}

\newcommand{\Afour}[1]{
\begin{tikzpicture}[baseline=-2]
\draw (0.6,0.3) -- (-0.6,0.3);
\draw (0.6,0.1) -- (-0.6,0.1);
\draw (0.6,-0.1) -- (-0.6,-0.1);
\draw (0.6,-0.3) -- (-0.6,-0.3);
\filldraw [fill=white] (0,0) circle [radius=4mm] ;
\node (O) at (0,0) {$#1$} ;
\end{tikzpicture}
}

\newcommand{\AmixL}{
\begin{tikzpicture}[baseline=-2]
\draw (0.7,-0.15) -- (-0.7,-0.15);
\draw  [decorate, decoration={snake}] (-0.7,0.15) -- (0, 0.15) ;
\draw  [decorate, decoration={snake}] (0.7,0.15) -- (0, 0.15) ;
\filldraw [fill=white] (0,0) circle [radius=4mm] ;
\draw ($(0,0)+(90:0.4)$) to [out = -90, in = 170] ($(0,0)+(-10:0.4)$) ;
\node (O1) at (0.2,0.2) {$-$} ;
\node (O2) at (-0.05,-0.05) {$+$} ;
\end{tikzpicture}}

\newcommand{\AmixR}{
\begin{tikzpicture}[baseline=-2]
\draw (0.7,-0.15) -- (-0.7,-0.15);
\draw  [decorate, decoration={snake}] (-0.7,0.15) -- (0, 0.15) ;
\draw  [decorate, decoration={snake}] (0.7,0.15) -- (0, 0.15) ;
\filldraw [fill=white] (0,0) circle [radius=4mm] ;
\draw ($(0,0)+(90:0.4)$) to [out = -90, in = 10] ($(0,0)+(190:0.4)$) ;
\node (O1) at (-0.2,0.2) {$+$} ;
\node (O2) at (0.05,-0.05) {$-$} ;
\end{tikzpicture}
}

\newcommand{\Auns}[1]{
\begin{tikzpicture}[baseline=-2]
\draw (0.7,-0.15) -- (-0.7,-0.15);
\draw  [decorate, decoration={snake}] (-0.7,0.15) -- (0, 0.15) ;
\draw  [decorate, decoration={snake}] (0.7,0.15) -- (0, 0.15) ;
\filldraw [fill=white] (0,0) circle [radius=4mm] ;
\node at (0,0) {$#1$} ;
\end{tikzpicture}
}

\newcommand{\Pole}[5]{
\draw (#1,#2) -- (#3,#4);
\filldraw [fill=white] ($(#1 /2,#2 /2)+(#3 /2,#4 /2)$) circle [radius=1.2mm] ;
\node at ($(#1 /2,#2 /2)+(#3 /2,#4 /2)$) {\scalebox{0.7}{$#5$}};
}

\begin{document}
\baselineskip=12pt

%<<<<<<<<<<<<< TITLE >>>>>>>>>>>>>>>%
%%
\preprint{YITP-22-153}
\title{Unitarity and unstable-particle scattering amplitudes
}
%%
%<<<<<<<<<<<<< AUTHOR >>>>>>>>>>>>>>>%
%%
\author{Katsuki Aoki}
%\email{katsuki.aoki@yukawa.kyoto-u.ac.jp}
\affiliation{
Center for Gravitational Physics and Quantum Information, Yukawa Institute for Theoretical Physics, Kyoto University, 606-8502, Kyoto, Japan
}

%<<<<<<<<<<<<< DATE >>>>>>>>>>>>>>>%
\date{\today}

%======================================%
%<<<<<<<<<<<<< ABSTRACT >>>>>>>>>>>>>>>%
%======================================%
\begin{abstract}
Unitarity is the fundamental property of the S-matrix while its usage for a scattering of unstable particles has been subtle as unstable particles do not appear in the asymptotic states. Defining unstable-particle amplitudes as residues of a higher-point amplitude at an appropriate complex pole, we find unitarity equations for the 2-to-2 unstable-particle amplitudes from unitarity and analyticity of stable-particle scattering amplitudes. The unstable-particle unitarity equations take a form analogous to those of the stable-particle amplitudes when the in and out states are chosen to be complex-conjugate positions. In particular, as in the optical theorem, we find a positivity constraint on a discontinuity of the amplitudes in a positive region of the momentum transfer variable.
\end{abstract}

\maketitle

\section{Introduction}
We are in the era of the revival of the S-matrix theory of the 1960s. The physical requirements such as unitarity and causality provide powerful consistency conditions on the S-matrix, called positivity bounds, and they can be used to constrain low-energy physics without detailed knowledge of high-energy physics~\cite{Pham:1985cr, Ananthanarayan:1994hf, Adams:2006sv}. In the lack of understanding ultimate laws of nature, the positivity bounds provide one of the most promising ways to investigate new physics in a model-independent way, and recently there has been a lot of progress such as sharpening the bounds~\cite{Bellazzini:2016xrt, deRham:2017imi, deRham:2017avq, deRham:2017zjm, Bellazzini:2020cot, Tolley:2020gtv, Caron-Huot:2020cmc, Sinha:2020win, Arkani-Hamed:2020blm, Li:2021lpe, Chiang:2021ziz, Du:2021byy, Chowdhury:2021ynh, Bellazzini:2021oaj, Chiang:2022ltp} and extending the bounds to gravitational systems~\cite{Hamada:2018dde, Bellazzini:2019xts, Alberte:2020jsk, Tokuda:2020mlf, Herrero-Valea:2020wxz, Caron-Huot:2021rmr, Bern:2021ppb, Caron-Huot:2022ugt, Chiang:2022jep, Herrero-Valea:2022lfd} and systems without Lorentz invariance~\cite{Baumann:2015nta, Grall:2021xxm, Aoki:2021ffc, Creminelli:2022onn}.

The S-matrix is a probability amplitude connecting the infinite past and the infinite future. When a theory contains unstable particles as well as stable particles, the asymptotic states are spanned by the stable particles only, and the unstable particles do not appear in the asymptotic states~\cite{Veltman:1963th}. A question is then whether we can apply the S-matrix constraints to a ``scattering'' of unstable particles. Unstable particles appear in many contexts. Most of the known particles, either elementary ones or composite ones, have finite decay widths~\cite{ParticleDataGroup:2020ssz}. New particles in physics beyond the Standard Model and in quantum gravity would be unstable unless they are protected by symmetry. If we cannot apply the S-matrix arguments to the unstable particles, this gives a strong limitation in the availability of the S-matrix theory. Having said that, unstable particles and stable particles may be indistinguishable when the lifetime is sufficiently longer than the timescale of interest. At least in an approximate sense, ``scattering amplitudes'' of the unstable particles should be constrained by the physical requirements. This would be a reason why the subtlety associated with the unstable particles has not been seriously studied. Yet, we need to quantify the requirements with a proper definition of unstable-particle amplitudes to make definite predictions. 
Even tiny corrections are important in investigating quantum gravity constraints~\cite{Alberte:2020bdz, Aoki:2021ckh, Noumi:2021uuv, Alberte:2021dnj, Noumi:2022zht}.

A definition of the unstable-particle amplitudes was suggested in the 1960s ~(see~\cite{Eden:1966dnq} and references therein). In the S-matrix theory, physical amplitudes are identified with boundary values of analytic functions with singularities implied by unitarity. A stable-particle exchange leads to a pole whose residue is factorized by amplitudes involving this particle; that is, a lower-point amplitude is embedded in a higher-point amplitude. Analogously, a lower-point unstable-particle scattering amplitude can be defined by a residue of a higher-point stable-particle amplitude at a complex pole corresponding to the unstable particle. Then, we can discuss their properties although there is still an ambiguity in the definition regarding the choice of the complex pole as we will discuss later.

In a weakly-coupled system, one may use the perturbation theory to compute unstable-particle amplitudes. Stable-particle amplitudes and unstable-particle amplitudes exhibit different behaviours at the loop level as new singularities such as anomalous thresholds and external-mass singularities~\cite{nakanishi1963external, Nakanishi:1974wm, Hannesdottir:2022bmo} arise when the mass of an external state is extrapolated to an unstable region. 
Nonetheless, in general, the perturbation series does not necessarily converge and a resummation is required. We need to make sure whether perturbative calculations correctly capture the properties followed by the general requirements such as unitarity. It is desirable to understand the general properties of unstable-particle amplitudes {\it ab initio} without relying on the perturbation theory.

We thus revisit the S-matrix theory~\cite{Eden:1966dnq} whose underlying idea is making use of unitarity and analyticity of a higher-point amplitude to analyse an embedded lower-point amplitude. We first explain our assumptions and their immediate consequences in Sec.~\ref{sec:S-matrix} by following~\cite{Eden:1966dnq}. In Sec.~\ref{sec:unitarity}, we derive the unitarity equations for the unstable-particle amplitudes from those for higher-point stable-particle amplitudes. We discuss the properties of the unstable-particle amplitudes based on the obtained unitarity equations. The unitarity equations are derived in a negative momentum transfer $t<0$, where $t$ is one of the Mandelstam variables, in Sec.~\ref{sec:unitarity} while they are extended to a finite positive $t$ in Sec.~\ref{sec:extended}. In Sec.~\ref{sec:optical}, we especially focus on the sign of a discontinuity of the unstable-particle amplitudes and show that the sign is fixed in a positive region of the momentum transfer variable.
We conclude in Sec.~\ref{sec:summary} with a summary and discussions. In Appendix~\ref{sec:stable_unitarity}, we briefly review the stable-particle unitarity equations and we study a triangle Feynman diagram as a perturbative validation of our results in Appendix~\ref{sec:triangle}.

\section{S-matrix theory}
\label{sec:S-matrix}

For simplicity, we focus on a theory composed of a stable scalar field $\varphi$ with the mass $\mu$ and an unstable scalar field $A$ with the pole mass $M$ in four dimensions. 
By the use of the connected part of the S-matrix, the $n$-to-$n'$ scattering amplitude and its Hermitian conjugate are defined by
\begin{align}
\langle \{p' \} |S| \{ p \} \rangle_c &= -i(2\pi)^4 \delta^{(4)}(p'_{\rm tot} - p_{\rm tot}) \Amp^{(+)}_{n'n}
\,, \label{defAmp} \\
\langle \{p' \} |S^{\dagger} |\{ p \} \rangle_c &= i(2\pi)^4 \delta^{(4)}(p'_{\rm tot} - p_{\rm tot}) \Amp^{(-)}_{n'n}
\,,
\end{align}
with $\Amp^{(-)}_{n'n}=(\Amp^{(+)}_{nn'})^*$ where $\{p\}$ and $\{p'\}$ are the sets of the external four-momenta of the initial and final states with the total four-momenta $p_{\rm tot}$ and $p'_{\rm tot}$, respectively. We use the $(-,+,+,+)$ convention and omit the Lorentz indices for notational simplicity. We again emphasize that the unstable particle does not appear in the asymptotic states and its existence can be seen as a complex pole with ${\rm Im} M^2<0$ as we will explain later. Hence, $|\{ p\}\rangle$ only involves the stable particles. Note that the amplitude is defined with a minus sign in \eqref{defAmp} to follow the convention of Chapter 4 of~\cite{Eden:1966dnq}; accordingly, the imaginary part of the forward amplitude is negative rather than positive.

Having defined the amplitudes, we require:
\begin{enumerate}
\item \label{axiom1} Lorentz invariance: The amplitudes are given by functions of Lorentz-invariant variables.
\item \label{axiom2} Unitarity: $SS^{\dagger}=S^{\dagger}S=1$.
\item \label{axiom3} Analyticity: The physical amplitude and its Hermitian conjugate are real boundary values of the same analytic function with singularities inferred from unitarity. In particular, the unitarity equations are supposed to be the sums of discontinuities across individual thresholds.
\end{enumerate}
This analytic property is called Hermitian analyticity which can be proved under a weaker condition on analyticity (or causality) at least for the 2-to-2 amplitude in a gapped system~\cite{Eden:1966dnq}. We do not prove Hermitian analyticity of higher-point amplitudes in the present paper and simply employ it as our analyticity postulate. Let us below explain our assumptions in order. See~\cite{Eden:1966dnq} and references therein for details.

Lorentz invariance concludes that the $n$-to-$n'$ amplitudes are functions of $3(n+n')-10$ independent inner products of external momenta which we collectively denote by $s_A$.\footnote{In general, the amplitudes are decomposed into scalar and pseudoscalar parts. The pseudoscalar part is absent for $n=n'=2$ in four dimensions since the conservation law leaves three independent momenta and no pseudoscalar can be formed by three vectors. Since we are interested in 2-to-2 subamplitudes embedded in higher-point amplitudes, the pseudoscalar part of higher-point amplitudes would be irrelevant to our discussions.} It is convenient to use the variables
\begin{align}
s_{ijk\cdots}:=-p_{ijk\cdots}^2=-(\pm p_i \pm p_j \pm p_k \pm \cdots)^2
\end{align}
where $p_i$ refers to both in and out momenta with the $(+)$ sign for in momenta and with the $(-)$ sign for out momenta, respectively. In particular, the total energy variable is denoted by $s=s_{12\cdots}=-p_{\rm tot}^2$ unless otherwise stated.\footnote{The variables $s,t$, and $u$ will be used to denote the Mandelstam variables of embedded 2-to-2 amplitudes. The relations between $\{s,t,u\}$ and the energy variables of a higher-point amplitude depend on how the 2-to-2 amplitude is embedded. In most cases of the present paper, the total energy variable of the higher-point amplitude agrees with that of the embedded 2-to-2 amplitude.} The variables $s_{ij\cdots}$ are subject to constraints in the physical region. Together with the positive frequency conditions, the physical region constraints correspond to the Gram determinants (or the Cayley–Menger determinants) to have appropriate signs. Let us write the set of indices $i_1\cdots i_p$ as $I$, $i_{p+1}\cdots i_{p+q}$ as $J$, $i_1\cdots i_p i_{p+1} \cdots i_{p+q}$ as $IJ$ and so on. Considering timelike vectors $p_I,p_J, $ and $p_K$, the physical region constraints on their inner products are given by
\begin{align}
\begin{vmatrix}
0& 1 & 1 & 1 \\
1 & 0 & s_J & s_I \\
1 & s_J & 0 & s_{IJ} \\
1 & s_I & s_{IJ} & 0 
\end{vmatrix}
&>0
\,,
\label{phys1}
\\
\begin{vmatrix}
0 & 1 & 1 & 1 & 1 \\
1 & 0 & s_J & s_{JK} & s_I \\
1 & s_J & 0 & s_K & s_{IJ} \\
1 & s_{JK} & s_K & 0 & s_{IJK} \\
1 & s_I & s_{IJ} & s_{IJK} & 0 
\end{vmatrix}
&>0
\,.
\label{phys2}
\end{align}
Lorentz transformation can also be used to interchange the in momenta and the out momenta of the $n=n'=2$ amplitude, concluding the symmetry condition
\begin{align}
\langle p'_2 p'_1 |S|p_1 p_2\rangle = \langle p_2 p_1 |S| p'_1 p'_2\rangle
\,.
\label{timereversal22}
\end{align}

In the following, we adopt the diagrammatic notation of \cite{olive1964exploration,Eden:1966dnq}:
\begin{align}
\begin{split}
{}_{n'}
\begin{tikzpicture}[baseline=-2]
\draw (0.8,0.3) -- (-0.8,0.3);
\draw (0.8,0.1) -- (-0.8,0.1);
\draw [dotted] (0.8,-0.1) -- (-0.8,-0.1);
\draw (0.8,-0.3) -- (-0.8,-0.3);
\filldraw [fill=white] (0,0) circle [radius=5mm] ;
\node (O) at (0,0) {$\pm$} ;
\end{tikzpicture}
 {}_{n}
 &=\Amp^{(\pm)}_{n'n}
 \,, \\
 \text{each internal line} &= -2\pi i \theta(q^0) \delta(q^2+\mu^2)
 \,, \\
 \text{each loop} &= \frac{i}{(2\pi)^4} \int \D^4 k
 \,, 
 \\
 \text{$n$ lines joining two bubbles} &=\text{a symmetry factor } \frac{1}{n!}\,,
 \end{split}
 \label{rules}
\end{align}
where $q$ is the four-momentum of an internal line which is determined by external momenta $p$ and loop momenta $k$ according to the conservation law.
Then, the unitarity equations, which are consequences of $SS^{\dagger}=1$, of the 2-to-2 amplitude and the 3-to-3 amplitudes in $(3\mu)^2 \leq s < (4\mu)^2$ are written as
\begin{align}
\Atwo{+} - \Atwo{-}
&=
\begin{tikzpicture}[baseline=-2]
\draw (-1,0.15) -- (1,0.15);
\draw (-1,-0.15) -- (1,-0.15);
\filldraw [fill=white] (-0.5,0) circle [radius=4mm] ;
\node (O1) at (-0.5,0) {$+$} ;
\filldraw [fill=white] (0.5,0) circle [radius=4mm] ;
\node (O2) at (0.5,0) {$-$} ;
\end{tikzpicture}
+
\begin{tikzpicture}[baseline=-2]
\draw (-1,0.15) -- (-0.5,0.15);
\draw (-1,-0.15) -- (-0.5,-0.15);
\draw (1,0.15) -- (0.5,0.15);
\draw (1,-0.15) -- (0.5,-0.15);
\draw (0.5,0.2) -- (-0.5,0.2);
\draw (0.5,0) -- (-0.5,0);
\draw (0.5,-0.2) -- (-0.5,-0.2);
\filldraw [fill=white] (-0.5,0) circle [radius=4mm] ;
\node (O1) at (-0.5,0) {$+$} ;
\filldraw [fill=white] (0.5,0) circle [radius=4mm] ;
\node (O2) at (0.5,0) {$-$} ;
\end{tikzpicture}
\label{unitarity22}
\,, \\
\Athree{+} - \Athree{-}
&=
\begin{tikzpicture}[baseline=-2]
\draw (-1,0.2) -- (-0.5,0.2);
\draw (-1,0) -- (-0.5,0);
\draw (-1,-0.2) -- (-0.5,-0.2);
\draw (1,0.2) -- (0.5,0.2);
\draw (1,0) -- (0.5,0);
\draw (1,-0.2) -- (0.5,-0.2);
\draw (-0.5,0.1) -- (0.5,0.1) ;
\draw (-0.5,-0.1) -- (0.5,-0.1) ;
\filldraw [fill=white] (-0.5,0) circle [radius=4mm] ;
\node (O1) at (-0.5,0) {$+$} ;
\filldraw [fill=white] (0.5,0) circle [radius=4mm] ;
\node (O2) at (0.5,0) {$-$} ;
\end{tikzpicture}
+
\begin{tikzpicture}[baseline=-2]
\draw (-1,0.2) -- (1,0.2);
\draw (-1,0) -- (1,0);
\draw (-1,-0.2) -- (1,-0.2);
\filldraw [fill=white] (-0.5,0) circle [radius=4mm] ;
\node (O1) at (-0.5,0) {$+$} ;
\filldraw [fill=white] (0.5,0) circle [radius=4mm] ;
\node (O2) at (0.5,0) {$-$} ;
\end{tikzpicture}
\nn
&+\sum 
\begin{tikzpicture}[baseline=-2]
\draw (-1,0.2) -- (0.6,0.2);
\draw (-1,0) -- (0.6,0);
\draw (-1,-0.2) -- (0.6,-0.2);
\filldraw [fill=white] (-0.5,0) circle [radius=4mm] ;
\node (O1) at (-0.5,0) {$+$} ;
\filldraw [fill=white] (0.3,0.1) circle [radius=2mm] ;
\node (O2) at (0.3,0.1) {$-$} ;
\end{tikzpicture}
+\sum 
\begin{tikzpicture}[baseline=-2]
\draw (1,0.2) -- (-0.6,0.2);
\draw (1,0) -- (-0.6,0);
\draw (1,-0.2) -- (-0.6,-0.2);
\filldraw [fill=white] (0.5,0) circle [radius=4mm] ;
\node (O1) at (0.5,0) {$-$} ;
\filldraw [fill=white] (-0.3,0.1) circle [radius=2mm] ;
\node (O2) at (-0.3,0.1) {$+$} ;
\end{tikzpicture}
\nn
&+ \sum
\begin{tikzpicture}[baseline=-2]
\draw (0.6,0.2) -- (-0.6,0.2);
\draw (0.6,0) -- (0.3,0);
\draw (-0.6,0) -- (-0.3,0);
\draw (-0.3,0.1) -- (0.3,-0.1);
\draw (0.6,-0.2) -- (-0.6,-0.2);
\filldraw [fill=white] (-0.3,0.1) circle [radius=2mm] ;
\node (O1) at (-0.3,0.1) {$+$} ;
\filldraw [fill=white] (0.3,-0.1) circle [radius=2mm] ;
\node (O2) at (0.3,-0.1) {$-$} ;
\end{tikzpicture}
\,,
\label{unitarity33}
\end{align}
where the summations are over the possible connected diagrams with different choices of the particles. The unitarity equation only involves the solid lines representing the stable $\varphi$-particle because of the absence of the unstable particles in the asymptotic states~\cite{Veltman:1963th} (see also~\cite{Denner:2014zga}).
As the total energy increases, the number of internal lines increases while the structure of the equations is the same in the 2-to-2 and 3-to-3 unitarity equations (see Appendix~\ref{sec:stable_unitarity}).
According to \eqref{timereversal22}, the l.h.s.~of the 2-to-2 unitarity equation represents the imaginary part,
$2i{\rm Im}\Amp_{22}=
\begin{tikzpicture}[baseline=-2]
\draw (0.3,0.1) --  (-0.3,0.1);
\draw  (0.3,-0.1) -- (-0.3,-0.1);
\filldraw [fill=white] (0,0) circle [radius=2mm] ;
\node (O) at (0,0) {$+$} ;
\end{tikzpicture}
-
\begin{tikzpicture}[baseline=-2]
\draw (0.3,0.1) --  (-0.3,0.1);
\draw  (0.3,-0.1) -- (-0.3,-0.1);
\filldraw [fill=white] (0,0) circle [radius=2mm] ;
\node (O) at (0,0) {$-$} ;
\end{tikzpicture}
$.

The amplitudes $\Amp^{(\pm)}_{n'n}$ can be analytically continued from the physical region into the complex plane with the relation $\Amp^{(-)}_{n'n}(s_A)=[\Amp^{(+)}_{nn'}(s_A^*)]^*$ in the complex domain. 
Hermitian analyticity states that $\Amp_{n'n}^{(\pm)}$ are opposite boundary values of the same analytic functions $\Amp_{n'n}$,\footnote{When the amplitude satisfies the symmetry condition \eqref{timereversal22}, Hermitian analyticity is reduced to real analyticity; see e.g.,~\cite{Miramontes:1999gd} for their distinctions.} which we may write
\begin{align}
\Amp^{(\pm)}_{n'n}(s_A) = \lim_{\varepsilon \to 0+} \Amp_{n'n}(s_A \pm i \varepsilon)
\,.
\label{HermitianA}
\end{align}
The directions to approach the real axis may depend on the choice of the variables. The precise directions to approach the real axis have to be chosen to coincide with the causal direction for the $(+)$ amplitude and the anti-causal direction for the $(-)$ amplitude, respectively (see~\cite{Hannesdottir:2022bmo} for a recent discussion).
An immediate consequence of \eqref{HermitianA} is that the unitarity equation represents the total discontinuity of the analytically-continued amplitude $\Amp_{n'n}$. For example, the 2-to-2 unitarity equation is understood as the discontinuity between $s+i\varepsilon$ and $s-i\varepsilon$,
$\Disc_s \Amp_{22}=
\begin{tikzpicture}[baseline=-2]
\draw (0.3,0.1) --  (-0.3,0.1);
\draw  (0.3,-0.1) -- (-0.3,-0.1);
\filldraw [fill=white] (0,0) circle [radius=2mm] ;
\node (O) at (0,0) {$+$} ;
\end{tikzpicture}
-
\begin{tikzpicture}[baseline=-2]
\draw (0.3,0.1) --  (-0.3,0.1);
\draw  (0.3,-0.1) -- (-0.3,-0.1);
\filldraw [fill=white] (0,0) circle [radius=2mm] ;
\node (O) at (0,0) {$-$} ;
\end{tikzpicture}
$. The first and second terms on the r.h.s.~of \eqref{unitarity22} are the discontinuities associated with the two-particle intermediate states and the three-particle states, respectively.
The unitarity equations of higher-point amplitudes are more involved. As we stated in the assumption \ref{axiom3}, we shall assume that the unitarity equations are decomposed into discontinuities across individual energy variables~\cite{olive1965unitarity}. The first line of the r.h.s.~of \eqref{unitarity33} is regarded as the discontinuity in the total energy variable $s$ while the second line of \eqref{unitarity33} is understood as the 2-particle discontinuities in subenergy variables
\begin{align}
\begin{split}
\begin{tikzpicture}[baseline=-2]
\draw (0.6,0.2) -- (-0.6,0.2);
\draw (0.6,0) -- (-0.6,0);
\draw (0.6,-0.2) -- (-0.6,-0.2);
\filldraw [fill=white] (0,0) circle [radius=4mm] ;
\draw ($(0,0)+(90:0.4)$) to [out = -90, in = 170] ($(0,0)+(-10:0.4)$) ;
\node (O) at (0.2,0.2) {$+$} ;
\end{tikzpicture}
-
\begin{tikzpicture}[baseline=-2]
\draw (0.6,0.2) -- (-0.6,0.2);
\draw (0.6,0) -- (-0.6,0);
\draw (0.6,-0.2) -- (-0.6,-0.2);
\filldraw [fill=white] (0,0) circle [radius=4mm] ;
\draw ($(0,0)+(90:0.4)$) to [out = -90, in = 170] ($(0,0)+(-10:0.4)$) ;
\node (O) at (0.2,0.2) {$-$} ;
\end{tikzpicture}
&=
\begin{tikzpicture}[baseline=-2]
\draw (1,0.2) -- (-0.5,0.2);
\draw (1,0) -- (-0.5,0);
\draw (1,-0.2) -- (-0.5,-0.2);
\filldraw [fill=white] (0,0) circle [radius=4mm] ;
\draw ($(0,0)+(90:0.4)$) to [out = -90, in = 170] ($(0,0)+(-10:0.4)$) ;
\filldraw [fill=white] (0.7,0.1) circle [radius=2mm];
\node (O1) at (0.2,0.2) {$+$} ;
\node (O2) at (0.7,0.1) {$-$} ;
\end{tikzpicture}
=
\begin{tikzpicture}[baseline=-2]
\draw (1,0.2) -- (-0.5,0.2);
\draw (1,0) -- (-0.5,0);
\draw (1,-0.2) -- (-0.5,-0.2);
\filldraw [fill=white] (0,0) circle [radius=4mm] ;
\draw ($(0,0)+(90:0.4)$) to [out = -90, in = 170] ($(0,0)+(-10:0.4)$) ;
\filldraw [fill=white] (0.7,0.1) circle [radius=2mm];
\node (O1) at (0.2,0.2) {$-$} ;
\node (O2) at (0.7,0.1) {$+$} ;
\end{tikzpicture}
\,,  \\
\begin{tikzpicture}[baseline=-2]
\draw (0.6,0.2) -- (-0.6,0.2);
\draw (0.6,0) -- (-0.6,0);
\draw (0.6,-0.2) -- (-0.6,-0.2);
\filldraw [fill=white] (0,0) circle [radius=4mm] ;
\draw ($(0,0)+(90:0.4)$) to [out = -90, in = 10] ($(0,0)+(190:0.4)$) ;
\node (O) at (-0.2,0.2) {$+$} ;
\end{tikzpicture}
-
\begin{tikzpicture}[baseline=-2]
\draw (0.6,0.2) -- (-0.6,0.2);
\draw (0.6,0) -- (-0.6,0);
\draw (0.6,-0.2) -- (-0.6,-0.2);
\filldraw [fill=white] (0,0) circle [radius=4mm] ;
\draw ($(0,0)+(90:0.4)$) to [out = -90, in = 10] ($(0,0)+(190:0.4)$) ;
\node (O) at (-0.2,0.2) {$-$} ;
\end{tikzpicture}
&=
\begin{tikzpicture}[baseline=-2]
\draw (-1,0.2) -- (0.5,0.2);
\draw (-1,0) -- (0.5,0);
\draw (-1,-0.2) -- (0.5,-0.2);
\filldraw [fill=white] (0,0) circle [radius=4mm] ;
\draw ($(0,0)+(90:0.4)$) to [out = -90, in = 10] ($(0,0)+(190:0.4)$) ;
\filldraw [fill=white] (-0.7,0.1) circle [radius=2mm];
\node (O1) at (-0.2,0.2) {$-$} ;
\node (O2) at (-0.7,0.1) {$+$} ;
\end{tikzpicture}
=
\begin{tikzpicture}[baseline=-2]
\draw (-1,0.2) -- (0.5,0.2);
\draw (-1,0) -- (0.5,0);
\draw (-1,-0.2) -- (0.5,-0.2);
\filldraw [fill=white] (0,0) circle [radius=4mm] ;
\draw ($(0,0)+(90:0.4)$) to [out = -90, in = 10] ($(0,0)+(190:0.4)$) ;
\filldraw [fill=white] (-0.7,0.1) circle [radius=2mm];
\node (O1) at (-0.2,0.2) {$+$} ;
\node (O2) at (-0.7,0.1) {$-$} ;
\end{tikzpicture}
\,,
\end{split}
\label{discsub}
\end{align}
where the labels $(\pm)$ only refer to the ways of approach of the specified subenergy variable, namely $s_{12}=-(p_1+p_2)^2$ and $s_{45}=-(p_4+p_5)^2$ with the label 
\scalebox{0.6}{
\begin{tikzpicture}[baseline=-2]
\node (a) at (1,0.3) {1};
\node (b) at (1,0) {2};
\node (c) at (1,-0.3) {3};
\node (d) at (-1,0.3) {4};
\node (e) at (-1,0) {5};
\node (f) at (-1,-0.3) {6};
\draw (a) -- (d);
\draw (b) -- (e);
\draw (c) -- (f);
\filldraw [fill=white] (0,0) circle [radius=5mm] ;
\node (O) at (0,0) {} ;
\end{tikzpicture}
}. The last term of \eqref{unitarity33} is the sum of the one-particle singularities in cross-energy variables, e.g.,~$s_{145}=-(p_1-p_4-p_5)^2$.

\begin{figure}[t]
\centering
\begin{tikzpicture}
\draw[->] (-1.5,0) -- (4,0);
\coordinate (P1) at (3,-0.8) node at (P1) [right] {$P$} ;
\coordinate (P2) at (3,0.8) node at (P2) [right] {$P'$} ;
\fill[red] (P1) circle (2pt) (P2) circle (2pt);
\node (S) at (3.8,1.1) {$s$};
\draw  (3.6,1.3) |- (4,0.9);
\draw[->, dashed] (-1,0.2) -- (0,0.2) ;
\draw[dashed] (0.05,0.2) [rounded corners] -- (1,0.2) -- (1.4,0) ;
\draw[->, dotted, red] (1.4,0) -- (2.9,-0.75);
\draw[->, dashed] (-1,-0.2) -- (0,-0.2) ;
\draw[dashed] (0.05,-0.2) [rounded corners] -- (1,-0.2) -- (1.4,0) ;
\draw[->, dotted, red] (1.4,0) -- (2.9, 0.75);
\draw[->, dashed] (-1.2,0) |- (2.8,-0.8) ;
\node (e1) at (0,0.4) {$+i\varepsilon$};
\node (e2) at (0,-0.4) {$-i\varepsilon$};
\draw[red, decoration = {zigzag,segment length = 3mm, amplitude = 1mm},decorate] (0,0)--(4,0);
\draw[red] (-0.1,-0.1) -- (0.1,0.1) ;
\draw[red] (-0.1,0.1) -- (0.1,-0.1) ;
\node at (1, 0.4) {$(a)$};
\node at (1, -0.4) {$(b)$};
\node at (-1.5,-0.4) {$(c)$};
\end{tikzpicture}
\caption{The positions of the complex pole $P$ at $s=M^2$ and the shadow pole at $s=(M^2)^*$ where the symbol $\times$ denotes the branch point. The black dashed lines are the paths on the physical sheet while the red dotted lines are on the unphysical sheet. Depending on the path, we do or do not reach the complex poles: the paths $(a)$ and $(b)$ reach the complex pole $M^2$ and the shadow pole $(M^2)^*$ whereas the path $(c)$ stays on the physical sheet on which no complex poles exist.}
\label{fig:complexpole}
\end{figure}
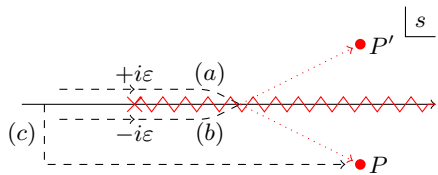

A resonance of an unstable particle can be explained by a complex pole $s=M^2$ residing on the unphysical sheet that is reached from the physical region by going down through the branch cut (see Fig.~\ref{fig:complexpole}). The existence of a complex pole is indeed predicted by unitarity~\cite{gunson1960unstable}. Hermitian analyticity implies that another pole, called a shadow pole, also exists at the complex-conjugate position $s=(M^2)^*$ which is reached from the opposite boundary by going up through the cut. 
%\footnote{It is necessary to decide which poles to discuss, as there may be other shadow poles on different sheets. Throughout, we shall refer to $s=M^2$ and $s=(M^2)^*$ as the pole that lies closest to the physical region, namely the complex pole responsible for the resonance, and the pole at its complex conjugate position, respectively. }
Therefore, when one continues the $(\pm)$ amplitudes from a real $s$ to $s=M^2$ by going down, the analytic continuation of $\Amp^{(+)}_{22}$ has a singularity while $\Amp^{(-)}_{22}$ is regular since $s$ stays on the physical sheet for $\Amp^{(-)}_{22}$ [see the path $(c)$]. Conversely, when one moves $s$ up to reach $s=(M^2)^*$, $\Amp^{(+)}_{22}$ is regular whereas $\Amp^{(-)}_{22}$ is singular, respectively. 

We then analytically continue the 3-to-3 amplitude by moving the subenergy variable $s_{12}=-(p_1+p_2)^2$ from the upper-half plane and investigate the neighbourhood of the complex pole $s_{12}=M^2$,
\begin{align}
\begin{tikzpicture}[baseline=-2]
\draw (0.6,0.2) -- (-0.6,0.2);
\draw (0.6,0) -- (-0.6,0);
\draw (0.6,-0.2) -- (-0.6,-0.2);
\filldraw [fill=white] (0,0) circle [radius=4mm] ;
\draw ($(0,0)+(90:0.4)$) to [out = -90, in = 170] ($(0,0)+(-10:0.4)$) ;
\node (O) at (0.2,0.2) {$+$} ;
\end{tikzpicture}
\sim
\begin{tikzpicture}[baseline=-2]
\draw (0,0.2) -- (-0.6,0.2);
\draw (0,0) -- (-0.6,0);
\draw (1.5,-0.2) -- (-0.6,-0.2);
\draw (1.5, 0.2) -- (1,0.2);
\draw (1.5, 0) -- (1, 0);
\draw  [decorate, decoration={snake}] (1,0.1) -- (0, 0.1) ;
\filldraw [fill=white] (0,0) circle [radius=4mm] ;
\filldraw [fill=white] (1.1, 0.1) circle [radius=2mm] ;
\draw ($(0,0)+(90:0.4)$) to [out = -90, in = 170] ($(0,0)+(-10:0.4)$) ;
\node (O) at (0.2,0.2) {$+$} ;
\node (O2) at (1.1,0.1) {$+$} ;
\end{tikzpicture}
\,, 
\label{Amp33unstable1}
\end{align}
where we add the label $(+)$ to cover the $1$ and $2$ external lines only since we have only specified the direction of the analytic continuation of the variable $s_{12}$. The symbol $\sim$ means that we have picked out the singular terms and the wavy line connecting to the $(+)$ bubble in the right diagram is understood as the pole factor $(s_{12}-M^2)^{-1}$. The existence of the singularity at $s_{12}=M^2$ is deduced from \eqref{discsub} since the small $(+)$ bubble has a pole at $s_{12}=M^2$ when the 2-to-2 amplitude does. On the other hand, we can consider a different path of the analytic continuation to reach the shadow pole
\begin{align}
\begin{tikzpicture}[baseline=-2]
\draw (0.6,0.2) -- (-0.6,0.2);
\draw (0.6,0) -- (-0.6,0);
\draw (0.6,-0.2) -- (-0.6,-0.2);
\filldraw [fill=white] (0,0) circle [radius=4mm] ;
\draw ($(0,0)+(90:0.4)$) to [out = -90, in = 170] ($(0,0)+(-10:0.4)$) ;
\node (O) at (0.2,0.2) {$-$} ;
\end{tikzpicture}
\sim
\begin{tikzpicture}[baseline=-2]
\draw (0,0.2) -- (-0.6,0.2);
\draw (0,0) -- (-0.6,0);
\draw (1.5,-0.2) -- (-0.6,-0.2);
\draw (1.5, 0.2) -- (1,0.2);
\draw (1.5, 0) -- (1, 0);
\draw  [decorate, decoration={snake}] (1,0.1) -- (0, 0.1) ;
\filldraw [fill=white] (0,0) circle [radius=4mm] ;
\filldraw [fill=white] (1.1, 0.1) circle [radius=2mm] ;
\draw ($(0,0)+(90:0.4)$) to [out = -90, in = 170] ($(0,0)+(-10:0.4)$) ;
\node (O) at (0.2,0.2) {$-$} ;
\node (O2) at (1.1,0.1) {$-$} ;
\end{tikzpicture}
\,,
\label{Amp33unstable2}
\end{align}
in which the wavy line connecting to the $(-)$ bubble should be understood as $[s_{12}-(M^2)^*]^{-1}$. The residue is computed by considering an integration contour encircling the pole according to Cauchy's residue theorem. The 2-to-3 amplitude with an external unstable particle is then defined by the residue at either $s_{12}=M^2$ or $s_{12}=(M^2)^*$ after subtracting the three-point coupling constant
\scalebox{0.6}{\begin{tikzpicture}[baseline=-2]
\draw (1.5, 0.3) -- (1,0.3);
\draw (1.5, 0) -- (1, 0);
\draw  [decorate, decoration={snake}] (1,0.15) -- (0.4, 0.15) ;
\filldraw [fill=white] (1.1, 0.15) circle [radius=2.5mm] ;
\end{tikzpicture}}.
Similarly, we can define the 2-to-2 amplitudes representing a ``scattering'' of a stable particle and an unstable particle which are diagrammatically denoted by
\begin{align}
\begin{tikzpicture}[baseline=-2]
\draw  [decorate, decoration={snake}] (-0.7,0.15) -- (0, 0.15) ;
\draw  [decorate, decoration={snake}] (0.7,0.15) -- (0, 0.15) ;
\draw (0.7,-0.15) -- (-0.7,-0.15);
\filldraw [fill=white] (0, 0.0) circle [radius=4mm] ;
\draw ($(0,0)+(85:0.4)$) to [out = -90, in = 180] ($(0,0)+(0:0.4)$) ;
\draw ($(0,0)+(95:0.4)$) to [out = -90, in = 0] ($(0,0)+(180:0.4)$) ;
\node (O) at (0.24,0.2) {$+$} ;
\node (O) at (-0.24,0.2) {$+$} ;
\end{tikzpicture}
\,, \quad
\begin{tikzpicture}[baseline=-2]
\draw  [decorate, decoration={snake}] (-0.7,0.15) -- (0, 0.15) ;
\draw  [decorate, decoration={snake}] (0.7,0.15) -- (0, 0.15) ;
\draw (0.7,-0.15) -- (-0.7,-0.15);
\filldraw [fill=white] (0, 0.0) circle [radius=4mm] ;
\draw ($(0,0)+(85:0.4)$) to [out = -90, in = 180] ($(0,0)+(0:0.4)$) ;
\draw ($(0,0)+(95:0.4)$) to [out = -90, in = 0] ($(0,0)+(180:0.4)$) ;
\node (O) at (0.24,0.2) {$-$} ;
\node (O) at (-0.24,0.2) {$+$} ;
\end{tikzpicture}
\,.
\label{unstableamplitudes}
\end{align}
We stress that the definition of the unstable-particle amplitude is not unique as it depends on the choice of the complex pole. In the literature (e.g.,~\cite{Hannesdottir:2022bmo}), the unstable-particle amplitudes are defined by the all-$(+)$ amplitudes [the former one of \eqref{unstableamplitudes}] because $M^2$ is the pole approached from the causal direction. However, there is no need to use the all-$(+)$ amplitudes to investigate the S-matrix constraints since we are interested in the analytically-continued amplitudes which no longer describe physical scattering processes. In fact, as we will show, unitarity equations take a simpler form in the case of the mixed type [the latter one of \eqref{unstableamplitudes}] and the mixed-type would be more useful for studying unitarity constraints.

\section{Unitarity equations}
\label{sec:unitarity}
\subsection{$A\varphi \to A \varphi$}
\label{sec:Aphi}
Using the above properties, let us find the unitarity equation of the unstable-particle scattering. We start with the unitarity equation of the 3-to-3 amplitude and consider its residue at the complex poles of the subenergy variables $s_{12}$ and $s_{45}$ with the label
\scalebox{0.6}{
\begin{tikzpicture}[baseline=-2]
\node (a) at (1,0.3) {1};
\node (b) at (1,0) {2};
\node (c) at (1,-0.3) {3};
\node (d) at (-1,0.3) {4};
\node (e) at (-1,0) {5};
\node (f) at (-1,-0.3) {6};
\draw (a) -- (d);
\draw (b) -- (e);
\draw (c) -- (f);
\filldraw [fill=white] (0,0) circle [radius=5mm] ;
\node (O) at (0,0) {} ;
\end{tikzpicture}
}. The path of analytic continuation should be chosen to reach the correct sheet on which the complex pole exists (see Fig.~\ref{fig:complexpole}). Since the complex pole is responsible for the resonance, there should be paths from $\ReM\pm i\varepsilon $ to the poles $M^2$ and $(M^2)^*$ without a non-analytic change of the unitarity equation. In other words, we deal with the complex pole that is the singularity closest to the real boundaries. Hence, we first consider the unitarity equation in the vicinity of $s_{12}=s_{45}= \ReM $ and then continue it by complexifying $s_{12}$ and $s_{45}$. 

\begin{figure}[t]
\centering
 \includegraphics[width=0.8\linewidth]{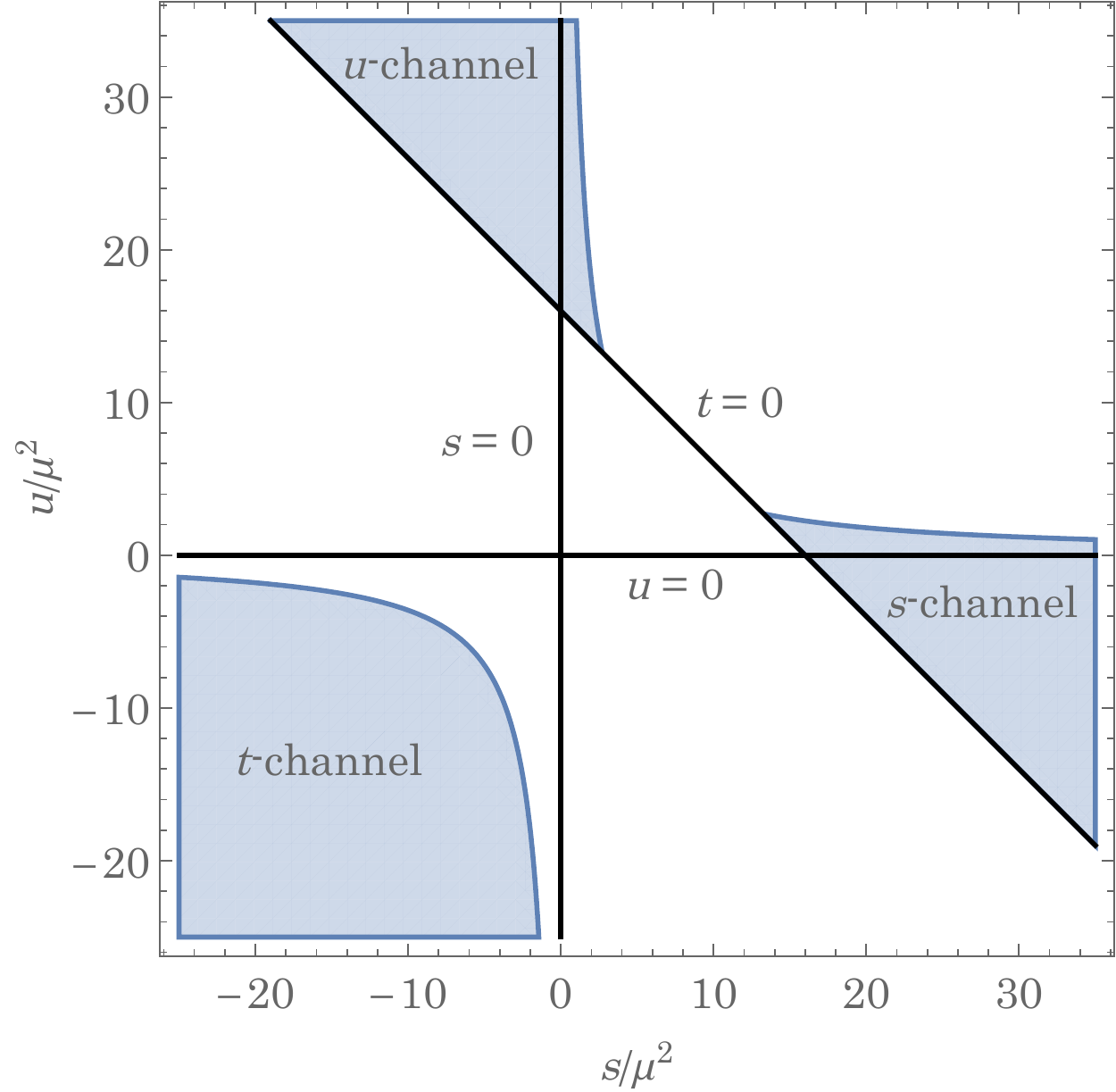}	
\caption{The physical regions in terms of $s,t,u$ for $s_{12}=s_{45}=7\mu^2$. The three isolated regions correspond to the physical regions for the $s$-channel process $123 \to 456$ (right), the $u$-channel process $126 \to 345$ (top) and the $t$-channel process $1245 \to 36$ (left), respectively.
}
\label{fig:physical}
\end{figure}

The variables $s=s_{123}=s_{456}, u=s_{126}=s_{345}$ and $t=s_{1245}=s_{36}$ play the role of the standard Mandelstam variables after extracting the residue where $s,t,u$ are subject to the constraint $s+t+u=2\mu^2+s_{12}+s_{45}$. We should understand the physical region in which the unitarity equations are first applied. For fixed $\{s_{12}, s_{45}\}$, the physical region of $\{s,t,u\}$ can be identified by using \eqref{phys1} and \eqref{phys2}, e.g.,~putting $I=12,J=3,K=45$ in \eqref{phys2}. The constraint is shown in Fig.~\ref{fig:physical} where the physical region of the process $123\to456$ (the $s$-channel process) corresponds to the right region. The variable $u$ can take a positive value even in the $s$-channel region; thus, the $u$-channel normal threshold can be present in the $s$-channel unitarity equation. On the other hand, $t$ is always negative in the $s$-channel region and the $t$-channel normal threshold would not appear. We thus regard the amplitude as a function of $s, u, s_{12}, s_{45}$ and variables that are irrelevant to the residue.

When $(2\mu)^2<\ReM<(3\mu)^2$, we can use \eqref{unitarity33} in $s<(4\mu)^2$. On the l.h.s.~of the unitarity equation~\eqref{unitarity33}, the variables $s_{12}$ and $s_{45}$ in
\scalebox{0.7}{
\begin{tikzpicture}[baseline=-2]
\draw (0.6,0.2) -- (-0.6,0.2);
\draw (0.6,0) -- (-0.6,0);
\draw (0.6,-0.2) -- (-0.6,-0.2);
\filldraw [fill=white] (0,0) circle [radius=4mm] ;
\node (O) at (0,0) {$+$} ;
\end{tikzpicture}}
and
\scalebox{0.7}{
\begin{tikzpicture}[baseline=-2]
\draw (0.6,0.2) -- (-0.6,0.2);
\draw (0.6,0) -- (-0.6,0);
\draw (0.6,-0.2) -- (-0.6,-0.2);
\filldraw [fill=white] (0,0) circle [radius=4mm] ;
\node (O) at (0,0) {$-$} ;
\end{tikzpicture}}
are situated in different positions on the Riemann surface. We use the discontinuity equations of the subenergies \eqref{discsub} to set them to the same position:
\begin{align}
&\quad
\begin{tikzpicture}[baseline=-2]
\draw (0.6,0.2) -- (-0.6,0.2);
\draw (0.6,0) -- (-0.6,0);
\draw (0.6,-0.2) -- (-0.6,-0.2);
\filldraw [fill=white] (0,0) circle [radius=4mm] ;
\draw ($(0,0)+(90:0.4)$) to [out = -90, in = 170] ($(0,0)+(-10:0.4)$) ;
\node (O1) at (0.2,0.2) {$-$} ;
\node (O2) at (-0.05,-0.05) {$+$} ;
\end{tikzpicture}
-
\begin{tikzpicture}[baseline=-2]
\draw (0.6,0.2) -- (-0.6,0.2);
\draw (0.6,0) -- (-0.6,0);
\draw (0.6,-0.2) -- (-0.6,-0.2);
\filldraw [fill=white] (0,0) circle [radius=4mm] ;
\draw ($(0,0)+(90:0.4)$) to [out = -90, in = 10] ($(0,0)+(190:0.4)$) ;
\node (O1) at (-0.2,0.2) {$+$} ;
\node (O2) at (0.05,-0.05) {$-$} ;
\end{tikzpicture}
\nn
&=
\begin{tikzpicture}[baseline=-2]
\draw (-1,0.2) -- (-0.5,0.2);
\draw (-1,0) -- (-0.5,0);
\draw (-1,-0.2) -- (-0.5,-0.2);
\draw (1,0.2) -- (0.5,0.2);
\draw (1,0) -- (0.5,0);
\draw (1,-0.2) -- (0.5,-0.2);
\draw (-0.5,0.1) -- (0.5,0.1) ;
\draw (-0.5,-0.1) -- (0.5,-0.1) ;
\filldraw [fill=white] (-0.5,0) circle [radius=4mm] ;
\node (O1) at (-0.5,0) {$+$} ;
\filldraw [fill=white] (0.5,0) circle [radius=4mm] ;
\node (O2) at (0.5,0) {$-$} ;
\end{tikzpicture}
+
\begin{tikzpicture}[baseline=-2]
\draw (-1,0.2) -- (1,0.2);
\draw (-1,0) -- (1,0);
\draw (-1,-0.2) -- (1,-0.2);
\filldraw [fill=white] (-0.5,0) circle [radius=4mm] ;
\node (O1) at (-0.5,0) {$+$} ;
\filldraw [fill=white] (0.5,0) circle [radius=4mm] ;
\node (O2) at (0.5,0) {$-$} ;
\end{tikzpicture}
\nn
&+\left( \sum 
\begin{tikzpicture}[baseline=-2]
\draw (-1,0.2) -- (0.6,0.2);
\draw (-1,0) -- (0.6,0);
\draw (-1,-0.2) -- (0.6,-0.2);
\filldraw [fill=white] (-0.5,0) circle [radius=4mm] ;
\node (O1) at (-0.5,0) {$+$} ;
\filldraw [fill=white] (0.3,0.1) circle [radius=2mm] ;
\node (O2) at (0.3,0.1) {$-$} ;
\end{tikzpicture}
\right)
-
\begin{tikzpicture}[baseline=-2]
\draw (-1,0.2) -- (0.6,0.2);
\draw (-1,0) -- (0.6,0);
\draw (-1,-0.2) -- (0.6,-0.2);
\filldraw [fill=white] (-0.5,0) circle [radius=4mm] ;
\node (O1) at (-0.5,0) {$+$} ;
\filldraw [fill=white] (0.3,0.1) circle [radius=2mm] ;
\node (O2) at (0.3,0.1) {$-$} ;
\end{tikzpicture}
\nn
&+\left(\sum 
\begin{tikzpicture}[baseline=-2]
\draw (1,0.2) -- (-0.6,0.2);
\draw (1,0) -- (-0.6,0);
\draw (1,-0.2) -- (-0.6,-0.2);
\filldraw [fill=white] (0.5,0) circle [radius=4mm] ;
\node (O1) at (0.5,0) {$-$} ;
\filldraw [fill=white] (-0.3,0.1) circle [radius=2mm] ;
\node (O2) at (-0.3,0.1) {$+$} ;
\end{tikzpicture}
\right)
-
\begin{tikzpicture}[baseline=-2]
\draw (1,0.2) -- (-0.6,0.2);
\draw (1,0) -- (-0.6,0);
\draw (1,-0.2) -- (-0.6,-0.2);
\filldraw [fill=white] (0.5,0) circle [radius=4mm] ;
\node (O1) at (0.5,0) {$-$} ;
\filldraw [fill=white] (-0.3,0.1) circle [radius=2mm] ;
\node (O2) at (-0.3,0.1) {$+$} ;
\end{tikzpicture}
+ \sum
\begin{tikzpicture}[baseline=-2]
\draw (0.6,0.2) -- (-0.6,0.2);
\draw (0.6,0) -- (0.3,0);
\draw (-0.6,0) -- (-0.3,0);
\draw (-0.3,0.1) -- (0.3,-0.1);
\draw (0.6,-0.2) -- (-0.6,-0.2);
\filldraw [fill=white] (-0.3,0.1) circle [radius=2mm] ;
\node (O1) at (-0.3,0.1) {$+$} ;
\filldraw [fill=white] (0.3,-0.1) circle [radius=2mm] ;
\node (O2) at (0.3,-0.1) {$-$} ;
\end{tikzpicture}
\nn
%%%%%%%%%%%%%%%%%%%%%%%%
%%%%%%%%%%%%%%%%%%%%%%%%
%%%%%%%%%%%%%%%%%%%%%%%%
%%%%%%%%%%%%%%%%%%%%%%%%
&=
\begin{tikzpicture}[baseline=-2]
\draw (-1,0.2) -- (-0.5,0.2);
\draw (-1,0) -- (-0.5,0);
\draw (-1,-0.2) -- (-0.5,-0.2);
\draw (1,0.2) -- (0.5,0.2);
\draw (1,0) -- (0.5,0);
\draw (1,-0.2) -- (0.5,-0.2);
\draw (-0.5,0.1) -- (0.5,0.1) ;
\draw (-0.5,-0.1) -- (0.5,-0.1) ;
\filldraw [fill=white] (-0.5,0) circle [radius=4mm] ;
\node (O1) at (-0.5,0) {$+$} ;
\filldraw [fill=white] (0.5,0) circle [radius=4mm] ;
\node (O2) at (0.5,0) {$-$} ;
\end{tikzpicture}
+
\begin{tikzpicture}[baseline=-2]
\draw (-1,0.2) -- (1,0.2);
\draw (-1,0) -- (1,0);
\draw (-1,-0.2) -- (1,-0.2);
\filldraw [fill=white] (-0.5,0) circle [radius=4mm] ;
\node (O1) at (-0.5,0) {$+$} ;
\filldraw [fill=white] (0.5,0) circle [radius=4mm] ;
\node (O2) at (0.5,0) {$-$} ;
\end{tikzpicture}
+
\begin{tikzpicture}[baseline=-2]
\draw (-0.8,0.2) -- (-0.4,0.2) ;
\draw (-0.8,0) -- (-0.4,0) ;
\draw (0.8,0.2) -- (0.4,0.2) ;
\draw (0.8,0) -- (0.4,0) ;
\draw (-0.4,0.1) -- (0.4,0.1) ;
\draw (-0.8,-0.3) -- (0.4,-0.15);
\draw (0.8,-0.3) -- (-0.4, -0.15);
\filldraw [fill=white] (-0.4,0.1) circle [radius=3mm] ;
\node (O1) at (-0.4,0.1) {$+$} ;
\filldraw [fill=white] (0.4,0.1) circle [radius=3mm] ;
\node (O2) at (0.4,0.1) {$-$} ;
\end{tikzpicture}
+R\,.
\label{unitarity33disc}
\end{align}
Here, $R$ denotes the diagrams that either the 1 and 2 lines or the 4 and 5 lines do not connect to a single $(\pm)$ bubble which may not possess the complex pole of either $s_{12}$ or $s_{45}$. 
As $\ReM$ or $s$ increases, the number of internal lines increases as long as kinematically allowed. Therefore, the general expression is
\begin{align}
&\quad
\begin{tikzpicture}[baseline=-2]
\draw (0.6,0.2) -- (-0.6,0.2);
\draw (0.6,0) -- (-0.6,0);
\draw (0.6,-0.2) -- (-0.6,-0.2);
\filldraw [fill=white] (0,0) circle [radius=4mm] ;
\draw ($(0,0)+(90:0.4)$) to [out = -90, in = 170] ($(0,0)+(-10:0.4)$) ;
\node (O1) at (0.2,0.2) {$-$} ;
\node (O2) at (-0.05,-0.05) {$+$} ;
\end{tikzpicture}
-
\begin{tikzpicture}[baseline=-2]
\draw (0.6,0.2) -- (-0.6,0.2);
\draw (0.6,0) -- (-0.6,0);
\draw (0.6,-0.2) -- (-0.6,-0.2);
\filldraw [fill=white] (0,0) circle [radius=4mm] ;
\draw ($(0,0)+(90:0.4)$) to [out = -90, in = 10] ($(0,0)+(190:0.4)$) ;
\node (O1) at (-0.2,0.2) {$+$} ;
\node (O2) at (0.05,-0.05) {$-$} ;
\end{tikzpicture}
\nn
&=
\sum_{a}
\begin{tikzpicture}[baseline=-2]
\draw (-1,0.2) -- (-0.5,0.2);
\draw (-1,0) -- (-0.5,0);
\draw (-1,-0.2) -- (-0.5,-0.2);
\draw (1,0.2) -- (0.5,0.2);
\draw (1,0) -- (0.5,0);
\draw (1,-0.2) -- (0.5,-0.2);
\draw (-0.5,0.3) -- (0.5,0.3) ;
\draw (-0.5,-0.3) -- (0.5,-0.3) ;
\draw (-0.5,0.1) -- (0.5, 0.1);
\draw[dotted] (-0.5, -0.1 ) -- (0.5,-0.1) ;
\filldraw [fill=white] (-0.5,0) circle [radius=4mm] ;
\node (O1) at (-0.5,0) {$+$} ;
\filldraw [fill=white] (0.5,0) circle [radius=4mm] ;
\node (O2) at (0.5,0) {$-$} ;
\node at (0,-0.45) {\scalebox{0.7}{$a$} };
\end{tikzpicture}
+\sum_{b}
\begin{tikzpicture}[baseline=-2]
\draw (-0.8,0.2) -- (-0.4,0.2) ;
\draw (-0.8,0) -- (-0.4,0) ;
\draw (0.8,0.2) -- (0.4,0.2) ;
\draw (0.8,0) -- (0.4,0) ;
\draw (-0.4,0.25) -- (0.4,0.25) ;
\draw (-0.4,0.15) -- (0.4,0.15);
\draw[dotted] (-0.4,0.05) -- (0.4,0.05) ;
\draw (-0.4,-0.05) -- (0.4,-0.05) ;
\draw (-0.8,-0.3) -- (0.4,-0.15);
\draw (0.8,-0.3) -- (-0.4, -0.15);
\filldraw [fill=white] (-0.4,0.1) circle [radius=3mm] ;
\node (O1) at (-0.4,0.1) {$+$} ;
\filldraw [fill=white] (0.4,0.1) circle [radius=3mm] ;
\node (O2) at (0.4,0.1) {$-$} ;
\node at (0,0.4) {\scalebox{0.7}{$b$} };
\end{tikzpicture}
+R
\nn
&=
\begin{tikzpicture}[baseline=-2]
\draw (-1,0.2) -- (-0.5,0.2);
\draw (-1,0) -- (-0.5,0);
\draw (-1,-0.2) -- (-0.5,-0.2);
\draw (1,0.2) -- (0.5,0.2);
\draw (1,0) -- (0.5,0);
\draw (1,-0.2) -- (0.5,-0.2);
\draw[line width=5pt] (-0.5,0) -- (0.5,0);
\filldraw [fill=white] (-0.5,0) circle [radius=4mm] ;
\node (O1) at (-0.5,0) {$+$} ;
\filldraw [fill=white] (0.5,0) circle [radius=4mm] ;
\node (O2) at (0.5,0) {$-$} ;
\end{tikzpicture}
+
\begin{tikzpicture}[baseline=-2]
\draw (-0.8,0.2) -- (-0.4,0.2) ;
\draw (-0.8,0) -- (-0.4,0) ;
\draw (0.8,0.2) -- (0.4,0.2) ;
\draw (0.8,0) -- (0.4,0) ;
\draw[line width=5pt] (-0.4, 0.1) -- (0.4,0.1);
\draw (-0.8,-0.3) -- (0.4,-0.15);
\draw (0.8,-0.3) -- (-0.4, -0.15);
\filldraw [fill=white] (-0.4,0.1) circle [radius=3mm] ;
\node (O1) at (-0.4,0.1) {$+$} ;
\filldraw [fill=white] (0.4,0.1) circle [radius=3mm] ;
\node (O2) at (0.4,0.1) {$-$} ;
\end{tikzpicture}
+R
\label{unitarity33n}
\end{align}
where $a,b$ are the number of the internal lines and the bold line is the shorthand for the summation.
We then move $s_{12}$ and $s_{45}$ by going up/down to reach $s_{12}=(M^2)^*$ and $s_{45}=M^2$ at which  $R$ may be regular in either $s_{12}$ or $s_{45}$,
leading to
\begin{align}
&\quad
\begin{tikzpicture}[baseline=-2]
\draw (1.1,-0.2) -- (-1.1,-0.2);
\draw (-1.1,0.2) -- (-0.8,0.2) ;
\draw (-1.1,0) -- (-0.8,0);
\draw  [decorate, decoration={snake}] (-0.8,0.1) -- (0, 0.1) ;
\draw (1.1,0.2) -- (0.8,0.2) ;
\draw (1.1,0) -- (0.8,0);
\draw  [decorate, decoration={snake}] (0.8,0.1) -- (0, 0.1) ;
\filldraw [fill=white] (0,0) circle [radius=4mm] ;
\filldraw [fill=white] (-0.8,0.1) circle [radius=2mm] ;
\filldraw [fill=white] (0.8,0.1) circle [radius=2mm] ;
\draw ($(0,0)+(90:0.4)$) to [out = -90, in = 170] ($(0,0)+(-10:0.4)$) ;
\node (O1) at (0.2,0.2) {$-$} ;
\node (O2) at (-0.05,-0.05) {$+$} ;
\node (O3) at (-0.8,0.1) {$+$} ;
\node (O4) at (0.8,0.1) {$-$} ;
\end{tikzpicture}
-
\begin{tikzpicture}[baseline=-2]
\draw (1.1,-0.2) -- (-1.1,-0.2);
\draw (-1.1,0.2) -- (-0.8,0.2) ;
\draw (-1.1,0) -- (-0.8,0);
\draw  [decorate, decoration={snake}] (-0.8,0.1) -- (0, 0.1) ;
\draw (1.1,0.2) -- (0.8,0.2) ;
\draw (1.1,0) -- (0.8,0);
\draw  [decorate, decoration={snake}] (0.8,0.1) -- (0, 0.1) ;
\filldraw [fill=white] (0,0) circle [radius=4mm] ;
\filldraw [fill=white] (-0.8,0.1) circle [radius=2mm] ;
\filldraw [fill=white] (0.8,0.1) circle [radius=2mm] ;
\draw ($(0,0)+(90:0.4)$) to [out = -90, in = 10] ($(0,0)+(190:0.4)$) ;
\node (O1) at (-0.2,0.2) {$+$} ;
\node (O2) at (0.05,-0.05) {$-$} ;
\node (O3) at (-0.8,0.1) {$+$} ;
\node (O4) at (0.8,0.1) {$-$} ;
\end{tikzpicture}
\nn
&\sim
\begin{tikzpicture}[baseline=-2]
\draw (-1.6,-0.2) -- (-0.5,-0.2);
\draw (1.6,-0.2) -- (0.5,-0.2);
\draw[line width=5pt] (-0.5, 0) -- (0.5,0);
\draw (-1.6,0.2) -- (-1.3,0.2) ;
\draw (-1.6,0) -- (-1.3,0);
\draw  [decorate, decoration={snake}] (-1.3,0.1) -- (-0.5, 0.1) ;
\draw (1.6,0.2) -- (1.3,0.2) ;
\draw (1.6,0) -- (1.3,0);
\draw  [decorate, decoration={snake}] (1.3,0.1) -- (0.5, 0.1) ;
\filldraw [fill=white] (-0.5,0) circle [radius=4mm] ;
\node (O1) at (-0.5,0) {$+$} ;
\filldraw [fill=white] (0.5,0) circle [radius=4mm] ;
\node (O2) at (0.5,0) {$-$} ;
\filldraw [fill=white] (-1.3,0.1) circle [radius=2mm] ;
\filldraw [fill=white] (1.3,0.1) circle [radius=2mm] ;
\node (O3) at (-1.3,0.1) {$+$} ;
\node (O4) at (1.3,0.1) {$-$} ;
\end{tikzpicture}
+
\begin{tikzpicture}[baseline=-2]
\draw (-1.4,0.2) -- (-1.1,0.2) ;
\draw (-1.4,0) -- (-1.1,0) ;
\draw (1.4,0.2) -- (1.1,0.2) ;
\draw (1.4,0) -- (1.1,0) ;
\draw[line width=5pt] (-0.4, 0.1) -- (0.4,0.1);
\draw (-1.4,-0.3) -- (0.4,-0.2);
\draw (1.4,-0.3) -- (-0.4, -0.2);
\draw  [decorate, decoration={snake}] (-0.4,0.1) -- (-1.1, 0.1) ;
\draw  [decorate, decoration={snake}] (0.4,0.1) -- (1.1, 0.1) ;
\filldraw [fill=white] (-0.4,0.1) circle [radius=3mm] ;
\node (O1) at (-0.4,0.1) {$+$} ;
\filldraw [fill=white] (0.4,0.1) circle [radius=3mm] ;
\node (O2) at (0.4,0.1) {$-$} ;
\filldraw [fill=white] (-1.1,0.1) circle [radius=2mm] ;
\node at (-1.1,0.1) {$+$} ;
\filldraw [fill=white] (1.1,0.1) circle [radius=2mm] ;
\node at (1.1,0.1) {$-$} ;
\end{tikzpicture}
\,. \label{unitarity33to22}
\end{align}
Cancelling the common factors, we obtain the unitarity equation of the unstable-particle scattering as the residue of the 3-to-3 unitarity equation:
\begin{align} 
 \AmixL - \AmixR
=
\begin{tikzpicture}[baseline=-2]
\draw (-1.2,-0.2) -- (-0.5,-0.2);
\draw (1.2,-0.2) -- (0.5,-0.2);
\draw[line width=5pt] (-0.5, 0) -- (0.5,0);
\draw  [decorate, decoration={snake}] (-1.2,0.1) -- (-0.5, 0.1) ;
\draw  [decorate, decoration={snake}] (1.2,0.1) -- (0.5, 0.1) ;
\filldraw [fill=white] (-0.5,0) circle [radius=4mm] ;
\node (O1) at (-0.5,0) {$+$} ;
\filldraw [fill=white] (0.5,0) circle [radius=4mm] ;
\node (O2) at (0.5,0) {$-$} ;
\end{tikzpicture}
+
\begin{tikzpicture}[baseline=-2]
\draw[line width=5pt] (-0.4, 0.1) -- (0.4,0.1);
\draw (-1,-0.3) -- (0.4,-0.2);
\draw (1,-0.3) -- (-0.4, -0.2);
\draw  [decorate, decoration={snake}] (-1, 0.1) -- (-0.4,0.1)  ;
\draw  [decorate, decoration={snake}] (1, 0.1) -- (0.4,0.1);
\filldraw [fill=white] (-0.4,0.1) circle [radius=3mm] ;
\node (O1) at (-0.4,0.1) {$+$} ;
\filldraw [fill=white] (0.4,0.1) circle [radius=3mm] ;
\node (O2) at (0.4,0.1) {$-$} ;
\end{tikzpicture}
\,.
\label{unitarityAphisu}
\end{align}
Note that thanks to choosing the complex conjugate positions $s_{12}=(M^2)^*$ and $s_{45}=M^2$, the constraint $s+t+u=2\mu^2+2\ReM$ is unchanged.\footnote{If there is a stable-particle pole in the $s_{12}$- and $s_{45}$-planes, one may continue \eqref{unitarity33disc} to approach the stable poles $s_{12}=s_{45}=\mu^2$. Since the stable-particle poles should reside on the physical sheet, the continuations from the upper-half plane and the lower-half plane reach the same pole, meaning that there is no distinction in the $(\pm)$ labels attached to a single line. Then, one could reproduce the 2-to-2 unitarity equation as the residue of the 3-to-3 unitarity equation \eqref{unitarity33disc} although the constraint $s+t+u=2\mu^2+s_{12}+s_{45}$ should be taken care of; $t$ takes different values before and after analytic continuation.}

Let us discuss what the unitarity equation \eqref{unitarityAphisu} evaluates. We recall that the $(+)$ amplitude is related to the $(-)$ amplitude via the Hermitian conjugation $\Amp^{(-)}_{n'n} = (\Amp^{(+)}_{nn'})^*$, reading
\begin{align}
\Amp^{(-)}_{33}(s,u,s_{12},s_{45},\cdots ) = [\Amp^{(+)}_{33}(s,u,s_{45}^*,s_{12}^*,\cdots) ]^*
\end{align}
for $s,u \in \mathbb{R}$, $s_{12},s_{45} \in \mathbb{C}$, and $\cdots$ are variables irrelevant to the residue. We have used that $s$ and $u$ are invariant under the replacement $\{p_1, p_2, p_3 \} \leftrightarrow \{p_4, p_5,p_6\}$ while $s_{12}$ and $s_{45}$ are interchanged with complex conjugation. When $s_{12}$ and $s_{45}$ are located at complex-conjugate positions, $s_{45}^*=s_{12}$, we find
\begin{align}
\Amp^{(-)}_{33}(s,u,s_{12},s_{12}^*,\cdots ) = [\Amp^{(+)}_{33}(s,u,s_{12},s_{12}^*,\cdots) ]^*
\,.
\end{align}
Therefore, the l.h.s.~of the unitarity equation \eqref{unitarityAphisu} is the imaginary part of the amplitude: 
\begin{align}
\AmixL - \AmixR
=
2i {\rm Im} \left( \AmixL \right)
\,.
\end{align}

The unitarity equation \eqref{unitarityAphisu} is also understood as the sum of the $s$-channel discontinuity and the $u$-channel discontinuity.
Let us write
\begin{align}
\AmixL
&=\Amp_{A\varphi \to A\varphi} (s_+ ,u_+)
\label{Aphiplus}
\,,\\
\AmixR
&=\Amp_{A\varphi \to A\varphi} (s_- ,u_-)
\,,
\label{Aphiminus}
\end{align}
where $s_{\pm}=s\pm i \varepsilon$ and $u_{\pm}=u\pm i \varepsilon$ and $\lim_{\varepsilon \to +0}$ is understood. According to our analyticity postulate (the assumption \ref{axiom3}), the unitarity equation \eqref{unitarityAphisu} can be decomposed into a couple of equations:
\begin{align}
\begin{tikzpicture}[baseline=-2]
\draw (-1.2,-0.2) -- (-0.5,-0.2);
\draw (1.2,-0.2) -- (0.5,-0.2);
\draw[line width=5pt] (-0.5, 0) -- (0.5,0);
\draw  [decorate, decoration={snake}] (-1.2,0.1) -- (-0.5, 0.1) ;
\draw  [decorate, decoration={snake}] (1.2,0.1) -- (0.5, 0.1) ;
\filldraw [fill=white] (-0.5,0) circle [radius=4mm] ;
\node (O1) at (-0.5,0) {$+$} ;
\filldraw [fill=white] (0.5,0) circle [radius=4mm] ;
\node (O2) at (0.5,0) {$-$} ;
\end{tikzpicture}
&=
\Amp_{A\varphi \to A\varphi} (s_+,u_+)
-\Amp_{A\varphi \to A\varphi} (s_-,u_+)
\nn
&=
\Amp_{A\varphi \to A\varphi} (s_+,u_-)
-\Amp_{A\varphi \to A\varphi} (s_-,u_-)
\,,
\label{disc22s}
\\
\begin{tikzpicture}[baseline=-2]
\draw[line width=5pt] (-0.4, 0.1) -- (0.4,0.1);
\draw (-1,-0.3) -- (0.4,-0.2);
\draw (1,-0.3) -- (-0.4, -0.2);
\draw  [decorate, decoration={snake}] (-1, 0.1) -- (-0.4,0.1)  ;
\draw  [decorate, decoration={snake}] (1, 0.1) -- (0.4,0.1);
\filldraw [fill=white] (-0.4,0.1) circle [radius=3mm] ;
\node (O1) at (-0.4,0.1) {$+$} ;
\filldraw [fill=white] (0.4,0.1) circle [radius=3mm] ;
\node (O2) at (0.4,0.1) {$-$} ;
\end{tikzpicture}
&=
\Amp_{A\varphi \to A\varphi} (s_-,u_+)
-\Amp_{A\varphi \to A\varphi} (s_-,u_-)
\nn
&=\Amp_{A\varphi \to A\varphi} (s_+,u_+)
-\Amp_{A\varphi \to A\varphi} (s_+,u_-)
\,.
\label{disc22u}
\end{align}
This decomposition is related to the existence of a regular region. 
When $u$ is fixed in the region $u<4\mu^2, u\neq \mu^2$, the second term on r.h.s.~of \eqref{unitarityAphisu} is not kinematically allowed, leading to
\begin{align}
\begin{tikzpicture}[baseline=-2]
\draw (-1.2,-0.2) -- (-0.5,-0.2);
\draw (1.2,-0.2) -- (0.5,-0.2);
\draw[line width=5pt] (-0.5, 0) -- (0.5,0);
\draw  [decorate, decoration={snake}] (-1.2,0.1) -- (-0.5, 0.1) ;
\draw  [decorate, decoration={snake}] (1.2,0.1) -- (0.5, 0.1) ;
\filldraw [fill=white] (-0.5,0) circle [radius=4mm] ;
\node (O1) at (-0.5,0) {$+$} ;
\filldraw [fill=white] (0.5,0) circle [radius=4mm] ;
\node (O2) at (0.5,0) {$-$} ;
\end{tikzpicture}
&=
\Amp_{A\varphi \to A\varphi} (s_+,u_+)
-\Amp_{A\varphi \to A\varphi} (s_-,u_-)
\,.
\label{disc_smallu}
\end{align}
We then continue \eqref{disc22s} in the region $u<4\mu^2~(u\neq \mu^2)$ and compare it with \eqref{disc_smallu}. We conclude
\begin{align}
0&= 
\Amp_{A\varphi \to A\varphi} (s_-,u_+)
-\Amp_{A\varphi \to A\varphi} (s_-,u_-)
\nn
&=\Amp_{A\varphi \to A\varphi} (s_+,u_+)
-\Amp_{A\varphi \to A\varphi} (s_+,u_-)
\,,
\label{nodisc}
\end{align}
in the region $u<4\mu^2~(u\neq \mu^2)$, namely no discontinuity in $u$. The converse is also true: if \eqref{nodisc} is assumed in $u<4\mu^2,~u\neq \mu^2$, the continuation of \eqref{disc_smallu} to the region $u>4\mu^2$ leads to \eqref{disc22s} and \eqref{disc22u}.

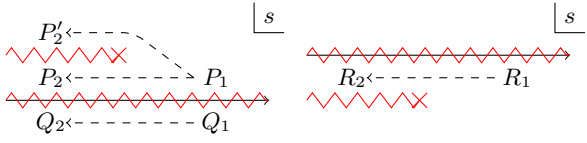
\begin{figure}[t]
\centering
\begin{tikzpicture}
\draw[->] (-1.5,0) -- (2,0);
\node at (2,1.1) {$s$};
\draw  (1.8,1.3) |- (2.2,0.9);
\draw[red, decoration = {zigzag,segment length = 3mm, amplitude = 1mm},decorate] (-1.5,0.6)--(0, 0.6);
\draw[red, decoration = {zigzag,segment length = 3mm, amplitude = 1mm},decorate] (-1.5,0)--(2,0);
\draw[red] ($(0,0.6)+(-0.1,-0.1)$) -- ($(0,0.6)+(0.1,0.1)$) ;
\draw[red] ($(0,0.6)+(-0.1,0.1)$) -- ($(0,0.6)+(0.1,-0.1)$) ;
\draw[->, dashed] (1,0.3) -- (-0.7,0.3);
\draw[->, dashed] (1,0.3) [rounded corners]-- (0.2,0.9) -- (-0.7,0.9);
\draw[->, dashed] (1,-0.3) -- (-0.7,-0.3);
\node at (1.3, 0.3) {$P_1$};
\node at (1.3, -0.3) {$Q_1$};
\node at (-0.9, 0.3) {$P_2$};
\node at (-0.9, -0.3) {$Q_2$};
\node at (-0.9, 0.9) {$P_2'$};
%%%%%%%%%%%%%%%%
%%%%%%%%%%%%%%%%
%%%%%%%%%%%%%%%%
\draw[->] ($(4,0) + (-1.5,0.6)$) -- ($(4,0) + (2,0.6)$);
\node at ($(4,0) + (2,1.1)$) {$s$};
\draw  ($(4,0) + (1.8,1.3)$) |- ($(4,0)+(2.2,0.9)$);
\draw[red, decoration = {zigzag,segment length = 3mm, amplitude = 1mm},decorate] ($(4,0) + (-1.5,0)$) --($(4,0) + (0, 0)$);
\draw[red, decoration = {zigzag,segment length = 3mm, amplitude = 1mm},decorate] ($(4,0) + (-1.5,0.6) $)--($(4,0) + (2,0.6) $);
\draw[red] ($(4,0) +(0,0)+(-0.1,-0.1)$) -- ($(4,0) +(0,0)+(0.1,0.1)$) ;
\draw[red] ($(4,0) +(0,0)+(-0.1,0.1)$) -- ($(4,0) +(0,0)+(0.1,-0.1)$) ;
\draw[->, dashed] ($(4,0) + (1,0.3)$) -- ($(4,0) + (-0.7,0.3) $);
\node at ($(4,0) + (1.3, 0.3)$) {$R_1$};
\node at ($(4,0) + (-0.9, 0.3)$) {$R_2$};
\end{tikzpicture}
\caption{Analytic structures (modulo poles) of $\Amp_{A\varphi \to A\varphi }(s,t_-)$ (left) and $\Amp_{A\varphi \to A\varphi }(s,t_+)$ (right) in the vicinity of the $s$-channel region. The zigzag line on the real axis is the $s$-channel branch cut while the zigzag lines above (left) and below (right) are the $u$-channel branch cut, respectively.}
\label{fig:s-plane}
\end{figure}

The amplitude is often regarded as a function of a single variable by fixing other variables. By using \eqref{disc22s} and \eqref{disc22u}, let us discuss the analytic structure of the fixed-$t$ amplitude $\Amp_{A\varphi\to A\varphi}(s)$ where $u$ is eliminated by the constraint $s+t+u=2\mu^2+2\ReM$. The $s$-channel singularity and the $u$-channel singularity coexist in the complex $s$-plane. When $t$ is fixed at $t_-=t_0-i\varepsilon'$ with $t_0<0$ and $\varepsilon'>0$, the analytic structure is given by Fig.~\ref{fig:s-plane} (left). In $s_1>2\ReM-2\mu^2-t_0$, the $u$-channel cut is absent and the discontinuity between $s_1+i\varepsilon$ ($P_1$ in Fig.~\ref{fig:s-plane}) and $s_1-i\varepsilon$ ($P_2$) is given by 
\scalebox{0.7}{
\begin{tikzpicture}[baseline=-2]
\draw (-1.2,-0.2) -- (-0.5,-0.2);
\draw (1.2,-0.2) -- (0.5,-0.2);
\draw[line width=5pt] (-0.5, 0) -- (0.5,0);
\draw  [decorate, decoration={snake}] (-1.2,0.1) -- (-0.5, 0.1) ;
\draw  [decorate, decoration={snake}] (1.2,0.1) -- (0.5, 0.1) ;
\filldraw [fill=white] (-0.5,0) circle [radius=4mm] ;
\node (O1) at (-0.5,0) {$+$} ;
\filldraw [fill=white] (0.5,0) circle [radius=4mm] ;
\node (O2) at (0.5,0) {$-$} ;
\end{tikzpicture}}.
We then continue the function to the region $s_2<2\ReM-2\mu^2-t_0$. Since ${\rm Im} u = -{\rm Im}(s+t)=-{\rm Im}s + \varepsilon'$, the lower-half $s$-plane always gives ${\rm Im} u>0$ (the path $Q_1 \to Q_2$). On the other hand, there exists a branch cut on the upper-half plane: the path $P_1 \to P_2~(0<{\rm Im}s<\varepsilon')$ corresponds to ${\rm Im}u>0$ and the path $P_1 \to P_2'~({\rm Im}s>\varepsilon')$ is ${\rm Im}u<0$. Hence, the discontinuity between $P_2'$ and $Q_2$ is given by
$\scalebox{0.7}{
\begin{tikzpicture}[baseline=-2]
\draw (-1.2,-0.2) -- (-0.5,-0.2);
\draw (1.2,-0.2) -- (0.5,-0.2);
\draw[line width=5pt] (-0.5, 0) -- (0.5,0);
\draw  [decorate, decoration={snake}] (-1.2,0.1) -- (-0.5, 0.1) ;
\draw  [decorate, decoration={snake}] (1.2,0.1) -- (0.5, 0.1) ;
\filldraw [fill=white] (-0.5,0) circle [radius=4mm] ;
\node (O1) at (-0.5,0) {$+$} ;
\filldraw [fill=white] (0.5,0) circle [radius=4mm] ;
\node (O2) at (0.5,0) {$-$} ;
\end{tikzpicture}}
-
\scalebox{0.7}{\begin{tikzpicture}[baseline=-2]
\draw[line width=5pt] (-0.4, 0.1) -- (0.4,0.1);
\draw (-1,-0.3) -- (0.4,-0.2);
\draw (1,-0.3) -- (-0.4, -0.2);
\draw  [decorate, decoration={snake}] (-1, 0.1) -- (-0.4,0.1)  ;
\draw  [decorate, decoration={snake}] (1, 0.1) -- (0.4,0.1);
\filldraw [fill=white] (-0.4,0.1) circle [radius=3mm] ;
\node (O1) at (-0.4,0.1) {$+$} ;
\filldraw [fill=white] (0.4,0.1) circle [radius=3mm] ;
\node (O2) at (0.4,0.1) {$-$} ;
\end{tikzpicture}}
$
which does not agree with \eqref{unitarityAphisu} due to the negative sign in the second term. A similar consideration gives the analytic structure for $t_+=t_0+i\varepsilon'$ as shown in Fig.~\ref{fig:s-plane} (right). The amplitude \scalebox{0.7}{\AmixL} is identified with the values of the analytically-continued amplitude $\Amp_{A\varphi \to A\varphi }(s,t_-)$ at $P_1$ and $P_2$ while \scalebox{0.7}{\AmixR} is the values of $\Amp_{A\varphi \to A\varphi }(s,t_+)$ at $R_1$ and $R_2$. In other words, \scalebox{0.7}{\AmixL} and \scalebox{0.7}{\AmixR} cannot be understood as opposite boundary values on one sheet.

\begin{figure}[t]
\centering
\begin{tikzpicture}
\draw[->] (-3,0) -- (3,0);
\node at (3,1.1) {$s$};
\draw  (2.8,1.3) |- (3.2,0.9);
\draw[red, decoration = {zigzag,segment length = 3mm, amplitude = 1mm},decorate] (-3,0.7)--(0, 0.7);
\draw[red, decoration = {zigzag,segment length = 3mm, amplitude = 1mm},decorate] (-1.5,0)--(3,0);
\draw[red] ($(0,0.7)+(-0.1,-0.1)$) -- ($(0,0.7)+(0.1,0.1)$) ;
\draw[red] ($(0,0.7)+(-0.1,0.1)$) -- ($(0,0.7)+(0.1,-0.1)$) ;
\draw[red] ($(-1.5,0)+(-0.1,-0.1)$) -- ($(-1.5,0)+(0.1,0.1)$) ;
\draw[red] ($(-1.5,0)+(-0.1,0.1)$) -- ($(-1.5,0)+(0.1,-0.1)$) ;
\draw[->, dashed] (1,0.3) [rounded corners] -- (-0.65, 0.3);
\draw[->, dashed] (-0.7,0.3) -- (-1.7,0.3) [rounded corners] -- (-1.9,0.1) [rounded corners] -- (-1.9,-0.1) [rounded corners] -- (-1.7,-0.3) -- (-0.75,-0.3);
\draw[->, dashed] (-0.7,-0.3) -- (0.95,-0.3) ;
\draw[dashed] (1,-0.3) -- (0.9,0);
\draw[->, dotted, red] (0.9,0) [rounded corners] -- (0.6, 1) -- (-0.65, 1);
\fill[black] (1,0.3) circle (1.5pt) (-0.7,0.3) circle (1.5pt) (-0.7,-0.3) circle (1.5pt) (1,-0.3) circle (1.5pt) (-0.7,1) circle (1.5pt);
\node at (1.2, 0.45) {$P_1$};
\node at (-0.95,0.45) {$P_2$};
\node at (-0.95,-0.5) {$Q_2$};
\node at (1.6, -0.5) {$Q_1=R_1$};
\node at (-0.95,1.15) {$R_2$};
\end{tikzpicture}
\caption{A path going round the $s$-channel branch point $(P_2\to Q_2)$ and the $u$-channel branch point $(Q_2 \to R_2)$. The black dashed curve and the red dotted curve are paths on the different sheets.}
\label{fig:s-plane2}
\end{figure}
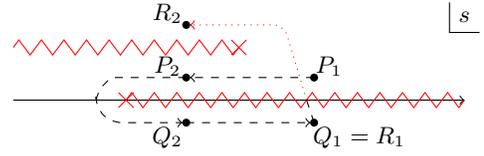

Let us investigate whether \scalebox{0.7}{\AmixL} and \scalebox{0.7}{\AmixR} can be still described by the same analytic function of the single variable.
We assume that the $s$-channel branch cut ends at $s=4\mu^2$ and the existence of a path drawn in Fig.~\ref{fig:s-plane2} for $t=t_-$. The points $P_1,P_2,Q_1,$ and $Q_2$ in Fig.~\ref{fig:s-plane2} are the same as the points in Fig.~\ref{fig:s-plane} (left). We go round the branch point $s=4\mu^2$ and reach $Q_1$ that corresponds to $s=s_-$ and $t=t_-$ with $s_1>2\ReM-2\mu^2-t_0$. Recall that $R_1$ in the right panel of Fig.~\ref{fig:s-plane} is at $s=s_-$ and $t=t_+$ with $s_1>2\ReM-2\mu^2-t_0$. Eq.~\eqref{nodisc} implies that there is no discontinuity in $t$ in the region $s_1>2\ReM-2\mu^2-t_0$. Therefore, $Q_1$ can be identified with $R_1$ under the limit $\varepsilon' \to 0$. Then, $R_2$ is reached by passing above the $u$-channel branch point as shown in Fig.~\ref{fig:s-plane2}. Note that we have only drawn the 2-particle thresholds in Fig.~\ref{fig:s-plane2} while other normal thresholds (and anomalous thresholds if any) have to be taken into account as well. $R_1 \to R_2$ should pass through the lower side of the $s$-channel branch points and the upper side of the $u$-channel branch points to be consistent with Fig.~\ref{fig:s-plane}.

Therefore, we should carefully discuss the unitarity equation to read the discontinuity in a single variable. Nonetheless, when one is interested in the discontinuity in the region $s>2\ReM-2\mu^2-t$, the r.h.s.~of the unitarity equation \eqref{unitarityAphi} is given by the $s$-channel discontinuity only. \scalebox{0.7}{\AmixL} and \scalebox{0.7}{\AmixR} are regarded as opposite boundary values, $P_1$ and $Q_1=R_1$ in Fig.~\ref{fig:s-plane2}. The unitarity equation simply reads
\begin{align}
\Disc_s \left( \AmixL \right)
=
\begin{tikzpicture}[baseline=-2]
\draw (-1.2,-0.2) -- (-0.5,-0.2);
\draw (1.2,-0.2) -- (0.5,-0.2);
\draw[line width=5pt] (-0.5, 0) -- (0.5,0);
\draw  [decorate, decoration={snake}] (-1.2,0.1) -- (-0.5, 0.1) ;
\draw  [decorate, decoration={snake}] (1.2,0.1) -- (0.5, 0.1) ;
\filldraw [fill=white] (-0.5,0) circle [radius=4mm] ;
\node (O1) at (-0.5,0) {$+$} ;
\filldraw [fill=white] (0.5,0) circle [radius=4mm] ;
\node (O2) at (0.5,0) {$-$} ;
\end{tikzpicture}
\,,
\label{unitarityAphi}
\end{align}
in $s>2\ReM-2\mu^2-t$.

If we postmultiply \eqref{unitarity33n} by
$
\scalebox{0.7}{
\begin{tikzpicture}[baseline=-2]
\draw (-0.3,0.2) -- (0.3,0.2);
\draw (-0.3,0) -- (0.3,0);
\draw (-0.3,-0.2) -- (0.3,-0.2);
\end{tikzpicture}}
+
\scalebox{0.7}{
\begin{tikzpicture}[baseline=-2]
\draw (-0.3,0.2) -- (0.3,0.2);
\draw (-0.3,0) -- (0.3,0);
\draw (-0.3,-0.2) -- (0.3,-0.2);
\filldraw [fill=white] (0,0.1) circle [radius=2mm] ;
\node at (0,0.1) {$+$};
\end{tikzpicture}}
$
and use the 2-particle discontinuity equations like \eqref{discsub} we obtain
\begin{align}
&\quad
\begin{tikzpicture}[baseline=-2]
\draw (0.6,0.2) -- (-0.6,0.2);
\draw (0.6,0) -- (-0.6,0);
\draw (0.6,-0.2) -- (-0.6,-0.2);
\filldraw [fill=white] (0,0) circle [radius=4mm] ;
\node (O2) at (0,0) {$+$} ;
\end{tikzpicture}
-
\begin{tikzpicture}[baseline=-2]
\draw (0.6,0.2) -- (-0.6,0.2);
\draw (0.6,0) -- (-0.6,0);
\draw (0.6,-0.2) -- (-0.6,-0.2);
\filldraw [fill=white] (0,0) circle [radius=4mm] ;
\draw ($(0,0)+(95:0.4)$) to [out = -90, in = 10] ($(0,0)+(190:0.4)$) ;
\draw ($(0,0)+(85:0.4)$) to [out = -90, in = 170] ($(0,0)+(-10:0.4)$) ;
\node at (-0.2,0.2) {$+$} ;
\node at (0.2,0.2) {$+$} ;
\node at (0,-0.1) {$-$} ;
\end{tikzpicture}
\nn
&=
\begin{tikzpicture}[baseline=-2]
\draw (-1,0.2) -- (-0.5,0.2);
\draw (-1,0) -- (-0.5,0);
\draw (-1,-0.2) -- (-0.5,-0.2);
\draw (1,0.2) -- (0.5,0.2);
\draw (1,0) -- (0.5,0);
\draw (1,-0.2) -- (0.5,-0.2);
\draw[line width=5pt] (-0.5,0) -- (0.5,0);
\filldraw [fill=white] (-0.5,0) circle [radius=4mm] ;
\node (O1) at (-0.5,0) {$+$} ;
\filldraw [fill=white] (0.5,0) circle [radius=4mm] ;
\node (O2) at (0.495,-0.05) {$-$} ;
\draw ($(0.5,0)+(90:0.4)$) to [out = -90, in = 170] ($(0.5,0)+(-10:0.4)$) ;
\node at (0.7,0.2) {$+$};
\end{tikzpicture}
+
\begin{tikzpicture}[baseline=-2]
\draw (-0.8,0.2) -- (-0.4,0.2) ;
\draw (-0.8,0) -- (-0.4,0) ;
\draw (0.8,0.2) -- (0.4,0.2) ;
\draw (0.8,0) -- (0.4,0) ;
\draw[line width=5pt] (-0.4, 0.1) -- (0.4,0.1);
\draw (-0.8,-0.3) -- (0.4,-0.15);
\draw (0.8,-0.3) -- (-0.4, -0.15);
\filldraw [fill=white] (-0.4,0.1) circle [radius=3mm] ;
\node (O1) at (-0.4,0.1) {$+$} ;
\filldraw [fill=white] (0.4,0.1) circle [radius=3mm] ;
\node (O2) at (0.3,0.1) {$-$} ;
\draw ($(0.4,0.1)+(60:0.3)$) to [out = -130, in = 130] ($(0.4,0.1)+(-60:0.3)$) ;
\node at (0.57,0.1) {$+$};
\end{tikzpicture}
+
\begin{tikzpicture}[baseline=-2]
\draw (0.6,0.2) -- (-0.6,0.2);
\draw (0.6,0) -- (-0.6,0);
\draw (0.6,-0.2) -- (-0.6,-0.2);
\filldraw [fill=white] (0,0) circle [radius=4mm] ;
\node (O2) at (0,0) {$R$} ;
\end{tikzpicture}
\left(
\begin{tikzpicture}[baseline=-2]
\draw (-0.3,0.2) -- (0.3,0.2);
\draw (-0.3,0) -- (0.3,0);
\draw (-0.3,-0.2) -- (0.3,-0.2);
\end{tikzpicture}
+
\begin{tikzpicture}[baseline=-2]
\draw (-0.3,0.2) -- (0.3,0.2);
\draw (-0.3,0) -- (0.3,0);
\draw (-0.3,-0.2) -- (0.3,-0.2);
\filldraw [fill=white] (0,0.1) circle [radius=2mm] ;
\node at (0,0.1) {$+$};
\end{tikzpicture}
\right)
\label{33unitarityplus}
\end{align}
where 
\scalebox{0.7}{\begin{tikzpicture}[baseline=-2]
\draw (0.6,0.2) -- (-0.6,0.2);
\draw (0.6,0) -- (-0.6,0);
\draw (0.6,-0.2) -- (-0.6,-0.2);
\filldraw [fill=white] (0,0) circle [radius=4mm] ;
\node (O2) at (0,0) {$R$} ;
\end{tikzpicture}}
is the diagrammatic notation for $R$. The last term includes, for instance, a triangle diagram
\begin{align}
\begin{tikzpicture}[baseline=-2]
\draw (0.6,0.2) -- (-0.6,0.2);
\draw (0.6,0) -- (0.3,0);
\draw (-0.6,0) -- (-0.3,0);
\draw (-0.3,0.1) -- (0.3,-0.1);
\draw (0.6,-0.2) -- (-0.6,-0.2);
\filldraw [fill=white] (-0.3,0.1) circle [radius=2mm] ;
\node (O1) at (-0.3,0.1) {$+$} ;
\filldraw [fill=white] (0.3,-0.1) circle [radius=2mm] ;
\node (O2) at (0.3,-0.1) {$-$} ;
\end{tikzpicture}
\times
\begin{tikzpicture}[baseline=-2]
\draw (-0.3,0.2) -- (0.3,0.2);
\draw (-0.3,0) -- (0.3,0);
\draw (-0.3,-0.2) -- (0.3,-0.2);
\filldraw [fill=white] (0,0.1) circle [radius=2mm] ;
\node at (0,0.1) {$+$};
\end{tikzpicture}
=
\begin{tikzpicture}[baseline=-2]
\draw (1,-0.2) -- (-1, -0.2) ;
\draw (1,0.2) -- (-1,0.2);
\draw (1,0) -- (0.6,0);
\draw (-1,0) -- (-0.6,0);
\draw (-0.6,0.1) -- (0,-0.1);
\draw (0.6,0.1) -- (0,-0.1);
\filldraw [fill=white] (-0.6,0.1) circle [radius=2mm] ;
\node at (-0.6,0.1) {$+$} ;
\filldraw [fill=white] (0,-0.1) circle [radius=2mm] ;
\node at (0,-0.1) {$-$} ;
\filldraw [fill=white] (0.6,0.1) circle [radius=2mm] ;
\node at (0.6,0.1) {$+$} ;
\end{tikzpicture}
\end{align}
which may contribute to the unitarity equation even after extracting the residue at $s_{12}=s_{45}=M^2$. \eqref{33unitarityplus} does not evaluate the imaginary part but can evaluate discontinuities. The form of the unitarity equation depends on the choice of the complex poles, suggesting that the analytic structures of
\scalebox{0.7}{\AmixL} and \scalebox{0.7}{\Auns{+}} are different. In Appendix~\ref{sec:triangle} we show that the analytic structure of the triangle Feynman diagram indeed depends on the positions of $s_{12}$ and $s_{45}$. 
One can also notice that even the first term on the r.h.s.~of \eqref{33unitarityplus}, which may represent the $s$-channel discontinuities, is not given by a sum of the products of $(+)$ amplitudes and $(-)$ amplitudes, differently from the standard unitarity equation. In general, the first term does not necessarily have a fixed sign.

We can also discuss the unitarity equation in crossed regions. The analysis in the $u$-channel region is the same as the previous discussion. Hence, we only consider the $t$-channel region starting with the $4\to 2$ unitarity equation 
\begin{align}
\begin{tikzpicture}[baseline=-2]
\draw (-0.6, 0.3) -- (0.6, 0.3);
\draw (0, 0.1) -- (0.6, 0.1);
\draw (0, -0.1) -- (0.6, -0.1);
\draw (-0.6, -0.3) -- (0.6, -0.3);
\filldraw [fill=white] (0,0) circle [radius=4mm] ;
\node at (0, 0) {$+$} ;
\end{tikzpicture}
-
\begin{tikzpicture}[baseline=-2]
\draw (-0.6, 0.3) -- (0.6, 0.3);
\draw (0, 0.1) -- (0.6, 0.1);
\draw (0, -0.1) -- (0.6, -0.1);
\draw (-0.6, -0.3) -- (0.6, -0.3);
\filldraw [fill=white] (0,0) circle [radius=4mm] ;
\node at (0, 0) {$-$} ;
\end{tikzpicture}
&=
\begin{tikzpicture}[baseline=-2]
\draw (-0.5, 0.3) -- (-1, 0.3);
\draw (-0.5, -0.3) -- (-1, -0.3);
\draw (0.5, 0.3) -- (1, 0.3);
\draw (0.5, 0.1) -- (1, 0.1);
\draw (0.5, -0.1) -- (1, -0.1);
\draw (0.5, -0.3) -- (1, -0.3);
\draw[line width=5pt] (-0.5, 0) -- (0.5, 0);
\filldraw [fill=white] (-0.5, 0) circle [radius=4mm] ;
\node at (-0.5 , 0) {$+$} ;
\filldraw [fill=white] (0.5, 0) circle [radius=4mm] ;
\node at (0.5, 0) {$-$} ;
\end{tikzpicture}
+
\begin{tikzpicture}[baseline=-2]
\draw (-0.5, 0.3) -- (-1, 0.3);
\draw (-0.5, -0.3) -- (-1, -0.3);
\draw (0.2, 0.3) -- (0.5, 0.3);
\draw (0.2, 0.1) -- (0.5, 0.1);
\draw (-0.5, -0.1) -- (0.5, -0.1);
\draw (-0.5, -0.3) -- (0.5, -0.3);
\draw[line width=5pt] (-0.5, 0.2) -- (0.3, 0.2);
\filldraw [fill=white] (-0.5, 0) circle [radius=4mm] ;
\node at (-0.5 , 0) {$+$} ;
\filldraw [fill=white] (0.2, 0.2) circle [radius=2mm] ;
\node at (0.2, 0.2) {$-$} ;
\end{tikzpicture}
\nn
&
+
\begin{tikzpicture}[baseline=-2]
\draw (-0.5, 0.3) -- (-1, 0.3);
\draw (-0.5, -0.3) -- (-1, -0.3);
\draw (0.2, -0.3) -- (0.5, -0.3);
\draw (0.2, -0.1) -- (0.5, -0.1);
\draw (-0.5, 0.1) -- (0.5, 0.1);
\draw (-0.5, 0.3) -- (0.5, 0.3);
\draw[line width=5pt] (-0.5, -0.2) -- (0.3, -0.2);
\filldraw [fill=white] (-0.5, 0) circle [radius=4mm] ;
\node at (-0.5 , 0) {$+$} ;
\filldraw [fill=white] (0.2, -0.2) circle [radius=2mm] ;
\node at (0.2, -0.2) {$-$} ;
\end{tikzpicture}
-
\begin{tikzpicture}[baseline=-2]
\draw (-0.5, 0.3) -- (-1, 0.3);
\draw (-0.5, -0.3) -- (-1, -0.3);
\draw (0.2, -0.3) -- (0.5, -0.3);
\draw (0.2, -0.1) -- (0.5, -0.1);
\draw[line width=5pt] (-0.5, -0.2) -- (0.3, -0.2);
\draw (0.2, 0.3) -- (0.5, 0.3);
\draw (0.2, 0.1) -- (0.5, 0.1);
\draw[line width=5pt] (-0.5, 0.2) -- (0.3, 0.2);
\filldraw [fill=white] (-0.5, 0) circle [radius=4mm] ;
\node at (-0.5 , 0) {$+$} ;
\filldraw [fill=white] (0.18, 0.2) circle [radius=1.8mm] ;
\node at (0.2, 0.2) {$-$} ;
\filldraw [fill=white] (0.18, -0.2) circle [radius=1.8mm] ;
\node at (0.2, -0.2) {$-$} ;
\end{tikzpicture}
+R
\,,
\end{align}
where the lines are labelled as
\begin{align*}
\begin{tikzpicture}[baseline=-2]
\draw (-0.6, 0.3) -- (0.6, 0.3);
\draw (0, 0.1) -- (0.6, 0.1);
\draw (0, -0.1) -- (0.6, -0.1);
\draw (-0.6, -0.3) -- (0.6, -0.3);
\filldraw [fill=white] (0,0) circle [radius=4mm] ;
\node at (0.7, 0.3) {$\scalebox{0.7}{1}$};
\node at (0.7, 0.1) {$\scalebox{0.7}{2}$};
\node at (0.7, -0.1) {$\scalebox{0.7}{4}$};
\node at (0.7, -0.3) {$\scalebox{0.7}{5}$};
\node at (-0.7, 0.3) {$\scalebox{0.7}{3}$};
\node at (-0.7, -0.3) {$\scalebox{0.7}{6}$};
\end{tikzpicture}
\,.
\end{align*}
Using the 2-particle discontinuity equations twice, we find~\cite{olive1965unitarity, Eden:1966dnq}
\begin{align}
&\quad
\begin{tikzpicture}[baseline=-2]
\draw (0.6,0.3) -- (-0.6,0.3);
\draw (0.6,0.1) -- (-0.6,0.1);
\draw (0.6,-0.1) -- (-0.6,-0.1);
\draw (0.6,-0.3) -- (-0.6,-0.3);
\filldraw [fill=white] (0,0) circle [radius=4mm] ;
\draw ($(0,0)+(90:0.4)$) to [out = -90, in = 180] ($(0,0)+(0:0.4)$) ;
\draw ($(0,0)+(-90:0.4)$) to [out = 90, in = 180] ($(0,0)+(0:0.4)$) ;
\node (O1) at (0.2,0.2) {$+$} ;
\node (O2) at (0.2,-0.2) {$+$} ;
\end{tikzpicture}
-
\begin{tikzpicture}[baseline=-2]
\draw (0.6,0.3) -- (-0.6,0.3);
\draw (0.6,0.1) -- (-0.6,0.1);
\draw (0.6,-0.1) -- (-0.6,-0.1);
\draw (0.6,-0.3) -- (-0.6,-0.3);
\filldraw [fill=white] (0,0) circle [radius=4mm] ;
\draw ($(0,0)+(90:0.4)$) to [out = -90, in = 180] ($(0,0)+(0:0.4)$) ;
\draw ($(0,0)+(-90:0.4)$) to [out = 90, in = 180] ($(0,0)+(0:0.4)$) ;
\node (O1) at (0.2,0.2) {$-$} ;
\node (O2) at (0.2,-0.2) {$-$} ;
\end{tikzpicture}
\nn
&=
\begin{tikzpicture}[baseline=-2]
\draw (-1,0.3) -- (-0.5,0.3);
\draw (-1,0.1) -- (-0.5,0.1);
\draw (-1,-0.1) -- (-0.5,-0.1);
\draw (-1,-0.3) -- (-0.5,-0.3);
\draw (0.6,0.3) -- (0.3,0.3);
\draw (0.6,0.1) -- (0.3,0.1);
\draw (0.6,-0.1) -- (-0.5,-0.1);
\draw (0.6,-0.3) -- (-0.5,-0.3);
\draw[line width=5pt] (-0.5,0.2) -- (0.3,0.2) ;
\filldraw [fill=white] (-0.5,0) circle [radius=4mm] ;
\draw ($(-0.5,0)+(90:0.4)$) to [out = -90, in = 180] ($(-0.5,0)+(0:0.4)$) ;
\draw ($(-0.5,0)+(-90:0.4)$) to [out = 90, in = 180] ($(-0.5,0)+(0:0.4)$) ;
\node (O1) at (-0.3,0.2) {$+$} ;
\node (O2) at (-0.3,-0.2) {$+$} ;
\filldraw [fill=white] (0.3,0.2) circle [radius=1.8mm] ;
\node (O2) at (0.3,0.2) {$-$} ;
\end{tikzpicture}
+
\begin{tikzpicture}[baseline=-2]
\draw (-1,0.3) -- (-0.5,0.3);
\draw (-1,0.1) -- (-0.5,0.1);
\draw (-1,-0.1) -- (-0.5,-0.1);
\draw (-1,-0.3) -- (-0.5,-0.3);
\draw (0.6,0.3) -- (-0.5,0.3);
\draw (0.6,0.1) -- (-0.5,0.1);
\draw (0.6,-0.1) -- (0.3,-0.1);
\draw (0.6,-0.3) -- (0.3,-0.3);
\draw[line width=5pt] (-0.5,-0.2) -- (0.3,-0.2) ;
\filldraw [fill=white] (-0.5,0) circle [radius=4mm] ;
\draw ($(-0.5,0)+(90:0.4)$) to [out = -90, in = 180] ($(-0.5,0)+(0:0.4)$) ;
\draw ($(-0.5,0)+(-90:0.4)$) to [out = 90, in = 180] ($(-0.5,0)+(0:0.4)$) ;
\node (O1) at (-0.3,0.2) {$+$} ;
\node (O2) at (-0.3,-0.2) {$+$} ;
\filldraw [fill=white] (0.3,-0.2) circle [radius =1.8mm] ;
\node (O3) at (0.3,-0.2) {$-$} ;
\end{tikzpicture}
-
\begin{tikzpicture}[baseline=-2]
\draw (-1,0.3) -- (-0.5,0.3);
\draw (-1,0.1) -- (-0.5,0.1);
\draw (-1,-0.1) -- (-0.5,-0.1);
\draw (-1,-0.3) -- (-0.5,-0.3);
\draw (0.6,0.3) -- (0.3,0.3);
\draw (0.6,0.1) -- (0.3,0.1);
\draw (0.6,-0.1) -- (0.3,-0.1);
\draw (0.6,-0.3) -- (0.3,-0.3);
\draw[line width=5pt] (-0.5,0.2) -- (0.3,0.2) ;
\draw[line width=5pt] (-0.5,-0.2) -- (0.3,-0.2) ;
\filldraw [fill=white] (-0.5,0) circle [radius=4mm] ;
\draw ($(-0.5,0)+(90:0.4)$) to [out = -90, in = 180] ($(-0.5,0)+(0:0.4)$) ;
\draw ($(-0.5,0)+(-90:0.4)$) to [out = 90, in = 180] ($(-0.5,0)+(0:0.4)$) ;
\node (O1) at (-0.3,0.2) {$+$} ;
\node (O2) at (-0.3,-0.2) {$+$} ;
\filldraw [fill=white] (0.3,0.2) circle [radius=1.8mm] ;
\filldraw [fill=white] (0.3,-0.2) circle [radius =1.8mm] ;
\node (O2) at (0.3,0.2) {$-$} ;
\node (O3) at (0.3,-0.2) {$-$} ;
\end{tikzpicture}
\,,
\label{discsub2}
\end{align}
which can be used to move $s_{12}$ and $s_{45}$ to the lower-half plane:
\begin{align}
\begin{tikzpicture}[baseline=-2]
\draw (-0.6, 0.3) -- (0.6, 0.3);
\draw (0, 0.1) -- (0.6, 0.1);
\draw (0, -0.1) -- (0.6, -0.1);
\draw (-0.6, -0.3) -- (0.6, -0.3);
\filldraw [fill=white] (0,0) circle [radius=4mm] ;
\node at (-0.1, 0) {$+$} ;
\draw ($(0,0)+(90:0.4)$) to [out = -90, in = 180] ($(0,0)+(0:0.4)$) ;
\draw ($(0,0)+(-90:0.4)$) to [out = 90, in = 180] ($(0,0)+(0:0.4)$) ;
\node (O1) at (0.2,0.2) {$-$} ;
\node (O2) at (0.2,-0.2) {$-$} ;
\end{tikzpicture}
-
\begin{tikzpicture}[baseline=-2]
\draw (-0.6, 0.3) -- (0.6, 0.3);
\draw (0, 0.1) -- (0.6, 0.1);
\draw (0, -0.1) -- (0.6, -0.1);
\draw (-0.6, -0.3) -- (0.6, -0.3);
\filldraw [fill=white] (0,0) circle [radius=4mm] ;
\node at (0, 0) {$-$} ;
\end{tikzpicture}
&=
\begin{tikzpicture}[baseline=-2]
\draw (-0.5, 0.3) -- (-1, 0.3);
\draw (-0.5, -0.3) -- (-1, -0.3);
\draw (0.5, 0.3) -- (1, 0.3);
\draw (0.5, 0.1) -- (1, 0.1);
\draw (0.5, -0.1) -- (1, -0.1);
\draw (0.5, -0.3) -- (1, -0.3);
\draw[line width=5pt] (-0.5, 0) -- (0.5, 0);
\filldraw [fill=white] (-0.5, 0) circle [radius=4mm] ;
\node at (-0.5 , 0) {$+$} ;
\filldraw [fill=white] (0.5, 0) circle [radius=4mm] ;
\node at (0.5, 0) {$-$} ;
\end{tikzpicture}
+R
\,.
\end{align}
Hence, the unitarity equation for $s_{12}=s_{45}=(M^2)^*$ takes a form similar to the standard 2-to-2 unitarity equation. 
On the other hand, one can replace $(-)$ with $(+)$ in the subenergy variable $s_{45}$ by following the discussion in \eqref{33unitarityplus}. One then finds the unitarity equation for
\scalebox{0.7}{
\begin{tikzpicture}[baseline=-2]
\draw (-0.6,-0.2) -- (0, -0.2) ;
\draw  [decorate, decoration={snake}] (0.7,-0.2) -- (0, -0.2) ;
\draw (-0.6,0.2) -- (0, 0.2) ;
\draw  [decorate, decoration={snake}] (0.7,0.2) -- (0, 0.2) ;
\filldraw [fill=white] (0,0) circle [radius=4mm] ;
\draw ($(0,0)+(90:0.4)$) to [out = -90, in = 180] ($(0,0)+(0:0.4)$) ;
\node (O1) at (0.2,0.2) {$-$} ;
\node (O3) at (-0.05,-0.05) {$+$} ;
\end{tikzpicture}}
that may have additional contributions from $R$.

%%%%%%%%%%%%%%%%%%%%%%%%%%%%%%
%%%%%%%%%%%%%%%%%%%%%%%%%%%%%%
%%%%%%%%%%%%%%%%%%%%%%%%%%%%%%
%%%%%%%%%%%%%%%%%%%%%%%%%%%%%%
%%%%%%%%%%%%%%%%%%%%%%%%%%%%%%
%%%%%%%%%%%%%%%%%%%%%%%%%%%%%%

\subsection{$AA\to AA$}
\label{sec:AA}
The previous analysis can be extended into amplitudes involving more unstable particles although the analysis becomes more complicated.
Let us briefly discuss the $AA\to AA$ amplitude which arises as a residue of the 4-to-4 amplitude
\begin{align*}
\begin{tikzpicture}[baseline=-2]
\draw (-0.6, 0.3) -- (0.6, 0.3);
\draw (-0.6, 0.1) -- (0.6, 0.1);
\draw (-0.6, -0.1) -- (0.6, -0.1);
\draw (-0.6, -0.3) -- (0.6, -0.3);
\filldraw [fill=white] (0,0) circle [radius=4mm] ;
\node at (0.7, 0.3) {$\scalebox{0.7}{1}$};
\node at (0.7, 0.1) {$\scalebox{0.7}{2}$};
\node at (0.7, -0.1) {$\scalebox{0.7}{3}$};
\node at (0.7, -0.3) {$\scalebox{0.7}{4}$};
\node at (-0.7, 0.3) {$\scalebox{0.7}{5}$};
\node at (-0.7, 0.1) {$\scalebox{0.7}{6}$};
\node at (-0.7, -0.1) {$\scalebox{0.7}{7}$};
\node at (-0.7, -0.3) {$\scalebox{0.7}{8}$};
\end{tikzpicture}
\,.
\end{align*}
When we set $s_{12}=s_{34}=s_{56}=s_{78}=\ReM$, the physical $s$-channel region is given by $s> 4\ReM, t< 0, u< 0$ where $s=s_{1234}=s_{5678}, t=s_{1256}=s_{3478}$ and $u=s_{1278}=s_{3456}$ with the constraint $s+t+u=s_{12}+s_{34}+s_{56}+s_{78}$.

For simplicity, we assume $(2\mu)^2 < \ReM<(3\mu)^2$. The 4-to-4 unitarity equation reads
\begin{align}
&
\Afour{+} - \Afour{-}
\nn
&=
\begin{tikzpicture}[baseline=-2]
\draw (-1,0.3) -- (-0.5,0.3);
\draw (-1,0.1) -- (-0.5,0.1);
\draw (-1,-0.1) -- (-0.5,-0.1);
\draw (-1,-0.3) -- (-0.5,-0.3);
\draw (1,0.3) -- (0.5,0.3);
\draw (1,0.1) -- (0.5,0.1);
\draw (1,-0.1) -- (0.5,-0.1);
\draw (1,-0.3) -- (0.5,-0.3);
\draw[line width=5pt] (-0.5,0) -- (0.5,0) ;
\filldraw [fill=white] (-0.5,0) circle [radius=4mm] ;
\node (O1) at (-0.5,0) {$+$} ;
\filldraw [fill=white] (0.5,0) circle [radius=4mm] ;
\node (O2) at (0.5,0) {$-$} ;
\end{tikzpicture}
+
\begin{tikzpicture}[baseline=-2]
\draw (-1,0.3) -- (0.6,0.3);
\draw (-1,0.1) -- (0.6,0.1);
\draw (-1,-0.1) -- (0.6,-0.1);
\draw (-1,-0.3) -- (0.6,-0.3);
\filldraw [fill=white] (-0.5,0) circle [radius=4mm] ;
\node (O1) at (-0.5,0) {$+$} ;
\filldraw [fill=white] (0.3,0.2) circle [radius=1.8mm] ;
\node (O2) at (0.3,0.2) {$-$} ;
\end{tikzpicture}
+
\begin{tikzpicture}[baseline=-2]
\draw (-1,0.3) -- (0.6,0.3);
\draw (-1,0.1) -- (0.6,0.1);
\draw (-1,-0.1) -- (0.6,-0.1);
\draw (-1,-0.3) -- (0.6,-0.3);
\filldraw [fill=white] (-0.5,0) circle [radius=4mm] ;
\node (O1) at (-0.5,0) {$+$} ;
\filldraw [fill=white] (0.3,-0.2) circle [radius =1.8mm] ;
\node (O3) at (0.3,-0.2) {$-$} ;
\end{tikzpicture}
-
\begin{tikzpicture}[baseline=-2]
\draw (-1,0.3) -- (0.6,0.3);
\draw (-1,0.1) -- (0.6,0.1);
\draw (-1,-0.1) -- (0.6,-0.1);
\draw (-1,-0.3) -- (0.6,-0.3);
\filldraw [fill=white] (-0.5,0) circle [radius=4mm] ;
\node (O1) at (-0.5,0) {$+$} ;
\filldraw [fill=white] (0.3,0.2) circle [radius=1.8mm] ;
\filldraw [fill=white] (0.3,-0.2) circle [radius =1.8mm] ;
\node (O2) at (0.3,0.2) {$-$} ;
\node (O3) at (0.3,-0.2) {$-$} ;
\end{tikzpicture}
\nn
&
+
\begin{tikzpicture}[baseline=-2]
\draw (1,0.3) -- (-0.6,0.3);
\draw (1,0.1) -- (-0.6,0.1);
\draw (1,-0.1) -- (-0.6,-0.1);
\draw (1,-0.3) -- (-0.6,-0.3);
\filldraw [fill=white] (0.5,0) circle [radius=4mm] ;
\node (O1) at (0.5,0) {$-$} ;
\filldraw [fill=white] (-0.3,0.2) circle [radius=1.8mm] ;
\node (O2) at (-0.3,0.2) {$+$} ;
\end{tikzpicture}
+
\begin{tikzpicture}[baseline=-2]
\draw (1,0.3) -- (-0.6,0.3);
\draw (1,0.1) -- (-0.6,0.1);
\draw (1,-0.1) -- (-0.6,-0.1);
\draw (1,-0.3) -- (-0.6,-0.3);
\filldraw [fill=white] (0.5,0) circle [radius=4mm] ;
\node (O1) at (0.5,0) {$-$} ;
\filldraw [fill=white] (-0.3,-0.2) circle [radius =1.8mm] ;
\node (O3) at (-0.3,-0.2) {$+$} ;
\end{tikzpicture}
+
\begin{tikzpicture}[baseline=-2]
\draw (1,0.3) -- (-0.6,0.3);
\draw (1,0.1) -- (-0.6,0.1);
\draw (1,-0.1) -- (-0.6,-0.1);
\draw (1,-0.3) -- (-0.6,-0.3);
\filldraw [fill=white] (0.5,0) circle [radius=4mm] ;
\node (O1) at (0.5,0) {$-$} ;
\filldraw [fill=white] (-0.3,0.2) circle [radius=1.8mm] ;
\filldraw [fill=white] (-0.3,-0.2) circle [radius =1.8mm] ;
\node (O2) at (-0.3,0.2) {$+$} ;
\node (O3) at (-0.3,-0.2) {$+$} ;
\end{tikzpicture}
+
\begin{tikzpicture}[baseline=-2]
\draw (0.65,0.3) -- (0.35,0.3);
\draw (0.65,0.1) -- (0.35,0.1);
\draw (0.65,-0.3) -- (0.35,-0.3);
\draw (0.65,-0.1) -- (0.35,-0.1);
\draw (-0.65,0.3) -- (-0.35,0.3);
\draw (-0.65,0.1) -- (-0.35,0.1);
\draw (-0.65,-0.3) -- (-0.35,-0.3);
\draw (-0.65,-0.1) -- (-0.35,-0.1);
\draw (0.35,0.2) -- (-0.35,0.2);
\draw (0.35,-0.2) -- (-0.35, -0.2);
\draw (0.35,0.2) -- (-0.35,-0.2);
\draw (-0.35,0.2) -- (0.35,-0.2);
\filldraw [fill=white] (0.35,0.2) circle [radius =1.8mm] ;
\filldraw [fill=white] (0.35,-0.2) circle [radius =1.8mm] ;
\filldraw [fill=white] (-0.35,0.2) circle [radius =1.8mm] ;
\filldraw [fill=white] (-0.35,-0.2) circle [radius =1.8mm] ;
\node at (0.35,0.2) {$-$} ;
\node at (0.35,-0.2) {$-$} ;
\node at (-0.35,0.2) {$+$} ;
\node at (-0.35,-0.2) {$+$} ;
\end{tikzpicture}
+R
\,,
\label{unitarity44}
\end{align}
where $(2\mu)^2 \leq s_{12},s_{34},s_{56},s_{78}<(3\mu)^2$ is assumed which restricts the number of lines connecting small bubbles. We then use \eqref{discsub2}-type discontinuity equations to set $s_{12}=s_{34}=\ReM-i\varepsilon$ and $s_{56}=s_{78}=\ReM + i \varepsilon$ in  \eqref{unitarity44}. There are many cancellations, yielding
\begin{align}
&
\begin{tikzpicture}[baseline=-2]
\draw (0.6,0.3) -- (-0.6,0.3);
\draw (0.6,0.1) -- (-0.6,0.1);
\draw (0.6,-0.1) -- (-0.6,-0.1);
\draw (0.6,-0.3) -- (-0.6,-0.3);
\filldraw [fill=white] (0,0) circle [radius=4mm] ;
\draw ($(0,0)+(90:0.4)$) to [out = -90, in = 180] ($(0,0)+(0:0.4)$) ;
\draw ($(0,0)+(-90:0.4)$) to [out = 90, in = 180] ($(0,0)+(0:0.4)$) ;
\node (O1) at (0.2,0.2) {$-$} ;
\node (O2) at (0.2,-0.2) {$-$} ;
\node (O3) at (-0.1,0) {$+$} ;
\end{tikzpicture}
-
\begin{tikzpicture}[baseline=-2]
\draw (0.6,0.3) -- (-0.6,0.3);
\draw (0.6,0.1) -- (-0.6,0.1);
\draw (0.6,-0.1) -- (-0.6,-0.1);
\draw (0.6,-0.3) -- (-0.6,-0.3);
\filldraw [fill=white] (0,0) circle [radius=4mm] ;
\draw ($(0,0)+(90:0.4)$) to [out = -90, in = 0] ($(0,0)+(180:0.4)$) ;
\draw ($(0,0)+(-90:0.4)$) to [out = 90, in = 0] ($(0,0)+(180:0.4)$) ;
\node (O1) at (-0.2,0.2) {$+$} ;
\node (O2) at (-0.2,-0.2) {$+$} ;
\node (O3) at (0.1,0) {$-$} ;
\end{tikzpicture}
=
\begin{tikzpicture}[baseline=-2]
\draw (-1,0.3) -- (-0.5,0.3);
\draw (-1,0.1) -- (-0.5,0.1);
\draw (-1,-0.1) -- (-0.5,-0.1);
\draw (-1,-0.3) -- (-0.5,-0.3);
\draw (1,0.3) -- (0.5,0.3);
\draw (1,0.1) -- (0.5,0.1);
\draw (1,-0.1) -- (0.5,-0.1);
\draw (1,-0.3) -- (0.5,-0.3);
\draw[line width=5pt] (-0.5,0) -- (0.5,0) ;
\filldraw [fill=white] (-0.5,0) circle [radius=4mm] ;
\node (O1) at (-0.5,0) {$+$} ;
\filldraw [fill=white] (0.5,0) circle [radius=4mm] ;
\node (O2) at (0.5,0) {$-$} ;
\end{tikzpicture}
+
\begin{tikzpicture}[baseline=-2]
\draw (0.65,0.3) -- (0.35,0.3);
\draw (0.65,0.1) -- (0.35,0.1);
\draw (0.65,-0.3) -- (0.35,-0.3);
\draw (0.65,-0.1) -- (0.35,-0.1);
\draw (-0.65,0.3) -- (-0.35,0.3);
\draw (-0.65,0.1) -- (-0.35,0.1);
\draw (-0.65,-0.3) -- (-0.35,-0.3);
\draw (-0.65,-0.1) -- (-0.35,-0.1);
\draw (0.35,0.2) -- (-0.35,0.2);
\draw (0.35,-0.2) -- (-0.35, -0.2);
\draw (0.35,0.2) -- (-0.35,-0.2);
\draw (-0.35,0.2) -- (0.35,-0.2);
\filldraw [fill=white] (0.35,0.2) circle [radius =1.8mm] ;
\filldraw [fill=white] (0.35,-0.2) circle [radius =1.8mm] ;
\filldraw [fill=white] (-0.35,0.2) circle [radius =1.8mm] ;
\filldraw [fill=white] (-0.35,-0.2) circle [radius =1.8mm] ;
\node at (0.35,0.2) {$-$} ;
\node at (0.35,-0.2) {$-$} ;
\node at (-0.35,0.2) {$+$} ;
\node at (-0.35,-0.2) {$+$} ;
\end{tikzpicture}
+R
\,.
\label{unitarity44_2}
\end{align}
There exists a crossed box diagram on the r.h.s.~of the unitarity equation \eqref{unitarity44_2}.

\begin{figure}[t]
\centering
 \includegraphics[width=0.8\linewidth]{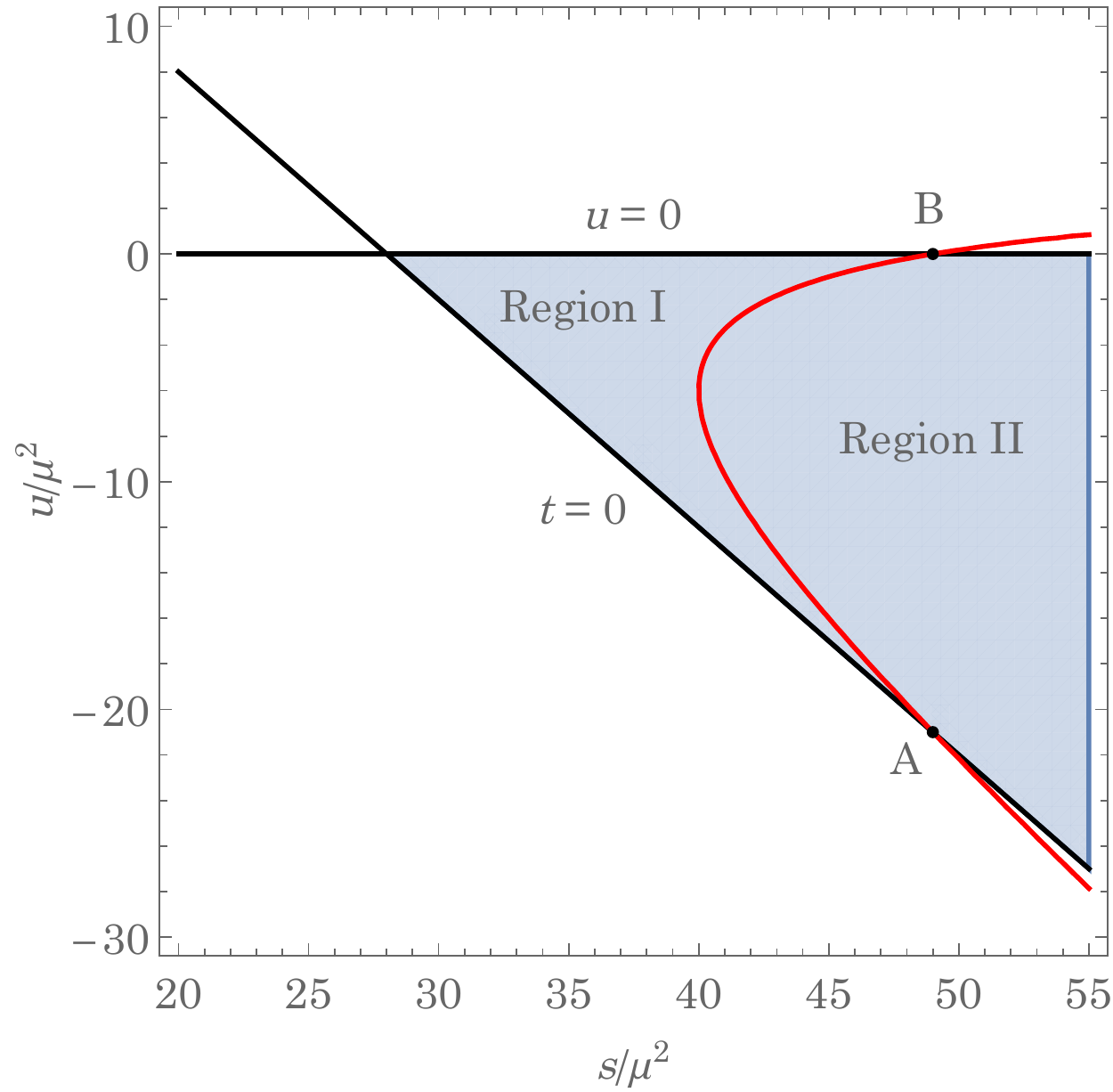}	
\caption{The physical region of the $s$-channel process is divided into region I and region II by the curve AB. Here, we set $\ReM=7\mu^2$.
}
\label{fig:boxsingularity}
\end{figure}

Let us discuss the kinematically-allowed region of the diagram
\scalebox{0.7}{
\begin{tikzpicture}[baseline=-2]
\draw (0.65,0.3) -- (0.35,0.3);
\draw (0.65,0.1) -- (0.35,0.1);
\draw (0.65,-0.3) -- (0.35,-0.3);
\draw (0.65,-0.1) -- (0.35,-0.1);
\draw (-0.65,0.3) -- (-0.35,0.3);
\draw (-0.65,0.1) -- (-0.35,0.1);
\draw (-0.65,-0.3) -- (-0.35,-0.3);
\draw (-0.65,-0.1) -- (-0.35,-0.1);
\draw (0.35,0.2) -- (-0.35,0.2);
\draw (0.35,-0.2) -- (-0.35, -0.2);
\draw (0.35,0.2) -- (-0.35,-0.2);
\draw (-0.35,0.2) -- (0.35,-0.2);
\filldraw [fill=white] (0.35,0.2) circle [radius =1.8mm] ;
\filldraw [fill=white] (0.35,-0.2) circle [radius =1.8mm] ;
\filldraw [fill=white] (-0.35,0.2) circle [radius =1.8mm] ;
\filldraw [fill=white] (-0.35,-0.2) circle [radius =1.8mm] ;
\node at (0.35,0.2) {$-$} ;
\node at (0.35,-0.2) {$-$} ;
\node at (-0.35,0.2) {$+$} ;
\node at (-0.35,-0.2) {$+$} ;
\end{tikzpicture}}
with $s_{12}=s_{34}=s_{56}=s_{78}=\ReM$. We name the internal lines as
\begin{align}
\begin{tikzpicture}[baseline=-2]
\draw (1,0.4) -- (-1,0.4);
\draw (1,-0.4) -- (-1,-0.4);
\draw (1,0.2) -- (0.55,0.2);
\draw (1,-0.2) -- (0.55,-0.2);
\draw (-1,0.2) -- (-0.55,0.2);
\draw (-1,-0.2) -- (-0.55,-0.2);
\draw (0.55,0.3) -- (-0.55,-0.3);
\draw (-0.55,0.3) -- (0.55,-0.3);
\filldraw [fill=white] (0.55,0.3) circle [radius = 2.5mm] ;
\filldraw [fill=white] (0.55,-0.3) circle [radius =2.5mm] ;
\filldraw [fill=white] (-0.55,0.3) circle [radius =2.5mm] ;
\filldraw [fill=white] (-0.55,-0.3) circle [radius =2.5mm] ;
\node at (0.55,0.3) {$-$} ;
\node at (0.55,-0.3) {$-$} ;
\node at (-0.55,0.3) {$+$} ;
\node at (-0.55,-0.3) {$+$} ;
\node at (0, 0.55) {$\scalebox{0.7}{1}$};
\node at (0.2, 0.25) {$\scalebox{0.7}{2}$};
\node at (0.2, -0.25) {$\scalebox{0.7}{3}$};
\node at (0, -0.55) {$\scalebox{0.7}{4}$};
\end{tikzpicture}
\,,
\end{align}
where each internal line is on the mass shell, $q_i^2=-\mu^2$. The time flows from right to left.
The energy-momentum conservation of each bubble gives
\begin{align}
\begin{cases}
p_{12}=q_1+q_2\,, \\
p_{34}=q_3+q_4\,, \\
p_{56}=q_1+q_3\,, \\
p_{78}=q_2+q_4\,, 
\end{cases}
\label{conservebox}
\end{align}
where $p_{12}=p_1+p_2$ and so on. Let us denote the Gram determinants of $q_i$ by
\begin{align}
G(q_1\cdots q_N)= {\rm det}(q_i \cdot q_j)\quad (i,j=1,\cdots N)
\,.
\end{align}
Then, the kinematically-allowed region, i.e., the vectors $q_i$ can be constructed to be real, is determined by
\begin{align}
G(q_1 q_2)<0 \,, \quad G(q_1 q_2 q_3)<0\,, \quad G(q_1q_2q_3q_4)<0
\,.
\end{align}
The first condition $G(q_1q_2)<0$ is satisfied by $s_{12}=\ReM>(2\mu)^2$. By using the conservation law \eqref{conservebox} and the on-shell conditions, the second and third conditions read
\begin{align}
\begin{split}
t<0\,, \qquad u&<0\,, \\
4\ReM(4\mu^2-\ReM)+ tu -4\mu^2(t+u)&<0
\,,
\end{split}
\end{align}
which is drawn as region I in Fig.~\ref{fig:boxsingularity}.
Therefore, the physical region $s> 4\ReM, t< 0, u< 0$ is divided into region I in which
\scalebox{0.7}{
\begin{tikzpicture}[baseline=-2]
\draw (0.65,0.3) -- (0.35,0.3);
\draw (0.65,0.1) -- (0.35,0.1);
\draw (0.65,-0.3) -- (0.35,-0.3);
\draw (0.65,-0.1) -- (0.35,-0.1);
\draw (-0.65,0.3) -- (-0.35,0.3);
\draw (-0.65,0.1) -- (-0.35,0.1);
\draw (-0.65,-0.3) -- (-0.35,-0.3);
\draw (-0.65,-0.1) -- (-0.35,-0.1);
\draw (0.35,0.2) -- (-0.35,0.2);
\draw (0.35,-0.2) -- (-0.35, -0.2);
\draw (0.35,0.2) -- (-0.35,-0.2);
\draw (-0.35,0.2) -- (0.35,-0.2);
\filldraw [fill=white] (0.35,0.2) circle [radius =1.8mm] ;
\filldraw [fill=white] (0.35,-0.2) circle [radius =1.8mm] ;
\filldraw [fill=white] (-0.35,0.2) circle [radius =1.8mm] ;
\filldraw [fill=white] (-0.35,-0.2) circle [radius =1.8mm] ;
\node at (0.35,0.2) {$-$} ;
\node at (0.35,-0.2) {$-$} ;
\node at (-0.35,0.2) {$+$} ;
\node at (-0.35,-0.2) {$+$} ;
\end{tikzpicture}}
is kinematically allowed and region II in which it is kinematically forbidden. They are divided by the curve AB which must correspond to a singularity curve. %Analysing the Landau equations~\cite{Landau:1959fi} of the crossed box diagram, one can indeed confirm that the boundary is a solution to the Landau equations.

In the high-energy limit (region II), the second term in \eqref{unitarity44_2} is kinematically forbidden and should be absent. By analytic continuation, we then obtain the standard form of the 2-to-2 unitarity equation for the $AA\to AA$ scattering as the residue at $s_{12}=s_{34}=(M^2)^*$ and $s_{56}=s_{78}=M^2$:
\begin{align}
\begin{tikzpicture}[baseline=-2]
\draw  [decorate, decoration={snake}] (-0.7,-0.2) -- (0, -0.2) ;
\draw  [decorate, decoration={snake}] (0.7,-0.2) -- (0, -0.2) ;
\draw  [decorate, decoration={snake}] (-0.7,0.2) -- (0, 0.2) ;
\draw  [decorate, decoration={snake}] (0.7,0.2) -- (0, 0.2) ;
\filldraw [fill=white] (0,0) circle [radius=4mm] ;
\draw ($(0,0)+(90:0.4)$) to [out = -90, in = 180] ($(0,0)+(0:0.4)$) ;
\draw ($(0,0)+(-90:0.4)$) to [out = 90, in = 180] ($(0,0)+(0:0.4)$) ;
\node (O1) at (0.2,0.2) {$-$} ;
\node (O2) at (0.2,-0.2) {$-$} ;
\node (O3) at (-0.1,0) {$+$} ;
\end{tikzpicture}
-
\begin{tikzpicture}[baseline=-2]
\draw  [decorate, decoration={snake}] (-0.7,-0.2) -- (0, -0.2) ;
\draw  [decorate, decoration={snake}] (0.7,-0.2) -- (0, -0.2) ;
\draw  [decorate, decoration={snake}] (-0.7,0.2) -- (0, 0.2) ;
\draw  [decorate, decoration={snake}] (0.7,0.2) -- (0, 0.2) ;
\filldraw [fill=white] (0,0) circle [radius=4mm] ;
\draw ($(0,0)+(90:0.4)$) to [out = -90, in = 0] ($(0,0)+(180:0.4)$) ;
\draw ($(0,0)+(-90:0.4)$) to [out = 90, in = 0] ($(0,0)+(180:0.4)$) ;
\node (O1) at (-0.2,0.2) {$+$} ;
\node (O2) at (-0.2,-0.2) {$+$} ;
\node (O3) at (0.1,0) {$-$} ;
\end{tikzpicture}
=
\begin{tikzpicture}[baseline=-2]
\draw[line width=5pt] (-0.5, 0 ) -- (0.5,0) ;
\draw  [decorate, decoration={snake}] (-1.2,0.2) -- (-0.5, 0.2) ;
\draw  [decorate, decoration={snake}] (1.2,0.2) -- (0.5, 0.2) ;
\draw  [decorate, decoration={snake}] (-1.2,-0.2) -- (-0.5, -0.2) ;
\draw  [decorate, decoration={snake}] (1.2,-0.2) -- (0.5, -0.2) ;
\filldraw [fill=white] (-0.5,0) circle [radius=4mm] ;
\node (O1) at (-0.5,0) {$+$} ;
\filldraw [fill=white] (0.5,0) circle [radius=4mm] ;
\node (O2) at (0.5,0) {$-$} ;
\end{tikzpicture}
\,,
\label{unitarityAA}
\end{align}
where the l.h.s.~represents the discontinuity across the real $s$-axis and the imaginary part
\begin{align}
\begin{tikzpicture}[baseline=-2]
\draw  [decorate, decoration={snake}] (-0.7,-0.2) -- (0, -0.2) ;
\draw  [decorate, decoration={snake}] (0.7,-0.2) -- (0, -0.2) ;
\draw  [decorate, decoration={snake}] (-0.7,0.2) -- (0, 0.2) ;
\draw  [decorate, decoration={snake}] (0.7,0.2) -- (0, 0.2) ;
\filldraw [fill=white] (0,0) circle [radius=4mm] ;
\draw ($(0,0)+(90:0.4)$) to [out = -90, in = 180] ($(0,0)+(0:0.4)$) ;
\draw ($(0,0)+(-90:0.4)$) to [out = 90, in = 180] ($(0,0)+(0:0.4)$) ;
\node (O1) at (0.2,0.2) {$-$} ;
\node (O2) at (0.2,-0.2) {$-$} ;
\node (O3) at (-0.1,0) {$+$} ;
\end{tikzpicture}
-
\begin{tikzpicture}[baseline=-2]
\draw  [decorate, decoration={snake}] (-0.7,-0.2) -- (0, -0.2) ;
\draw  [decorate, decoration={snake}] (0.7,-0.2) -- (0, -0.2) ;
\draw  [decorate, decoration={snake}] (-0.7,0.2) -- (0, 0.2) ;
\draw  [decorate, decoration={snake}] (0.7,0.2) -- (0, 0.2) ;
\filldraw [fill=white] (0,0) circle [radius=4mm] ;
\draw ($(0,0)+(90:0.4)$) to [out = -90, in = 0] ($(0,0)+(180:0.4)$) ;
\draw ($(0,0)+(-90:0.4)$) to [out = 90, in = 0] ($(0,0)+(180:0.4)$) ;
\node (O1) at (-0.2,0.2) {$+$} ;
\node (O2) at (-0.2,-0.2) {$+$} ;
\node (O3) at (0.1,0) {$-$} ;
\end{tikzpicture}
&=
\Disc_s\left( 
\begin{tikzpicture}[baseline=-2]
\draw  [decorate, decoration={snake}] (-0.7,-0.2) -- (0, -0.2) ;
\draw  [decorate, decoration={snake}] (0.7,-0.2) -- (0, -0.2) ;
\draw  [decorate, decoration={snake}] (-0.7,0.2) -- (0, 0.2) ;
\draw  [decorate, decoration={snake}] (0.7,0.2) -- (0, 0.2) ;
\filldraw [fill=white] (0,0) circle [radius=4mm] ;
\draw ($(0,0)+(90:0.4)$) to [out = -90, in = 180] ($(0,0)+(0:0.4)$) ;
\draw ($(0,0)+(-90:0.4)$) to [out = 90, in = 180] ($(0,0)+(0:0.4)$) ;
\node (O1) at (0.2,0.2) {$-$} ;
\node (O2) at (0.2,-0.2) {$-$} ;
\node (O3) at (-0.1,0) {$+$} ;
\end{tikzpicture}
\right)
\nn
&=2i{\rm Im}\left(
\begin{tikzpicture}[baseline=-2]
\draw  [decorate, decoration={snake}] (-0.7,-0.2) -- (0, -0.2) ;
\draw  [decorate, decoration={snake}] (0.7,-0.2) -- (0, -0.2) ;
\draw  [decorate, decoration={snake}] (-0.7,0.2) -- (0, 0.2) ;
\draw  [decorate, decoration={snake}] (0.7,0.2) -- (0, 0.2) ;
\filldraw [fill=white] (0,0) circle [radius=4mm] ;
\draw ($(0,0)+(90:0.4)$) to [out = -90, in = 180] ($(0,0)+(0:0.4)$) ;
\draw ($(0,0)+(-90:0.4)$) to [out = 90, in = 180] ($(0,0)+(0:0.4)$) ;
\node (O1) at (0.2,0.2) {$-$} ;
\node (O2) at (0.2,-0.2) {$-$} ;
\node (O3) at (-0.1,0) {$+$} ;
\end{tikzpicture}
\right)
\,.
\end{align}
On the other hand, a careful analysis is required in region~I because of the additional contribution
\scalebox{0.7}{
\begin{tikzpicture}[baseline=-2]
\draw (0.65,0.3) -- (0.35,0.3);
\draw (0.65,0.1) -- (0.35,0.1);
\draw (0.65,-0.3) -- (0.35,-0.3);
\draw (0.65,-0.1) -- (0.35,-0.1);
\draw (-0.65,0.3) -- (-0.35,0.3);
\draw (-0.65,0.1) -- (-0.35,0.1);
\draw (-0.65,-0.3) -- (-0.35,-0.3);
\draw (-0.65,-0.1) -- (-0.35,-0.1);
\draw (0.35,0.2) -- (-0.35,0.2);
\draw (0.35,-0.2) -- (-0.35, -0.2);
\draw (0.35,0.2) -- (-0.35,-0.2);
\draw (-0.35,0.2) -- (0.35,-0.2);
\filldraw [fill=white] (0.35,0.2) circle [radius =1.8mm] ;
\filldraw [fill=white] (0.35,-0.2) circle [radius =1.8mm] ;
\filldraw [fill=white] (-0.35,0.2) circle [radius =1.8mm] ;
\filldraw [fill=white] (-0.35,-0.2) circle [radius =1.8mm] ;
\node at (0.35,0.2) {$-$} ;
\node at (0.35,-0.2) {$-$} ;
\node at (-0.35,0.2) {$+$} ;
\node at (-0.35,-0.2) {$+$} ;
\end{tikzpicture}}.

\section{Extended unitarity}
\label{sec:extended}
In the previous section, the unstable-particle unitarity equations are derived in the physical regions of the Mandelstam variables $\{s,t,u\}$, especially in a negative $t$ region. On the other hand, the same unitarity equations are expected to hold even outside the physical region, called extended unitarity, and this is indeed the case in the stable-particle amplitude~\cite{Eden:1966dnq}. In this section, let us show that the unstable-particle unitarity equations hold even in a finite positive $t$. The positive $t$ extension will be used to find a positivity constraint on the unstable-particle amplitudes in the next section.

The positive $t$ extension requires analytic continuation. However, we do not know the analytic structure in a positive $t$. %In fact, we need the unitarity equation in $t>0$ to understand the analytic structure in a positive $t$. 
On the other hand, the analytic structure in the subenergy variable, especially the existence of the complex poles, can be deduced from the knowledge of 2-to-2 amplitudes and the two-particle discontinuity equations \eqref{discsub}. Hence, we first adjust subenergy variables so that the physical region can include a positive $t$ and then continue the unitarity equation in the subenergy variables.

Let us first consider the $AA\to AA$ amplitude. In Sec.~\ref{sec:AA}, we set $s_{12}=s_{34}=s_{56}=s_{78}=\ReM$ in the amplitude 
\begin{align*}
\begin{tikzpicture}[baseline=-2]
\draw (-0.6, 0.3) -- (0.6, 0.3);
\draw (-0.6, 0.1) -- (0.6, 0.1);
\draw (-0.6, -0.1) -- (0.6, -0.1);
\draw (-0.6, -0.3) -- (0.6, -0.3);
\filldraw [fill=white] (0,0) circle [radius=4mm] ;
\node at (0.7, 0.3) {$\scalebox{0.7}{1}$};
\node at (0.7, 0.1) {$\scalebox{0.7}{2}$};
\node at (0.7, -0.1) {$\scalebox{0.7}{3}$};
\node at (0.7, -0.3) {$\scalebox{0.7}{4}$};
\node at (-0.7, 0.3) {$\scalebox{0.7}{5}$};
\node at (-0.7, 0.1) {$\scalebox{0.7}{6}$};
\node at (-0.7, -0.1) {$\scalebox{0.7}{7}$};
\node at (-0.7, -0.3) {$\scalebox{0.7}{8}$};
\end{tikzpicture}
\end{align*}
and then continued the unitarity equation in the subenergy variables because $\ReM \pm i\varepsilon$ are the points closest to the complex poles in the real boundaries. 
However, as shown in Fig.~\ref{fig:extended}, we can still reach the correct sheet on which the complex pole exists (or does not exist) even starting from the vicinity of $\ReM$. For the unstable particle with $(2\mu)^2<\ReM<(3\mu)^2$, we can consider the 4-to-4 unitarity equation in $(2\mu)^2<s_{12}, s_{34}, s_{56}, s_{78}<(3\mu)^2$ (if there is no anomalous threshold) and then analytically continue the unitarity equation in $\{ s_{12}, s_{34}, s_{56}, s_{78}\}$ with fixed $s$ and $t$. The physical region of $s$ and $t$ is deformed thanks to the change of the initial input of $\{s_{12}, s_{34}, s_{56}, s_{78}\}$ and then the unitarity equation can be extended away from the negative $t$. For instance, let us consider
\begin{align}
s_{12}=s_{78}=\ReM + \delta s\,, \quad
s_{34}=s_{56}=\ReM - \delta s
\,.
\end{align}
Then, the physical region constraint of the $s$-channel region is given by
\begin{align}
\begin{split}
&s>2\ReM\left(1+\sqrt{1 -(\delta s/\ReM)^2 } \right)
\,, \\
&4\ReM-s < t < \frac{4\delta s^2}{s}
\,,
\end{split}
\end{align}
which includes a positive $t$ region for $\delta s \neq 0$.

\begin{figure}[t]
\centering
\begin{tikzpicture}
\draw[->] (-1.5,0) -- (2,0);
\node at (2.1,1.3) {$s_{12}$};
\draw  (1.8,1.5) |- (2.2,1.1);
\draw[red, decoration = {zigzag,segment length = 3mm, amplitude = 1mm},decorate] (0.6,0.25)--(2, 0.25);
\draw[red, decoration = {zigzag,segment length = 3mm, amplitude = 1mm},decorate] (-1.5,0)--(2,0);
\draw[red] ($(0.6,0.25)+(-0.1,-0.1)$) -- ($(0.6,0.25)+(0.1,0.1)$) ;
\draw[red] ($(0.6,0.25)+(-0.1,0.1)$) -- ($(0.6,0.25)+(0.1,-0.1)$) ;
\draw[dashed] (-1.3,0.5) -- (-1.3,0);
\draw[->, dotted, red] (-1.3,0) -- (-1.3,-1);
\draw[dashed] (-0.6,0.5) -- (-0.8,0);
\draw[->, dotted, red] (-0.8,0) -- (-1.2,-1);
\draw[dashed] (1,0.5) [rounded corners] -- (0.6,0.7) -- (-0.1,0);
\draw[->, dotted, red]  (-0.1, 0) -- (-1.1,-1);
\draw[dashed] (2, 0.5) -- (1.5, 0.25);
\draw[->, dash dot, blue]  (1.5, 0.25) -- (-1,-1);
\node at (-1.3, 0.7) {$(a)$};
\node at (-0.6, 0.7) {$(b)$};
\node at (1.2, 0.7) {$(c)$};
\node at (2, 0.7) {$(d)$};
\fill[red] (-1.3,-1.1) circle (2pt);
%%%%%%%%%%%%%%%%
%%%%%%%%%%%%%%%%
%%%%%%%%%%%%%%%%
\draw[->] ($(4,0) + (-1.5,0.5)$) -- ($(4,0) + (2,0.5)$);
\node at ($(4,0) + (2.1, 1.3)$) {$s_{12}$};
\draw  ($(4,0) + (1.8,1.5)$) |- ($(4,0)+(2.2,1.1)$);
\draw[red, decoration = {zigzag,segment length = 3mm, amplitude = 1mm},decorate] ($(4,0) + (0.6,0.25)$) --($(4,0) + (2, 0.25)$);
\draw[red, decoration = {zigzag,segment length = 3mm, amplitude = 1mm},decorate] ($(4,0) + (-1.5,0.5) $)--($(4,0) + (2,0.5) $);
\draw[red] ($(4,0) +(0.6,0.25)+(-0.1,-0.1)$) -- ($(4,0) +(0.6,0.25)+(0.1,0.1)$) ;
\draw[red] ($(4,0) +(0.6,0.25)+(-0.1,0.1)$) -- ($(4,0) +(0.6,0.25)+(0.1,-0.1)$) ;
\draw[->, dashed] ($(4,0) + (-1.3,0)$) -- ($(4,0) + (-1.3,-1) $);
\draw[->, dashed] ($(4,0) + (-0.6,0)$) -- ($(4,0) + (-1.2,-1) $);
\draw[dashed] ($(4,0) + (1, 0)$) -- ($(4,0) + (0.8, 0.25) $);
\draw[->, dash dot, blue] ($(4,0) + (0.8,0.25)$) [rounded corners] -- ($(4,0) + (0.6,0.5)$) -- ($(4,0) + (-1.1,-1)$);
\draw[->, dashed] ($(4,0) + (2,0)$) -- ($(4,0) + (-1,-1) $);
\node at ($(4,0) + (-1.3, 0.2)$) {$(a)$};
\node at ($(4,0) + (-0.6, 0.2)$) {$(b)$};
\node at ($(4,0) + (0.2, -0.2)$) {$(c)$};
\node at ($(4,0) + (2, -0.2)$) {$(d)$};
\fill[red] ($(4,0) + (-1.3,-1.1)$) circle (2pt);
\end{tikzpicture}
\caption{Several paths to approach $s_{12}=M^2$ starting at ${\rm Im}s_{12}=+\varepsilon$, corresponding to the $(+)$ amplitudes (left) and at ${\rm Im}s_{12}=-\varepsilon$ for the $(-)$ amplitudes (right) where the different types of the curves run on different sheets. In (a) and (b), the $+i\varepsilon$ paths correctly reach the complex pole while the $-i\varepsilon$ paths stay on the physical sheet. On the other hand, we need to bypass the branch point by adding a positive imaginary part in (c) or a negative imaginary part in (d). Then, we cannot reach the correct sheet in (c)  and (d). }
\label{fig:extended}
\end{figure}

We then discuss the $A\varphi\to A\varphi$ amplitude. Since the 3 and 6 external lines of
\scalebox{0.6}{
\begin{tikzpicture}[baseline=-2]
\node (a) at (1,0.3) {1};
\node (b) at (1,0) {2};
\node (c) at (1,-0.3) {3};
\node (d) at (-1,0.3) {4};
\node (e) at (-1,0) {5};
\node (f) at (-1,-0.3) {6};
\draw (a) -- (d);
\draw (b) -- (e);
\draw (c) -- (f);
\filldraw [fill=white] (0,0) circle [radius=5mm] ;
\node (O) at (0,0) {} ;
\end{tikzpicture}}
are fixed to be on-shell, the physical region of the 3-to-3 amplitude is always given by $t=s_{36}<0$ in the $s$-channel region even when we change the values of $s_{12}$ and $s_{45}$. Instead, let us consider the 4-to-4 amplitude and extract an embedded $A\varphi\to A\varphi$ amplitude at $s_{12}=(M^2)^*, s_{56}=M^2,$ and $s_{348}=\mu^2$, that is,
\begin{align}
\begin{tikzpicture}[baseline=-2]
\draw (-0.6, 0.3) -- (0.6, 0.3);
\draw (-0.6, 0.1) -- (0.6, 0.1);
\draw (-0.6, -0.1) -- (0.6, -0.1);
\draw (-0.6, -0.3) -- (0.6, -0.3);
\filldraw [fill=white] (0,0) circle [radius=4mm] ;
\node at (0.7, 0.3) {$\scalebox{0.7}{1}$};
\node at (0.7, 0.1) {$\scalebox{0.7}{2}$};
\node at (0.7, -0.1) {$\scalebox{0.7}{3}$};
\node at (0.7, -0.3) {$\scalebox{0.7}{4}$};
\node at (-0.7, 0.3) {$\scalebox{0.7}{5}$};
\node at (-0.7, 0.1) {$\scalebox{0.7}{6}$};
\node at (-0.7, -0.1) {$\scalebox{0.7}{7}$};
\node at (-0.7, -0.3) {$\scalebox{0.7}{8}$};
\end{tikzpicture}
\sim
\begin{tikzpicture}[baseline=-2]
\draw (-1, 0.3) -- (-0.75, 0.3);
\draw (-1, 0.1) -- (-0.75, 0.1);
\draw (1, 0.3) -- (0.75, 0.3);
\draw (1, 0.1) -- (0.75, 0.1);
\draw (-1, -0.25) -- (0, -0);
\draw (1, -0.25) -- (0.75, -0.25);
\draw (-1, -0.45) -- (1, -0.45);
\draw  [decorate, decoration={snake}] (-0.75,0.2) -- (0.75, 0.2) ;
\Pole{0.75}{-0.35}{0.1}{-0.05}{}
\filldraw [fill=white] (0,0.1) circle [radius=3mm] ;
\filldraw [fill=white] (-0.75,0.2) circle [radius=1.5mm] ;
\filldraw [fill=white] (0.75,0.2) circle [radius=1.5mm] ;
\filldraw [fill=white] (0.75,-0.35) circle [radius=1.5mm] ;
\end{tikzpicture}
\,,
\label{22from44}
\end{align}
where the line
$\begin{tikzpicture}[baseline=-2]
\Pole{-0.3}{0}{0.3}{0}{\pm}
\end{tikzpicture}
= 1/(-q^2-\mu^2 \pm i\varepsilon)$ is the pole factor of the stable particle, satisfying
\begin{align}
\begin{tikzpicture}[baseline=-2]
\Pole{-0.3}{0}{0.3}{0}{+}
\end{tikzpicture}
-
\begin{tikzpicture}[baseline=-2]
\Pole{-0.3}{0}{0.3}{0}{-}
\end{tikzpicture}
=
\begin{tikzpicture}[baseline=-2]
\draw (-0.3,0) -- (0.3,0);
\end{tikzpicture}
\,.
\label{polefactor}
\end{align}
The existence of a simple pole at $s_{348}=\mu^2$ is implied by unitarity~\cite{Eden:1966dnq}.
Hence, we consider the 4-to-4 unitarity equation and pick up the diagrams having the singular structure of \eqref{22from44}. Such a diagram should satisfy (i) the 1 and 2 momenta connect to a $(-)$ bubble, (ii) the 5 and 6 momenta connect to a $(+)$ bubble, and (iii) the 3, 4, and 8 momenta connect to a single bubble. The condition (iii) includes diagrams with the structure
\begin{align*}
\begin{tikzpicture}[baseline=-2]
\draw (1.4, -0.25) -- (0.75, -0.25);
\draw (0.75, -0.35) -- (0.2, -0.05);
\draw (0.0, -0.45) -- (1.4, -0.45);
\filldraw [fill=white] (0.75,-0.35) circle [radius=1.5mm] ;
\filldraw [fill=white] (1.15,-0.35) circle [radius=1.5mm] ;
\filldraw [fill=white] (0.45,-0.2) circle [radius=1.2mm] ;
\node at (1.5, -0.25) {$\scalebox{0.7}{3}$};
\node at (1.5, -0.45) {$\scalebox{0.7}{4}$};
\node at (-0.1, -0.45) {$\scalebox{0.7}{8}$};
\end{tikzpicture}
\end{align*}
due to the conservation in the right bubble. We also note that the diagrams that the 1 and 2 (or 5 and 6) momenta connect to a $(+)$ [or $(-)$] bubble can be discarded as far as the paths like (a) and (b) of Fig.~\ref{fig:extended} are concerned.\footnote{In Sec.~\ref{sec:unitarity}, we have aligned the sign of the bubbles by using the two-particle discontinuity equations before analytic continuation. This would be more illustrative in understanding how to extract the unstable-particle unitarity equations. On the other hand, when considering the paths like (a) and (b) of Fig.~\ref{fig:extended}, it is also possible to analytically continue the unitarity equation first and then use the discontinuity equations. The latter approach can simplify calculations because we can immediately ignore diagrams that the 1 and 2 (or 5 and 6) momenta do not connect to a $(-)$ [or $(+)$] bubble.} Sorting out the relevant terms according to (i), (ii), and (iii), the 4-to-4 unitarity equation is given by the following form (see Appendix~\ref{sec:stable_unitarity})
\begin{align}
0&=
\begin{tikzpicture}[baseline=-2]
\draw (1,0.3) -- (-0.6,0.3);
\draw (1,0.1) -- (-0.6,0.1);
\draw (1,-0.1) -- (-0.6,-0.1);
\draw (1,-0.3) -- (-0.6,-0.3);
\filldraw [fill=white] (0.5,0) circle [radius=4mm] ;
\node (O1) at (0.5,0) {$-$} ;
\filldraw [fill=white] (-0.3,0.2) circle [radius=1.8mm] ;
\node (O2) at (-0.3,0.2) {$+$} ;
\end{tikzpicture}
+
\begin{tikzpicture}[baseline=-2]
\draw (1,0.3) -- (0.5,0.3);
\draw (1,0.1) -- (0.5,0.1);
\draw (1,-0.1) -- (0.5,-0.1);
\draw (-0.35,0.3) -- (-0.75,0.3);
\draw (-0.35,0.1) -- (-0.75,0.1);
\draw (-0.35,-0.1) -- (-0.75,-0.1);
\draw (1,-0.3) -- (-0.75,-0.3);
\draw[line width=5pt] (0.5, 0.1 ) -- (-0.35,0.1) ;
\filldraw [fill=white] (0.5,0) circle [radius=4mm] ;
\node (O1) at (0.5,0) {$-$} ;
\filldraw [fill=white] (-0.35,0.1) circle [radius=3mm] ;
\node (O2) at (-0.35,0.1) {$+$} ;
\end{tikzpicture}
\nn
&+
\begin{tikzpicture}[baseline=-2]
\draw (-1,0.3) -- (0.6,0.3);
\draw (-1,0.1) -- (0.6,0.1);
\draw (-1,-0.1) -- (0.6,-0.1);
\draw (-1,-0.3) -- (0.6,-0.3);
\filldraw [fill=white] (-0.5,0) circle [radius=4mm] ;
\node (O1) at (-0.5,0) {$+$} ;
\filldraw [fill=white] (0.3,0.2) circle [radius=1.8mm] ;
\node (O2) at (0.3,0.2) {$-$} ;
\end{tikzpicture}
-
\begin{tikzpicture}[baseline=-2]
\draw (0.7,0.3) -- (-0.75,0.3);
\draw (0.7,0.1) -- (-0.75,0.1);
\draw (0.7,-0.1) -- (-0.75,-0.1);
\draw (0.7,-0.3) -- (-0.75,-0.3);
\filldraw [fill=white] (0.4,0.2) circle [radius=1.8mm] ;
\filldraw [fill=white] (0.4,-0.2) circle [radius=1.8mm] ;
\filldraw [fill=white] (-0.35,0.1) circle [radius=3mm] ;
\node at (-0.35,0.1) {$+$} ;
\node at (0.4,0.2) {$-$} ;
\node at (0.4,-0.2) {$-$} ;
\end{tikzpicture}
-
\begin{tikzpicture}[baseline=-2]
\draw (-1,0.3) -- (0.6,0.3);
\draw (-1,0.1) -- (0.6,0.1);
\draw (-1,-0.1) -- (0.6,-0.1);
\draw (-1,-0.3) -- (0.6,-0.3);
\filldraw [fill=white] (-0.5,0) circle [radius=4mm] ;
\node at (-0.5,0) {$+$} ;
\filldraw [fill=white] (0.3,0.2) circle [radius=1.8mm] ;
\filldraw [fill=white] (0.3,-0.2) circle [radius =1.8mm] ;
\node at (0.3,0.2) {$-$} ;
\node at (0.3,-0.2) {$-$} ;
\end{tikzpicture}
\nn
&+
\begin{tikzpicture}[baseline=-2]
\draw (0.4,0.3) -- (0.7,0.3);
\draw (0.4,0.1) -- (0.7,0.1);
\draw (-0.4,0.3) -- (-0.7,0.3);
\draw (-0.4,0.1) -- (-0.7,0.1);
\draw (-0.7, -0.2) -- (0.4, 0.1);
\draw (-0.7, -0.4) -- (-0.5, 0.1);
\draw (0.4,0.25) -- (-0.4,0.25);
\draw (-0.5, 0.25) -- (0.7,-0.2);
\draw (-0.5, 0.05) -- (0.7,-0.4);
\filldraw [fill=white] (0.4,0.2) circle [radius=2mm] ;
\filldraw [fill=white] (-0.4,0.2) circle [radius =2mm] ;
\node at (0.4,0.2) {$-$} ;
\node at (-0.4,0.2) {$+$} ;
\end{tikzpicture}
-
\begin{tikzpicture}[baseline=-2]
\draw (0.4,0.3) -- (0.7,0.3);
\draw (0.4,0.1) -- (0.7,0.1);
\draw (-0.4,0.3) -- (-0.7,0.3);
\draw (-0.4,0.1) -- (-0.7,0.1);
\draw (-0.7, -0.2) -- (0.4, 0.1);
\draw (-0.7, -0.4) -- (-0.5, 0.1);
\draw (0.4,0.25) -- (-0.4,0.25);
\draw (-0.5, 0.25) -- (0.4,-0.2);
\draw (-0.5, 0.05) -- (0.4,-0.4);
\draw (0.4, -0.2) -- (0.7,-0.2);
\draw (0.4, -0.4) -- (0.7,-0.4);
\filldraw [fill=white] (0.4,0.2) circle [radius=2mm] ;
\filldraw [fill=white] (-0.4,0.2) circle [radius =2mm] ;
\filldraw [fill=white] (0.4,-0.3) circle [radius=2mm] ;
\node at (0.4,0.2) {$-$} ;
\node at (0.4,-0.3) {$-$} ;
\node at (-0.4,0.2) {$+$} ;
\end{tikzpicture}
-
\begin{tikzpicture}[baseline=-2]
\draw (0.4,0.3) -- (0.7,0.3);
\draw (0.4,0.1) -- (0.7,0.1);
\draw (-0.4,0.3) -- (-0.7,0.3);
\draw (-0.4,0.1) -- (-0.7,0.1);
\draw (-0.7, -0.2) -- (0.4, 0.1);
\draw (-0.7, -0.4) -- (0.4, -0.4);
\draw (0.4,0.25) -- (-0.4,0.25);
\draw (-0.5, 0.15) -- (0.4,-0.3);
\draw (0.4, -0.2) -- (0.7,-0.2);
\draw (0.4, -0.4) -- (0.7,-0.4);
\filldraw [fill=white] (0.4,0.2) circle [radius=2mm] ;
\filldraw [fill=white] (-0.4,0.2) circle [radius =2mm] ;
\filldraw [fill=white] (0.4,-0.3) circle [radius=2mm] ;
\node at (0.4,0.2) {$-$} ;
\node at (0.4,-0.3) {$-$} ;
\node at (-0.4,0.2) {$+$} ;
\end{tikzpicture}
+R
\nn
%%%%%%%%%%%%%%%%%%%%%%%%%%%%%
%%%%%%%%%%%%%%%%%%%%%%%%%%%%%
%%%%%%%%%%%%%%%%%%%%%%%%%%%%%
%%%%%%%%%%%%%%%%%%%%%%%%%%%%%
&\sim
\begin{tikzpicture}[baseline=-2]
\draw (1.5,0.3) -- (-0.5,0.3);
\draw (1.5,0.1) -- (-0.5,0.1);
\draw (0.5,-0.1) -- (-0.5,-0.1);
\draw (1.5,-0.5) -- (-0.5,-0.5);
\draw (1.5,-0.3) -- (1.2,-0.3);
\draw (1.5,-0.5) -- (1.2,-0.5);
\Pole{0.6}{0}{1.2}{-0.4}{-}
\filldraw [fill=white] (0.5,0.1) circle [radius=3mm] ;
\node at (0.5,0.1) {$-$} ;
\filldraw [fill=white] (-0.2,0.2) circle [radius=1.8mm] ;
\node at (-0.2,0.2) {$+$} ;
\filldraw [fill=white] (1.2,-0.4) circle [radius=1.8mm] ;
\node at (1.2,-0.4) {$-$} ;
\end{tikzpicture}
+
\begin{tikzpicture}[baseline=-2]
\draw (1.35,0.3) -- (0.5,0.3);
\draw (1.35,0.1) -- (0.5,0.1);
\draw (-0.35,0.3) -- (-0.75,0.3);
\draw (-0.35,0.1) -- (-0.75,0.1);
\draw (-0.35,-0.1) -- (-0.75,-0.1);
\draw (1.35,-0.5) -- (-0.75,-0.5);
\draw (1.35,-0.3) -- (1.05,-0.3);
\draw (1.35,-0.5) -- (1.05,-0.5);
\draw[line width=5pt] (0.35, 0.1 ) -- (-0.35,0.1) ;
\Pole{0.45}{0.1}{1.05}{-0.4}{-}
\filldraw [fill=white] (0.35,0.1) circle [radius=3mm] ;
\node at (0.35,0.1) {$-$} ;
\filldraw [fill=white] (-0.35,0.1) circle [radius=3mm] ;
\node at (-0.35,0.1) {$+$} ;
\filldraw [fill=white] (1.05,-0.4) circle [radius=1.8mm] ;
\node at (1.05,-0.4) {$-$} ;
\end{tikzpicture}
\nn
&+
\begin{tikzpicture}[baseline=-2]
\draw (0.7,0.3) -- (-0.75,0.3);
\draw (0.7,0.1) -- (-0.75,0.1);
\draw (-0.35,-0.1) -- (-0.75,-0.1);
\draw (0.7,-0.5) -- (-0.75,-0.5);
\draw (0.7,-0.3) -- (0.4,-0.3);
\Pole{-0.35}{0}{0.4}{-0.4}{+}
\filldraw [fill=white] (0.4,0.2) circle [radius=1.8mm] ;
\node at (0.4,0.2) {$-$} ;
\filldraw [fill=white] (0.4,-0.4) circle [radius=1.8mm] ;
\node at (0.4,-0.4) {$+$} ;
\filldraw [fill=white] (-0.35,0.1) circle [radius=3mm] ;
\node at (-0.35,0.1) {$+$} ;
\end{tikzpicture}
-
\begin{tikzpicture}[baseline=-2]
\draw (0.7,0.3) -- (-0.75,0.3);
\draw (0.7,0.1) -- (-0.75,0.1);
\draw (-0.35,-0.1) -- (-0.75,-0.1);
\draw (0.7,-0.5) -- (-0.75,-0.5);
\draw (0.7,-0.3) -- (0.4,-0.3);
\draw (-0.35,0) -- (0.4,-0.4);
\filldraw [fill=white] (0.4,0.2) circle [radius=1.8mm] ;
\node at (0.4,0.2) {$-$} ;
\filldraw [fill=white] (0.4,-0.4) circle [radius=1.8mm] ;
\node at (0.4,-0.4) {$-$} ;
\filldraw [fill=white] (-0.35,0.1) circle [radius=3mm] ;
\node at (-0.35,0.1) {$+$} ;
\end{tikzpicture}
-
\begin{tikzpicture}[baseline=-2]
\draw (1.15,0.3) -- (-0.75,0.3);
\draw (1.15,0.1) -- (-0.75,0.1);
\draw (-0.35,-0.1) -- (-0.75,-0.1);
\draw (1.15,-0.5) -- (-0.75,-0.5);
\draw (1.15,-0.3) -- (0.4,-0.3);
\Pole{-0.35}{0}{0.4}{-0.4}{+}
\filldraw [fill=white] (0.4,0.2) circle [radius=1.8mm] ;
\node at (0.4,0.2) {$-$} ;
\filldraw [fill=white] (0.4,-0.4) circle [radius=1.8mm] ;
\node at (0.4,-0.4) {$+$} ;
\filldraw [fill=white] (0.85,-0.4) circle [radius=1.8mm] ;
\node at (0.85,-0.4) {$-$} ;
\filldraw [fill=white] (-0.35,0.1) circle [radius=3mm] ;
\node at (-0.35,0.1) {$+$} ;
\end{tikzpicture}
\nn
&+
\begin{tikzpicture}[baseline=-2]
\draw (0.4,0.3) -- (0.7,0.3);
\draw (0.4,0.1) -- (0.7,0.1);
\draw (-0.4,0.3) -- (-0.7,0.3);
\draw (-0.4,0.1) -- (-0.7,0.1);
\draw (-0.7, -0.2) -- (0.4, 0.1);
\draw (0.4,0.25) -- (-0.4,0.25);
\draw (0.4, -0.3) -- (0.7,-0.3);
\draw (-0.7, -0.5) -- (0.7, -0.5);
\Pole{-0.4}{0.1}{0.55}{-0.4}{+}
\filldraw [fill=white] (0.4,0.2) circle [radius=2mm] ;
\node at (0.4,0.2) {$-$} ;
\filldraw [fill=white] (-0.4,0.2) circle [radius =2mm] ;
\node at (-0.4,0.2) {$+$} ;
\filldraw [fill=white] (0.4,-0.4) circle [radius=2mm] ;
\node at (0.4,-0.4) {$+$} ;
\end{tikzpicture}
-
\begin{tikzpicture}[baseline=-2]
\draw (0.4,0.3) -- (1.2,0.3);
\draw (0.4,0.1) -- (1.2,0.1);
\draw (-0.4,0.3) -- (-0.7,0.3);
\draw (-0.4,0.1) -- (-0.7,0.1);
\draw (-0.7, -0.2) -- (0.4, 0.1);
\draw (0.4,0.25) -- (-0.4,0.25);
\draw (0.4, -0.3) -- (1.2,-0.3);
\draw (-0.7, -0.5) -- (1.2, -0.5);
\Pole{-0.4}{0.1}{0.55}{-0.4}{+}
\filldraw [fill=white] (0.4,0.2) circle [radius=2mm] ;
\node at (0.4,0.2) {$-$} ;
\filldraw [fill=white] (-0.4,0.2) circle [radius =2mm] ;
\node at (-0.4,0.2) {$+$} ;
\filldraw [fill=white] (0.4,-0.4) circle [radius=2mm] ;
\node at (0.4,-0.4) {$+$} ;
\filldraw [fill=white] (0.9,-0.4) circle [radius=2mm] ;
\node at (0.9,-0.4) {$-$} ;
\end{tikzpicture}
-
\begin{tikzpicture}[baseline=-2]
\draw (0.4,0.3) -- (0.7,0.3);
\draw (0.4,0.1) -- (0.7,0.1);
\draw (-0.4,0.3) -- (-0.7,0.3);
\draw (-0.4,0.1) -- (-0.7,0.1);
\draw (-0.7, -0.2) -- (0.4, 0.1);
\draw (0.4,0.25) -- (-0.4,0.25);
\draw (0.4, -0.3) -- (0.7,-0.3);
\draw (-0.7, -0.5) -- (0.7, -0.5);
\draw (-0.4, 0.1) -- (0.55, -0.4);
\filldraw [fill=white] (0.4,0.2) circle [radius=2mm] ;
\node at (0.4,0.2) {$-$} ;
\filldraw [fill=white] (-0.4,0.2) circle [radius =2mm] ;
\node at (-0.4,0.2) {$+$} ;
\filldraw [fill=white] (0.4,-0.4) circle [radius=2mm] ;
\node at (0.4,-0.4) {$-$} ;
\end{tikzpicture}
\label{unitarity44emb}
\end{align}
where $(2\mu)^2<s_{12},s_{34},s_{56}<(3\mu)^2$ is assumed for simplicity. We then use the 2-to-2 unitarity equation $\scalebox{0.7}{\Atwo{+}}-\scalebox{0.7}{\Atwo{-}}=
\scalebox{0.7}{\begin{tikzpicture}[baseline=-2]
\draw (-1,0.2) -- (-0.5,0.2);
\draw (-1,0) -- (-0.5,0);
\draw (-1,-0.2) -- (-0.5,-0.2);
\draw (1,0.2) -- (0.5,0.2);
\draw (1,0) -- (0.5,0);
\draw (1,-0.2) -- (0.5,-0.2);
\draw (-0.5,0.1) -- (0.5,0.1) ;
\draw (-0.5,-0.1) -- (0.5,-0.1) ;
\filldraw [fill=white] (-0.5,0) circle [radius=4mm] ;
\node (O1) at (-0.5,0) {$+$} ;
\filldraw [fill=white] (0.5,0) circle [radius=4mm] ;
\node (O2) at (0.5,0) {$-$} ;
\end{tikzpicture}}$ 
and \eqref{polefactor}, yielding
\begin{align}
-
\begin{tikzpicture}[baseline=-2]
\draw (0.7,0.3) -- (-0.75,0.3);
\draw (0.7,0.1) -- (-0.75,0.1);
\draw (-0.35,-0.1) -- (-0.75,-0.1);
\draw (0.7,-0.5) -- (-0.75,-0.5);
\draw (0.7,-0.3) -- (0.4,-0.3);
\Pole{-0.35}{0}{0.4}{-0.4}{-}
\filldraw [fill=white] (0.4,0.2) circle [radius=1.8mm] ;
\node at (0.4,0.2) {$-$} ;
\filldraw [fill=white] (0.4,-0.4) circle [radius=1.8mm] ;
\node at (0.4,-0.4) {$-$} ;
\filldraw [fill=white] (-0.35,0.1) circle [radius=3mm] ;
\node at (-0.35,0.1) {$+$} ;
\end{tikzpicture}
-
\begin{tikzpicture}[baseline=-2]
\draw (1.5,0.3) -- (-0.5,0.3);
\draw (1.5,0.1) -- (-0.5,0.1);
\draw (0.5,-0.1) -- (-0.5,-0.1);
\draw (1.5,-0.5) -- (-0.5,-0.5);
\draw (1.5,-0.3) -- (1.2,-0.3);
\draw (1.5,-0.5) -- (1.2,-0.5);
\Pole{0.6}{0}{1.2}{-0.4}{-}
\filldraw [fill=white] (0.5,0.1) circle [radius=3mm] ;
\node at (0.5,0.1) {$-$} ;
\filldraw [fill=white] (-0.2,0.2) circle [radius=1.8mm] ;
\node at (-0.2,0.2) {$+$} ;
\filldraw [fill=white] (1.2,-0.4) circle [radius=1.8mm] ;
\node at (1.2,-0.4) {$-$} ;
\end{tikzpicture}
\sim
\begin{tikzpicture}[baseline=-2]
\draw (1.35,0.3) -- (0.5,0.3);
\draw (1.35,0.1) -- (0.5,0.1);
\draw (-0.35,0.3) -- (-0.75,0.3);
\draw (-0.35,0.1) -- (-0.75,0.1);
\draw (-0.35,-0.1) -- (-0.75,-0.1);
\draw (1.35,-0.5) -- (-0.75,-0.5);
\draw (1.35,-0.3) -- (1.05,-0.3);
\draw (1.35,-0.5) -- (1.05,-0.5);
\draw[line width=5pt] (0.35, 0.1 ) -- (-0.35,0.1) ;
\Pole{0.45}{0.1}{1.05}{-0.4}{-}
\filldraw [fill=white] (0.35,0.1) circle [radius=3mm] ;
\node at (0.35,0.1) {$-$} ;
\filldraw [fill=white] (-0.35,0.1) circle [radius=3mm] ;
\node at (-0.35,0.1) {$+$} ;
\filldraw [fill=white] (1.05,-0.4) circle [radius=1.8mm] ;
\node at (1.05,-0.4) {$-$} ;
\end{tikzpicture}
+
\begin{tikzpicture}[baseline=-2]
\draw (0.4,0.3) -- (0.7,0.3);
\draw (0.4,0.1) -- (0.7,0.1);
\draw (-0.4,0.3) -- (-0.7,0.3);
\draw (-0.4,0.1) -- (-0.7,0.1);
\draw (-0.7, -0.2) -- (0.4, 0.1);
\draw (0.4,0.25) -- (-0.4,0.25);
\draw (0.4, -0.3) -- (0.7,-0.3);
\draw (-0.7, -0.5) -- (0.7, -0.5);
\Pole{-0.4}{0.1}{0.55}{-0.4}{-}
\filldraw [fill=white] (0.4,0.2) circle [radius=2mm] ;
\node at (0.4,0.2) {$-$} ;
\filldraw [fill=white] (-0.4,0.2) circle [radius =2mm] ;
\node at (-0.4,0.2) {$+$} ;
\filldraw [fill=white] (0.4,-0.4) circle [radius=2mm] ;
\node at (0.4,-0.4) {$-$} ;
\end{tikzpicture}
\,.
\label{22from44unitarity}
\end{align}
Furthermore, the two-particle discontinuity equations \eqref{discsub} imply
\begin{align}
\begin{tikzpicture}[baseline=-2]
\draw (-1,0.2) -- (0.6,0.2);
\draw (-1,0) -- (0.6,0);
\draw (-1,-0.2) -- (0.6,-0.2);
\filldraw [fill=white] (-0.5,0) circle [radius=4mm] ;
\node (O1) at (-0.5,0) {$+$} ;
\filldraw [fill=white] (0.3,0.1) circle [radius=2mm] ;
\node (O2) at (0.3,0.1) {$-$} ;
\end{tikzpicture}
&\sim
\begin{tikzpicture}[baseline=-2]
\draw (1.3,0.2) -- (1.6,0.2);
\draw (1.3,0) -- (1.6,0);
\draw (-0.8,0.2) -- (-1.1,0.2);
\draw (-0.8,0) -- (-1.1,0);
\draw (1.6,-0.2) -- (-1.1,-0.2);
\draw (0,0.2) -- (0.8,0.2);
\draw (0,0) -- (0.8,0);
\draw  [decorate, decoration={snake}] (-0.7,0.15) -- (0, 0.15) ;
\draw  [decorate, decoration={snake}] (0.8,0.15) -- (1.2, 0.15) ;
\filldraw [fill=white] (0,0) circle [radius=4mm] ;
\filldraw [fill=white] (0.7,0.1) circle [radius=2mm] ;
\filldraw [fill=white] (1.3,0.1) circle [radius=2mm] ;
\filldraw [fill=white] (-0.8,0.1) circle [radius=2mm] ;
\node  at (0,0) {$+$} ;
\node at (0.7,0.1) {$-$};
\node at (1.3,0.1) {$-$}; 
\node at (-0.8,0.1) {$+$};
\end{tikzpicture}
\sim -
\begin{tikzpicture}[baseline=-2]
\draw (1.1,-0.2) -- (-1.1,-0.2);
\draw (-1.1,0.2) -- (-0.8,0.2) ;
\draw (-1.1,0) -- (-0.8,0);
\draw  [decorate, decoration={snake}] (-0.8,0.1) -- (0, 0.1) ;
\draw (1.1,0.2) -- (0.8,0.2) ;
\draw (1.1,0) -- (0.8,0);
\draw  [decorate, decoration={snake}] (0.8,0.1) -- (0, 0.1) ;
\filldraw [fill=white] (0,0) circle [radius=4mm] ;
\filldraw [fill=white] (-0.8,0.1) circle [radius=2mm] ;
\filldraw [fill=white] (0.8,0.1) circle [radius=2mm] ;
\draw ($(0,0)+(90:0.4)$) to [out = -90, in = 170] ($(0,0)+(-10:0.4)$) ;
\node (O1) at (0.2,0.2) {$-$} ;
\node (O2) at (-0.05,-0.05) {$+$} ;
\node (O3) at (-0.8,0.1) {$+$} ;
\node (O4) at (0.8,0.1) {$-$} ;
\end{tikzpicture}
\,, \\
%%%%%%%%%%%%%%%%%%%%%%%%%%%%
%%%%%%%%%%%%%%%%%%%%%%%%%%%%
%%%%%%%%%%%%%%%%%%%%%%%%%%%%
%%%%%%%%%%%%%%%%%%%%%%%%%%%%
%%%%%%%%%%%%%%%%%%%%%%%%%%%%
\begin{tikzpicture}[baseline=-2]
\draw (1,0.2) -- (-0.6,0.2);
\draw (1,0) -- (-0.6,0);
\draw (1,-0.2) -- (-0.6,-0.2);
\filldraw [fill=white] (0.5,0) circle [radius=4mm] ;
\node (O1) at (0.5,0) {$-$} ;
\filldraw [fill=white] (-0.3,0.1) circle [radius=2mm] ;
\node (O2) at (-0.3,0.1) {$+$} ;
\end{tikzpicture}
&\sim
\begin{tikzpicture}[baseline=-2]
\draw (-1.3,0.2) -- (-1.6,0.2);
\draw (-1.3,0) -- (-1.6,0);
\draw (0.8,0.2) -- (1.1,0.2);
\draw (0.8,0) -- (1.1,0);
\draw (-1.6,-0.2) -- (1.1,-0.2);
\draw (0,0.2) -- (-0.8,0.2);
\draw (0,0) -- (-0.8,0);
\draw  [decorate, decoration={snake}] (0.7,0.15) -- (0, 0.15) ;
\draw  [decorate, decoration={snake}] (-0.8,0.15) -- (-1.2, 0.15) ;
\filldraw [fill=white] (0,0) circle [radius=4mm] ;
\filldraw [fill=white] (-0.7,0.1) circle [radius=2mm] ;
\filldraw [fill=white] (-1.3,0.1) circle [radius=2mm] ;
\filldraw [fill=white] (0.8,0.1) circle [radius=2mm] ;
\node at (0,0) {$-$} ;
\node at (-0.7,0.1) {$+$} ;
\node at (-1.3,0.1) {$+$}; 
\node at (0.8,0.1) {$-$};
\end{tikzpicture}
\sim
\begin{tikzpicture}[baseline=-2]
\draw (1.1,-0.2) -- (-1.1,-0.2);
\draw (-1.1,0.2) -- (-0.8,0.2) ;
\draw (-1.1,0) -- (-0.8,0);
\draw  [decorate, decoration={snake}] (-0.8,0.1) -- (0, 0.1) ;
\draw (1.1,0.2) -- (0.8,0.2) ;
\draw (1.1,0) -- (0.8,0);
\draw  [decorate, decoration={snake}] (0.8,0.1) -- (0, 0.1) ;
\filldraw [fill=white] (0,0) circle [radius=4mm] ;
\filldraw [fill=white] (-0.8,0.1) circle [radius=2mm] ;
\filldraw [fill=white] (0.8,0.1) circle [radius=2mm] ;
\draw ($(0,0)+(90:0.4)$) to [out = -90, in = 10] ($(0,0)+(190:0.4)$) ;
\node (O1) at (-0.2,0.2) {$+$} ;
\node (O2) at (0.05,-0.05) {$-$} ;
\node (O3) at (-0.8,0.1) {$+$} ;
\node (O4) at (0.8,0.1) {$-$} ;
\end{tikzpicture}
\,.
\end{align}
The l.h.s.~of \eqref{22from44unitarity} can be replaced with the mixed amplitudes. As a result, we obtain the 2-to-2 unstable-particle unitarity equation \eqref{unitarityAphisu} from the 4-to-4 unitarity equation of the stable particle. Here, we stress that the momentum transfer of the embedded $A\varphi \to A \varphi$ amplitude corresponds to $t=s_{1256}=s_{3478}$ which is not necessarily negative in the physical region of the 4-to-4 amplitude. For instance, when we consider
\begin{align}
s_{12}=\ReM+2\delta s\,, ~ s_{348}=\mu^2-2\delta s\,, ~ s_{56}=\ReM
\,,
\end{align}
the $s$-channel region in the large $s$ limit is given by
\begin{align}
-s+\mathcal{O}(s^0)<t<\frac{4\delta s^2}{s} + \mathcal{O}(s^{-2})
\,,
\end{align}
where $s=s_{12348}=s_{567}$ is the total energy variable of the embedded 3-to-3 amplitude.

All in all, we have obtained the unitarity equations which are applicable even away from the negative $t$ region. The unstable-particle unitarity equations are then obtained by extracting their residues as we did in the previous section. The validity of the unstable-particle unitarity equations can be extended to the region $0< t < 4\delta s^2/s$. The allowed value of $\delta s$ is determined by the condition for reaching the correct poles in analytic continuation.

%%%%%%%%%%%%%%%%%%%%%%%%%%%%%%%%%%%
%%%%%%%%%%%%%%%%%%%%%%%%%%%%%%%%%%%
%%%%%%%%%%%%%%%%%%%%%%%%%%%%%%%%%%%
%%%%%%%%%%%%%%%%%%%%%%%%%%%%%%%%%%%
%%%%%%%%%%%%%%%%%%%%%%%%%%%%%%%%%%%

\section{Optical theorem}
\label{sec:optical}
One of the important consequences of unitarity is the optical theorem which fixes the sign of the imaginary part (and the discontinuity multiplied by $i$). In our convention, the imaginary part of the 2-to-2 stable-particle amplitudes has to be negative in the forward limit. The optical theorem has been the basis of unitarity constraints in quantum field theory. Combined with the dispersion relation, the optical theorem provides strong consistency conditions on low-energy effective field theories called positivity bounds~\cite{Pham:1985cr, Ananthanarayan:1994hf, Adams:2006sv}. Motivated by this, let us investigate the sign of the r.h.s.~of the unitarity equations in the unstable-particle amplitudes.

We rename the external states as follows:
\begin{align}
\begin{tikzpicture}[baseline=-2]
\draw (0.7,-0.15) -- (-0.7,-0.15);
\draw  [decorate, decoration={snake}] (-0.7,0.15) -- (0, 0.15) ;
\draw  [decorate, decoration={snake}] (0.7,0.15) -- (0, 0.15) ;
\filldraw [fill=white] (0,0) circle [radius=4mm] ;
\draw ($(0,0)+(90:0.4)$) to [out = -90, in = 170] ($(0,0)+(-10:0.4)$) ;
\node (O1) at (0.2,0.2) {$-$} ;
\node (O2) at (-0.05,-0.05) {$+$} ;
\node (i1) at (0.9,0.15) {1};
\node (i2) at (0.9,-0.15) {2};
\node (o1) at (-0.9,0.15) {3};
\node (o2) at (-0.9,-0.15) {4};
\end{tikzpicture}
=\lim_{\varepsilon \to + 0} 
\Amp_{A \varphi \to A \varphi}(s+i\varepsilon,t)
\label{AmpAphi}
\end{align}
and
\begin{align}
\begin{tikzpicture}[baseline=-2]
\draw  [decorate, decoration={snake}] (-0.7,-0.2) -- (0, -0.2) ;
\draw  [decorate, decoration={snake}] (0.7,-0.2) -- (0, -0.2) ;
\draw  [decorate, decoration={snake}] (-0.7,0.2) -- (0, 0.2) ;
\draw  [decorate, decoration={snake}] (0.7,0.2) -- (0, 0.2) ;
\filldraw [fill=white] (0,0) circle [radius=4mm] ;
\draw ($(0,0)+(90:0.4)$) to [out = -90, in = 180] ($(0,0)+(0:0.4)$) ;
\draw ($(0,0)+(-90:0.4)$) to [out = 90, in = 180] ($(0,0)+(0:0.4)$) ;
\node (O1) at (0.2,0.2) {$-$} ;
\node (O2) at (0.2,-0.2) {$-$} ;
\node (O3) at (-0.1,0) {$+$} ;
\node (i1) at (0.9,0.2) {1};
\node (i2) at (0.9,-0.2) {2};
\node (o1) at (-0.9,0.2) {3};
\node (o2) at (-0.9,-0.2) {4};
\end{tikzpicture}
=\lim_{\varepsilon \to + 0}  \Amp_{AA\to A A}(s+i\varepsilon,t)
\,, \label{AmpAA}
\end{align}
where $s=-(p_1+p_2)^2, t=-(p_1-p_3)^2, u=-(p_1-p_4)^2$. Here, the variable $u$ is eliminated by using
\begin{align}
\begin{cases}
s+t+u=2\mu^2+2\ReM\quad &(A\varphi \to A \varphi)
\,, \\
s+t+u=4\ReM\quad &(AA \to AA)\,,
\end{cases}
\end{align}
and the amplitudes are regarded as a function of $s$ and $t$ at a sufficiently large $s$ so that the direction to approach the real $t$-axis does not matter.
The ``on-shell'' conditions of the external states should be understood as
\begin{align}
p_1^2=-(M^2)^*\,,~ p_2^2=-\mu^2\,,~ p_3^2=-M^2\,,~ p_4^2 = -\mu^2
\,,
\label{onshell}
\end{align}
for \eqref{AmpAphi} whereas
\begin{align}
p_1^2=p_2^2=(M^2)^*\,, \quad p_3^2=p_4^2=M^2\,.
\end{align}
for \eqref{AmpAA}, respectively.

As we have seen, when we consider a sufficiently large $s$, the unitarity equations \eqref{unitarity22}, \eqref{unitarityAphi}, and \eqref{unitarityAA} take the same form which all evaluate the discontinuity across the real $s$-axis. The r.h.s.~of the 2-to-2 unitarity equations are
\begin{align}
&-i\sumint \D \Pi \Amp^{(+)}_{a\to X_1 X_2}(p_3,p_4  | \{q \}) \Amp^{(-)}_{X_1 X_2 \to a}(\{q \} | p_1,p_2)
\label{unitaryrhs}
\end{align}
according to the rule \eqref{rules} where $X_{1,2}$ are either $\varphi$ or $A$ and
\begin{align}
\sumint\D \Pi &=\sum_{a=2} \frac{2\pi }{(2\pi)^{3(a-1)}a!} \int \prod_{i=1}^{a-1} \D^4 k_i \prod_{j=1}^a \theta(q^0_j)\delta(q_j^2+\mu^2) 
\nn
&=\sum_{a=2} \frac{1}{a!} \int \left[ \prod_{i=1}^a \frac{\D^3 {\bm q}_i}{(2\pi)^3} \frac{1}{2q^0_i} \right] (2\pi)^4 \delta^{(4)}(p_{\rm tot}-\sum_{j=1}^a q_j)
\,.
\end{align}
The arguments of the $(\pm)$ amplitudes are expressed by four-momenta; the momenta appearing on the left side of the bar are those of the out states while the right ones are the in states, respectively. The external momenta have to be complexified to satisfy the ``on-shell'' conditions.
The forward limit $p_1=p_3$ does not satisfy the ``on-shell'' conditions for unstable particles.
Let us instead consider 
\begin{align}
p_1 = p_3^*\,, \quad p_2 = p_4^*
\,.
\label{cforward}
\end{align}
The conservation $p_1+p_2=p_3+p_4$ implies that the total momentum $p_1+p_2$ is real under \eqref{cforward}. Then, the internal momenta $\{q\}$ can be real since only the kinematically-allowed internal lines appear in a given $s$. Therefore, by the use of $\Amp^{(-)}_{n'n}(s_A)=[\Amp^{(+)}_{nn'}(s_A^*)]^*$, \eqref{cforward} leads to
\begin{align}
\Amp^{(-)}_{X_1 X_2 \to a}(\{q \} | p_1,p_2) = [\Amp^{(+)}_{a\to X_1 X_2}(p_3, p_4  | \{q \}) ]^*
\,,
\label{complex_rel}
\end{align}
meaning that the integrand of the r.h.s.~is given by the modulus of the amplitude as in the forward limit. We then translate the condition \eqref{cforward} in terms of the Mandelstam variables. Since $p_1+p_2$ is real and timelike for $s>0$, we can move to the centre-of-mass frame in which
\begin{align}
p_1=(\sqrt{(m_{X_1}^2)^*+{\bm p}^2}, {\bm p})\,, \quad p_2=(\sqrt{(m_{X_2}^2)^*+{\bm p}^2}, -{\bm p})
\,,
\label{CM}
\end{align}
with a complexified three-momentum ${\bm p}={\bm p}_R +i {\bm p}_I$ where $m_{X_{1,2}}^2$ are either $M^2$ or $\mu^2$. The momentum conservation is manifest under \eqref{cforward} and \eqref{CM}. As for $X_1=X_2=X$, the energy conservation is solved by $2{\bm p}_R \cdot {\bm p}_I = {\rm Im}\, m_{X}^2$, leading to
\begin{align}
s&=4({\rm Re}\, m_{X}^2+{\bm p}_R^2-{\bm p}_I^2)
\,, \quad
t= 4{\bm p}_I^2
\,.
\end{align}
On the other hand, the generic solution to the energy conservation is not easily expressed for $X_1=A$ and $X_2=\varphi$. We thus present the result in the high-energy limit, $|{\bm p}_R| \gg |m_{X_{1,2}}|, |{\bm p}_I|$, in which $4{\bm p}_R \cdot {\bm p}_I \approx  {\rm Im}(m_{X_1}^2+m_{X_2}^2)$ and
\begin{align}
s&= 4{\bm p}_R^2 +\mathcal{O}({\bm p}_R^0) \,, \quad t = 4{\bm p}_I^2 +\mathcal{O}({\bm p}_R^{-2})
\,,
\label{cforwardst}
\end{align}
where we keep the imaginary parts of both masses for completeness.  
Note that the modulus of ${\bm p}_I$ is bounded from below to satisfy the energy conservation when the external state is unstable; accordingly, there exists a minimum value of $t$ which is given by
\begin{align}
t_{\rm min}(s) \approx \frac{[{\rm Im}(m_{X_1}^2 +  m_{X_2}^2)]^2}{s}
~\text{when}~
s \gg |m_{X_{1,2}}^2|
\end{align}
and in particular
\begin{align}
t_{\rm min}=\frac{1}{2}\left[ 4{\rm Re}m_X^2-s+\sqrt{16({\rm Im}m_X^2)^2+(s-4{\rm Re}m_X^2)^2} \right]
\end{align}
when $X_1=X_2=X$. As shown in Sec.~\ref{sec:extended}, the unitarity equations are extended to the positive momentum transfer region $t<t_*$ with $t_*(s) = 4\delta s^2/s + \mathcal{O}(s^{-2})$. Hence, as long as ${\rm Im}(m_{X_1}^2 +  m_{X_2}^2) < 2\delta s$ holds, we obtain the inequality 
\begin{align}
i \Disc_s \, \Amp_{X_1 X_2 \to X_1 X_2}(s,t) = \sum \hspace{-1.3em} \int \D \Pi |\Amp_{X_1X_2\to a}|^2 > 0
\,,
\label{opticaltheorem}
\end{align}
in $ 0\leq t_{\rm min}(s) \leq t < t_*(s)$ where $t_{\rm min}=0$ holds only if all external states are stable.\footnote{It may be possible to extend the positivity constraint \eqref{opticaltheorem} beyond $t_*$ by analytic continuation as long as analytic continuation does not require a distortion of the original integration hypercontour in the r.h.s.~of the unitarity equations. If the integration hypercontour is distorted, we can no longer use the relation \eqref{complex_rel} and the positivity is not guaranteed. In the stable-particle case, the positive $t$ extension of the positivity constraint up to the normal threshold is known; see e.g.~\cite{Martin:1965jj, deRham:2017avq}.}

%Therefore, as in the stable-particle case, the unstable-particle amplitudes satisfy the inequality \eqref{opticaltheorem} as long as the decay width is not so large ${\rm Im}(m_{X_1}^2 +  m_{X_2}^2) < 2\delta s$. This is consistent with the physical intuition that particles with small decay widths are indistinguishable from stable particles. On the other hand, we could not conclude the positivity for broad resonances ${\rm Im}(m_{X_1}^2 +  m_{X_2}^2) > 2\delta s$ at least by the present analysis. It would be interesting to explore whether the positivity constraint holds for such cases.

\section{Summary}
\label{sec:summary}
The unitarity equations are the fundamental equations in the S-matrix theory. We have derived the unitarity equations for unstable particles from the general properties of the stable-particle amplitudes, namely Lorentz invariance, unitarity, and analyticity. 
The unstable-particle amplitudes have been defined as residues of a higher-point stable-particle amplitude at a complex pole of the unstable particle. The unitarity equations depend on how we choose the complex poles. Since the unitarity equations are supposed to evaluate discontinuities, the analytic structure of the unstable-particle amplitudes should depend on the choice of the complex pole. Although we have not thoroughly discussed the analytic structure in the present paper, as demonstrated in Sec.~\ref{sec:Aphi}, our unitarity equations can be basic tools for investigating analytic properties and then establishing the S-matrix theory for unstable particles.

We have assumed Hermitian analyticity of higher-point amplitudes, while analyticity beyond the 2-to-2 amplitude has not been well understood. It is important to study analyticity from both perturbative and non-perturbative approaches and to confirm their validity. A precise understanding of the analytic structure of the unstable-particle amplitudes is indispensable to derive further general properties.

Meanwhile, it would be interesting to investigate applications. If the dispersion relation is proved (or assumed), the inequality \eqref{opticaltheorem} can lead to positivity bounds on unstable particles.\footnote{Note that we have studied the sign of $i \Disc_s \, \Amp_{X_1 X_2 \to X_1 X_2}$ only in a sufficiently large s; however, this would be sufficient for the positivity bounds because the low-energy part can be computed by using the knowledge of the low-energy theory and the only high-energy part of the discontinuity is used as a positivity constraint on the low-energy theory~\cite{Bellazzini:2016xrt, deRham:2017imi}.} We can study the bounds on not only the lightest state but also the whole spectra of a theory such as unstable particles in the (beyond) Standard Model or Regge states. Furthermore, the non-uniqueness of the unstable-particle amplitudes implies that there might be different sets of unitarity constraints from amplitudes with different choices of complex poles. We may obtain a strong consistency condition on the theory by combining all the possible unitarity constraints. We leave them for future studies.

\begin{acknowledgments}
We would like to thank Toshifumi Noumi, Ryo Saito, Sota Sato, Satoshi Shirai, Junsei Tokuda, and Masahito Yamazaki for useful discussions and comments.
The work of K.A. was supported in part by Grants-in-Aid from the Scientific Research Fund of the Japan Society for the Promotion of Science, No.~20K14468.
\end{acknowledgments}

%%%%%%%%%%%%%%%%%%%%%%%%%%%%%%%
%%%%%%%%%%%%%%%%%%%%%%%%%%%%%%%
%%%%%%%%%%%%%%%%%%%%%%%%%%%%%%%
%%%%%%%%%%%%%%%%%%%%%%%%%%%%%%%
%%%%%%%%%%%%%%%%%%%%%%%%%%%%%%%
%%%%%%%%%%%%%%%%%%%%%%%%%%%%%%%

\appendix
\section{Stable-particle unitarity equations}
\label{sec:stable_unitarity}
Let us briefly review the derivations of unitarity equations of higher-point amplitudes. We write
\begin{align}
\langle \{p' \} |S| \{ p \} \rangle &=
{}_{n'}
\begin{tikzpicture}[baseline=-2]
\draw (0.8,0.3) -- (-0.8,0.3);
\draw (0.8,0.1) -- (-0.8,0.1);
\draw [dotted] (0.8,-0.1) -- (-0.8,-0.1);
\draw (0.8,-0.3) -- (-0.8,-0.3);
\filldraw [fill=white] (0,0) circle [radius=5mm] ;
\node (O) at (0,0) {$S$} ;
\end{tikzpicture}
{}_{n}
\,, \\
\langle \{p' \} |S^{\dagger} |\{ p \} \rangle &= 
{}_{n'}
\begin{tikzpicture}[baseline=-2]
\draw (0.8,0.3) -- (-0.8,0.3);
\draw (0.8,0.1) -- (-0.8,0.1);
\draw [dotted] (0.8,-0.1) -- (-0.8,-0.1);
\draw (0.8,-0.3) -- (-0.8,-0.3);
\filldraw [fill=white] (0,0) circle [radius=5mm] ;
\node (O) at (0,0) {$S^{\dagger}$} ;
\end{tikzpicture}
{}_{n}\,.
\end{align}
Inserting the completeness relation $\sum |{q}\rangle \langle {q}|=1$
to $SS^{\dagger}=1$, the conditions for unitarity are given by
\begin{align}
\begin{tikzpicture}[baseline=-2]
\draw (-1,0.15) -- (-0.5,0.15);
\draw (-1,-0.15) -- (-0.5,-0.15);
\draw (1,0.15) -- (0.5,0.15);
\draw (1,-0.15) -- (0.5,-0.15);
\draw[line width=5pt] (-0.5,0) -- (0.5,0);
\filldraw [fill=white] (-0.5,0) circle [radius=4mm] ;
\node (O1) at (-0.5,0) {$S$} ;
\filldraw [fill=white] (0.5,0) circle [radius=4mm] ;
\node (O2) at (0.5,0) {$S^{\dagger}$} ;
\end{tikzpicture}
&=
\begin{tikzpicture}[baseline=-2]
\draw (-0.5, 0.15) -- (0.5, 0.15);
\draw (-0.5, -0.15) -- (0.5, -0.15);
\end{tikzpicture}
\,, \label{Stwo} \\
\begin{tikzpicture}[baseline=-2]
\draw (-1,0.2) -- (-0.5,0.2);
\draw (-1,0) -- (-0.5,0);
\draw (-1,-0.2) -- (-0.5,-0.2);
\draw (1,0.2) -- (0.5,0.2);
\draw (1,0) -- (0.5,0);
\draw (1,-0.2) -- (0.5,-0.2);
\draw[line width=5pt] (-0.5,0) -- (0.5,0);
\filldraw [fill=white] (-0.5,0) circle [radius=4mm] ;
\node (O1) at (-0.5,0) {$S$} ;
\filldraw [fill=white] (0.5,0) circle [radius=4mm] ;
\node (O2) at (0.5,0) {$S^{\dagger}$} ;
\end{tikzpicture}
&=
\begin{tikzpicture}[baseline=-2]
\draw (-0.5, 0.2) -- (0.5, 0.2);
\draw (-0.5, 0) -- (0.5, 0);
\draw (-0.5, -0.2) -- (0.5, -0.2);
\end{tikzpicture}
\,, \label{Sthree} \\
\begin{tikzpicture}[baseline=-2]
\draw (-1,0.3) -- (-0.5,0.3);
\draw (-1,0.1) -- (-0.5,0.1);
\draw (-1,-0.1) -- (-0.5,-0.1);
\draw (-1,-0.3) -- (-0.5,-0.3);
\draw (1,0.3) -- (0.5,0.3);
\draw (1,0.1) -- (0.5,0.1);
\draw (1,-0.1) -- (0.5,-0.1);
\draw (1,-0.3) -- (0.5,-0.3);
\draw[line width=5pt] (-0.5,0) -- (0.5,0);
\filldraw [fill=white] (-0.5,0) circle [radius=4mm] ;
\node (O1) at (-0.5,0) {$S$} ;
\filldraw [fill=white] (0.5,0) circle [radius=4mm] ;
\node (O2) at (0.5,0) {$S^{\dagger}$} ;
\end{tikzpicture}
&=
\begin{tikzpicture}[baseline=-2]
\draw (-0.5, 0.3) -- (0.5, 0.3);
\draw (-0.5, 0.1) -- (0.5, 0.1);
\draw (-0.5, -0.1) -- (0.5, -0.1);
\draw (-0.5, -0.3) -- (0.5, -0.3);
\end{tikzpicture}
\,, \label{Sfour}
\end{align}
up to $n=n'=4$ where the bold line is short for the summation up to the kinematically-allowed number of the states, e.g.,
\begin{align}
\begin{tikzpicture}[baseline=-2]
\draw (-1,0.15) -- (-0.5,0.15);
\draw (-1,-0.15) -- (-0.5,-0.15);
\draw (1,0.15) -- (0.5,0.15);
\draw (1,-0.15) -- (0.5,-0.15);
\draw[line width=5pt] (-0.5,0) -- (0.5,0);
\filldraw [fill=white] (-0.5,0) circle [radius=4mm] ;
\node (O1) at (-0.5,0) {$S$} ;
\filldraw [fill=white] (0.5,0) circle [radius=4mm] ;
\node (O2) at (0.5,0) {$S^{\dagger}$} ;
\end{tikzpicture}
=
\sum_{a}
\begin{tikzpicture}[baseline=-2]
\draw (-1,0.15) -- (-0.5,0.15);
\draw (-1,-0.15) -- (-0.5,-0.15);
\draw (1,0.15) -- (0.5,0.15);
\draw (1,-0.15) -- (0.5,-0.15);
\draw (-0.5,0.3) -- (0.5,0.3) ;
\draw (-0.5,-0.3) -- (0.5,-0.3) ;
\draw (-0.5,0.1) -- (0.5, 0.1);
\draw[dotted] (-0.5, -0.1 ) -- (0.5,-0.1) ;
\filldraw [fill=white] (-0.5,0) circle [radius=4mm] ;
\node (O1) at (-0.5,0) {$S$} ;
\filldraw [fill=white] (0.5,0) circle [radius=4mm] ;
\node (O2) at (0.5,0) {$S^{\dagger}$} ;
\node at (0,-0.45) {\scalebox{0.7}{$a$} };
\end{tikzpicture}
\,.
\end{align}
The S-matrix elements may be decomposed into disconnected parts and connected parts:
\begin{align}
\begin{tikzpicture}[baseline=-2]
\draw (-1,0.15) -- (-0.5,0.15);
\draw (-1,-0.15) -- (-0.5,-0.15);
\draw[line width=5pt] (-0.5,0) -- (0.1,0);
\filldraw [fill=white] (-0.5,0) circle [radius=4mm] ;
\node (O1) at (-0.5,0) {$S$} ;
\end{tikzpicture}
&=
\begin{tikzpicture}[baseline=-2]
\draw (-1,0.15) -- (-0.5, 0.15);
\draw (-1,-0.15) -- (-0.5, -0.15);
\end{tikzpicture}
+
\begin{tikzpicture}[baseline=-2]
\draw (-1,0.15) -- (-0.5,0.15);
\draw (-1,-0.15) -- (-0.5,-0.15);
\draw[line width=5pt] (-0.5,0) -- (0.1,0);
\filldraw [fill=white] (-0.5,0) circle [radius=4mm] ;
\node (O1) at (-0.5,0) {$+$} ;
\end{tikzpicture}
\,, \label{S2connected}
\\
\begin{tikzpicture}[baseline=-2]
\draw (-1,0.2) -- (-0.5,0.2);
\draw (-1,0) -- (-0.5,0);
\draw (-1,-0.2) -- (-0.5,-0.2);
\draw[line width=5pt] (-0.5,0) -- (0.1,0);
\filldraw [fill=white] (-0.5,0) circle [radius=4mm] ;
\node (O1) at (-0.5,0) {$S$} ;
\end{tikzpicture}
&=
\begin{tikzpicture}[baseline=-2]
\draw (-1,0.2) -- (-0.5,0.2);
\draw (-1,0) -- (-0.5,0);
\draw (-1,-0.2) -- (-0.5,-0.2);
\end{tikzpicture}
+
\sum
\begin{tikzpicture}[baseline=-2]
\draw (-1,0.2) -- (-0.5,0.2);
\draw (-1,0) -- (-0.5,0);
\draw (-1,-0.2) -- (-0.1,-0.2);
\draw[line width=5pt] (-0.5,0.1) -- (-0.1,0.1);
\filldraw [fill=white] (-0.5,0.1) circle [radius=2.5mm] ;
\node (O1) at (-0.5,0.1) {$+$} ;
\end{tikzpicture}
+
\begin{tikzpicture}[baseline=-2]
\draw (-1,0.2) -- (-0.5,0.2);
\draw (-1,0) -- (-0.5,0);
\draw (-1,-0.2) -- (-0.5,-0.2);
\draw[line width=5pt] (-0.5,0) -- (0.1,0);
\filldraw [fill=white] (-0.5,0) circle [radius=4mm] ;
\node (O1) at (-0.5,0) {$+$} ;
\end{tikzpicture}
\,, \\
\begin{tikzpicture}[baseline=-2]
\draw (-1,0.3) -- (-0.5,0.3);
\draw (-1,0.1) -- (-0.5,0.1);
\draw (-1,-0.1) -- (-0.5,-0.1);
\draw (-1,-0.3) -- (-0.5,-0.3);
\draw[line width=5pt] (-0.5,0) -- (0.1,0);
\filldraw [fill=white] (-0.5,0) circle [radius=4mm] ;
\node (O1) at (-0.5,0) {$S$} ;
\end{tikzpicture}
&=
\begin{tikzpicture}[baseline=-2]
\draw (-1,0.3) -- (-0.5,0.3);
\draw (-1,0.1) -- (-0.5,0.1);
\draw (-1,-0.1) -- (-0.5,-0.1);
\draw (-1,-0.3) -- (-0.5,-0.3);
\end{tikzpicture}
+
\sum
\begin{tikzpicture}[baseline=-2]
\draw (-1,0.3) -- (-0.5,0.3);
\draw (-1,0.1) -- (-0.5,0.1);
\draw (-1,-0.1) -- (-0.1,-0.1);
\draw (-1,-0.3) -- (-0.1,-0.3);
\draw[line width=5pt] (-0.5,0.2) -- (-0.1,0.2);
\filldraw [fill=white] (-0.5,0.2) circle [radius=2mm] ;
\node at (-0.5,0.2) {$+$} ;
\end{tikzpicture}
+
\sum
\begin{tikzpicture}[baseline=-2]
\draw (-1,0.3) -- (-0.5,0.3);
\draw (-1,0.1) -- (-0.5,0.1);
\draw (-1,-0.1) -- (-0.5,-0.1);
\draw (-1,-0.3) -- (-0.5,-0.3);
\draw[line width=5pt] (-0.5,0.2) -- (-0.1,0.2);
\filldraw [fill=white] (-0.5,0.2) circle [radius=1.8mm] ;
\node at (-0.5,0.2) {$+$} ;
\draw[line width=5pt] (-0.5,-0.2) -- (-0.1, -0.2);
\filldraw [fill=white] (-0.5, -0.2) circle [radius=1.8mm] ;
\node at (-0.5, -0.2) {$+$} ;
\end{tikzpicture}
\nn
&+
\sum
\begin{tikzpicture}[baseline=-2]
\draw (-1,0.3) -- (-0.5,0.3);
\draw (-1,0.1) -- (-0.5,0.1);
\draw (-1,-0.1) -- (-0.5,-0.1);
\draw (-1,-0.3) -- (-0,-0.3);
\draw[line width=5pt] (-0.5,0.1) -- (-0,0.1);
\filldraw [fill=white] (-0.5,0.1) circle [radius=3mm] ;
\node at (-0.5,0.1) {$+$} ;
\end{tikzpicture}
+
\begin{tikzpicture}[baseline=-2]
\draw (-1,0.3) -- (-0.5,0.3);
\draw (-1,0.1) -- (-0.5,0.1);
\draw (-1,-0.1) -- (-0.5,-0.1);
\draw (-1,-0.3) -- (-0.5,-0.3);
\draw[line width=5pt] (-0.5,0) -- (0.1,0);
\filldraw [fill=white] (-0.5,0) circle [radius=4mm] ;
\node at (-0.5,0) {$+$} ;
\end{tikzpicture}
\,, \\
%%%%%%%%%%%%%%%%%%%%%%%%%%%
%%%%%%%%%%%%%%%%%%%%%%%%%%%
%%%%%%%%%%%%%%%%%%%%%%%%%%%
%%%%%%%%%%%%%%%%%%%%%%%%%%%
%%%%%%%%%%%%%%%%%%%%%%%%%%%
\begin{tikzpicture}[baseline=-2]
\draw (1,0.15) -- (0.5,0.15);
\draw (1,-0.15) -- (0.5,-0.15);
\draw[line width=5pt] (0.5,0) -- (-0.1,0);
\filldraw [fill=white] (0.5,0) circle [radius=4mm] ;
\node (O1) at (0.5,0) {$S^{\dagger}$} ;
\end{tikzpicture}
&=
\begin{tikzpicture}[baseline=-2]
\draw (-1,0.15) -- (-0.5, 0.15);
\draw (-1,-0.15) -- (-0.5, -0.15);
\end{tikzpicture}
-
\begin{tikzpicture}[baseline=-2]
\draw (1,0.15) -- (0.5,0.15);
\draw (1,-0.15) -- (0.5,-0.15);
\draw[line width=5pt] (0.5,0) -- (-0.1,0);
\filldraw [fill=white] (0.5,0) circle [radius=4mm] ;
\node (O1) at (0.5,0) {$-$} ;
\end{tikzpicture}
\,, 
\\
\begin{tikzpicture}[baseline=-2]
\draw (1,0.2) -- (0.5,0.2);
\draw (1,0) -- (0.5,0);
\draw (1,-0.2) -- (0.5,-0.2);
\draw[line width=5pt] (0.5,0) -- (-0.1,0);
\filldraw [fill=white] (0.5,0) circle [radius=4mm] ;
\node (O1) at (0.5,0) {$S^{\dagger}$} ;
\end{tikzpicture}
&=
\begin{tikzpicture}[baseline=-2]
\draw (-1,0.2) -- (-0.5,0.2);
\draw (-1,0) -- (-0.5,0);
\draw (-1,-0.2) -- (-0.5,-0.2);
\end{tikzpicture}
-
\sum
\begin{tikzpicture}[baseline=-2]
\draw (1,0.2) -- (0.5,0.2);
\draw (1,0) -- (0.5,0);
\draw (1,-0.2) -- (0.1,-0.2);
\draw[line width=5pt] (0.5,0.1) -- (0.1,0.1);
\filldraw [fill=white] (0.5,0.1) circle [radius=2.5mm] ;
\node (O1) at (0.5,0.1) {$-$} ;
\end{tikzpicture}
-
\begin{tikzpicture}[baseline=-2]
\draw (1,0.2) -- (0.5,0.2);
\draw (1,0) -- (0.5,0);
\draw (1,-0.2) -- (0.5,-0.2);
\draw[line width=5pt] (0.5,0) -- (-0.1,0);
\filldraw [fill=white] (0.5,0) circle [radius=4mm] ;
\node (O1) at (0.5,0) {$-$} ;
\end{tikzpicture}
\,, \\
\begin{tikzpicture}[baseline=-2]
\draw (1,0.3) -- (0.5,0.3);
\draw (1,0.1) -- (0.5,0.1);
\draw (1,-0.1) -- (0.5,-0.1);
\draw (1,-0.3) -- (0.5,-0.3);
\draw[line width=5pt] (0.5,0) -- (-0.1,0);
\filldraw [fill=white] (0.5,0) circle [radius=4mm] ;
\node (O1) at (0.5,0) {$S^{\dagger}$} ;
\end{tikzpicture}
&=
\begin{tikzpicture}[baseline=-2]
\draw (-1,0.3) -- (-0.5,0.3);
\draw (-1,0.1) -- (-0.5,0.1);
\draw (-1,-0.1) -- (-0.5,-0.1);
\draw (-1,-0.3) -- (-0.5,-0.3);
\end{tikzpicture}
-
\sum
\begin{tikzpicture}[baseline=-2]
\draw (1,0.3) -- (0.5,0.3);
\draw (1,0.1) -- (0.5,0.1);
\draw (1,-0.1) -- (0.1,-0.1);
\draw (1,-0.3) -- (0.1,-0.3);
\draw[line width=5pt] (0.5,0.2) -- (0.1,0.2);
\filldraw [fill=white] (0.5,0.2) circle [radius=2mm] ;
\node at (0.5,0.2) {$-$} ;
\end{tikzpicture}
+
\sum
\begin{tikzpicture}[baseline=-2]
\draw (1,0.3) -- (0.5,0.3);
\draw (1,0.1) -- (0.5,0.1);
\draw (1,-0.1) -- (0.5,-0.1);
\draw (1,-0.3) -- (0.5,-0.3);
\draw[line width=5pt] (0.5,0.2) -- (0.1,0.2);
\filldraw [fill=white] (0.5,0.2) circle [radius=1.8mm] ;
\node at (0.5,0.2) {$-$} ;
\draw[line width=5pt] (0.5,-0.2) -- (0.1, -0.2);
\filldraw [fill=white] (0.5, -0.2) circle [radius=1.8mm] ;
\node at (0.5, -0.2) {$-$} ;
\end{tikzpicture}
\nn
&-
\sum
\begin{tikzpicture}[baseline=-2]
\draw (1,0.3) -- (0.5,0.3);
\draw (1,0.1) -- (0.5,0.1);
\draw (1,-0.1) -- (0.5,-0.1);
\draw (1,-0.3) -- (0,-0.3);
\draw[line width=5pt] (0.5,0.1) -- (0,0.1);
\filldraw [fill=white] (0.5,0.1) circle [radius=3mm] ;
\node at (0.5,0.1) {$-$} ;
\end{tikzpicture}
-
\begin{tikzpicture}[baseline=-2]
\draw (1,0.3) -- (0.5,0.3);
\draw (1,0.1) -- (0.5,0.1);
\draw (1,-0.1) -- (0.5,-0.1);
\draw (1,-0.3) -- (0.5,-0.3);
\draw[line width=5pt] (0.5,0) -- (-0.1,0);
\filldraw [fill=white] (0.5,0) circle [radius=4mm] ;
\node at (0.5,0) {$-$} ;
\end{tikzpicture}
\,.
\label{S4dconnected}
\end{align}
Here, the summations are over possible choices of the particles and the bold lines are used not to specify the number of lines explicitly. Note that the numbers of in/out states on the l.h.s.~and the r.h.s.~have to be the same: for instance, when the bold line in \eqref{S2connected} denotes three in states, the first term on r.h.s.~in \eqref{S2connected} is absent because there is no disconnected diagram with $n=3, n'=2$. We also note that in our notation \eqref{rules}, the $(\pm)$ bubbles denote the amplitudes while the connected parts of the S-matrix elements contain certain factors on top of the amplitudes. However, it turns out that the factors can be consistently omitted~\cite{olive1964exploration,Eden:1966dnq}, so we shall not write these factors for notational simplicity.

We then apply \eqref{S2connected}-\eqref{S4dconnected} to \eqref{Stwo}-\eqref{Sfour} and reorganize the equations. As for $n=n'=2$, we obtain
\begin{align}
\begin{tikzpicture}[baseline=-2]
\draw (-1,0.15) -- (-0.5, 0.15);
\draw (-1,-0.15) -- (-0.5, -0.15);
\end{tikzpicture}
&=
\left(
\begin{tikzpicture}[baseline=-2]
\draw (-1,0.15) -- (-0.5, 0.15);
\draw (-1,-0.15) -- (-0.5, -0.15);
\end{tikzpicture}
+
\begin{tikzpicture}[baseline=-2]
\draw (-1,0.15) -- (-0.5,0.15);
\draw (-1,-0.15) -- (-0.5,-0.15);
\draw[line width=5pt] (-0.5,0) -- (0.1,0);
\filldraw [fill=white] (-0.5,0) circle [radius=4mm] ;
\node (O1) at (-0.5,0) {$+$} ;
\end{tikzpicture}
\right)
\left(
\begin{tikzpicture}[baseline=-2]
\draw (-1,0.15) -- (-0.5, 0.15);
\draw (-1,-0.15) -- (-0.5, -0.15);
\end{tikzpicture}
-
\begin{tikzpicture}[baseline=-2]
\draw (1,0.15) -- (0.5,0.15);
\draw (1,-0.15) -- (0.5,-0.15);
\draw[line width=5pt] (0.5,0) -- (-0.1,0);
\filldraw [fill=white] (0.5,0) circle [radius=4mm] ;
\node (O1) at (0.5,0) {$-$} ;
\end{tikzpicture}
\right)
\nn
&=
\begin{tikzpicture}[baseline=-2]
\draw (-1,0.15) -- (-0.5, 0.15);
\draw (-1,-0.15) -- (-0.5, -0.15);
\end{tikzpicture}
+\Atwo{+}
-\Atwo{-}
-
\begin{tikzpicture}[baseline=-2]
\draw (-1,0.15) -- (-0.5,0.15);
\draw (-1,-0.15) -- (-0.5,-0.15);
\draw (1,0.15) -- (0.5,0.15);
\draw (1,-0.15) -- (0.5,-0.15);
\draw[line width=5pt] (-0.5,0) -- (0.5,0);
\filldraw [fill=white] (-0.5,0) circle [radius=4mm] ;
\node (O1) at (-0.5,0) {$+$} ;
\filldraw [fill=white] (0.5,0) circle [radius=4mm] ;
\node (O2) at (0.5,0) {$-$} ;
\end{tikzpicture}
\,.
\end{align}
The disconnected terms are cancelled, yielding
\begin{align}
\Atwo{+}
-\Atwo{-}
=
\begin{tikzpicture}[baseline=-2]
\draw (-1,0.15) -- (-0.5,0.15);
\draw (-1,-0.15) -- (-0.5,-0.15);
\draw (1,0.15) -- (0.5,0.15);
\draw (1,-0.15) -- (0.5,-0.15);
\draw[line width=5pt] (-0.5,0) -- (0.5,0);
\filldraw [fill=white] (-0.5,0) circle [radius=4mm] ;
\node (O1) at (-0.5,0) {$+$} ;
\filldraw [fill=white] (0.5,0) circle [radius=4mm] ;
\node (O2) at (0.5,0) {$-$} ;
\end{tikzpicture}
\,.
\end{align}
In the case of $n=n'=3$, the unitarity equation involves the product
\begin{align}
\left(
\sum
\begin{tikzpicture}[baseline=-2]
\draw (-1,0.2) -- (-0.5,0.2);
\draw (-1,0) -- (-0.5,0);
\draw (-1,-0.2) -- (-0.1,-0.2);
\draw[line width=5pt] (-0.5,0.1) -- (-0.1,0.1);
\filldraw [fill=white] (-0.5,0.1) circle [radius=2.5mm] ;
\node (O1) at (-0.5,0.1) {$+$} ;
\end{tikzpicture}
\right)
\left(
\sum
\begin{tikzpicture}[baseline=-2]
\draw (1,0.2) -- (0.5,0.2);
\draw (1,0) -- (0.5,0);
\draw (1,-0.2) -- (0.1,-0.2);
\draw[line width=5pt] (0.5,0.1) -- (0.1,0.1);
\filldraw [fill=white] (0.5,0.1) circle [radius=2.5mm] ;
\node (O1) at (0.5,0.1) {$-$} ;
\end{tikzpicture}
\right)
\end{align}
which can be expanded into the disconnected part and the connected part:
\begin{align}
\left(
\sum
\begin{tikzpicture}[baseline=-2]
\draw (-1,0.2) -- (-0.5,0.2);
\draw (-1,0) -- (-0.5,0);
\draw (-1,-0.2) -- (-0.1,-0.2);
\draw[line width=5pt] (-0.5,0.1) -- (-0.1,0.1);
\filldraw [fill=white] (-0.5,0.1) circle [radius=2.5mm] ;
\node (O1) at (-0.5,0.1) {$+$} ;
\end{tikzpicture}
\right)
\left(
\sum
\begin{tikzpicture}[baseline=-2]
\draw (1,0.2) -- (0.5,0.2);
\draw (1,0) -- (0.5,0);
\draw (1,-0.2) -- (0.1,-0.2);
\draw[line width=5pt] (0.5,0.1) -- (0.1,0.1);
\filldraw [fill=white] (0.5,0.1) circle [radius=2.5mm] ;
\node (O1) at (0.5,0.1) {$-$} ;
\end{tikzpicture}
\right)
=
\sum
\begin{tikzpicture}[baseline=-2]
\draw (-0.65,0.2) -- (-0.3,0.2);
\draw (-0.65,0) -- (-0.3,0);
\draw (0.65,0.2) -- (0.3,0.2);
\draw (0.65,0) -- (0.3,0);
\draw (-0.65,-0.2) -- (0.65,-0.2);
\draw[line width=5pt] (-0.3,0.1) -- (0.3,0.1);
\filldraw [fill=white] (-0.3,0.1) circle [radius=2.5mm] ;
\node at (-0.3,0.1) {$+$} ;
\filldraw [fill=white] (0.3,0.1) circle [radius=2.5mm] ;
\node at (0.3,0.1) {$-$} ;
\end{tikzpicture}
+
\sum
\begin{tikzpicture}[baseline=-2]
\draw (-0.65,0.2) -- (0.65,0.2);
\draw (-0.65,0) -- (-0.3,0);
\draw (0.65,0) -- (0.3,0);
\draw (-0.65,-0.2) -- (0.65,-0.2);
\draw[line width=5pt] (-0.3,0.1) -- (0.3,-0.1);
\filldraw [fill=white] (-0.3,0.1) circle [radius=2.5mm] ;
\node at (-0.3,0.1) {$+$} ;
\filldraw [fill=white] (0.3,-0.1) circle [radius=2.5mm] ;
\node at (0.3,-0.1) {$-$} ;
\end{tikzpicture}
\,.
\end{align}
In this way, we rearrange \eqref{Sthree} based on the connectedness structure and find
\begin{align}
0&=\left(
\begin{tikzpicture}[baseline=-2]
\draw (-1,0.2) -- (-0.5,0.2);
\draw (-1,0) -- (-0.5,0);
\draw (-1,-0.2) -- (-0.5,-0.2);
\end{tikzpicture} 
-
\begin{tikzpicture}[baseline=-2]
\draw (-1,0.2) -- (-0.5,0.2);
\draw (-1,0) -- (-0.5,0);
\draw (-1,-0.2) -- (-0.5,-0.2);
\end{tikzpicture}
\right)
\nn
&+\sum
\left(
\begin{tikzpicture}[baseline=-2]
\draw (-0.4,0.2) -- (0.4,0.2);
\draw (-0.4, 0) -- (0.4,0);
\draw (-0.4, -0.2) -- (0.4, -0.2);
\filldraw [fill=white] (0,0.1) circle [radius=2.5mm] ;
\node at (0,0.1) {$+$} ;
\end{tikzpicture}
-
\begin{tikzpicture}[baseline=-2]
\draw (-0.4,0.2) -- (0.4,0.2);
\draw (-0.4, 0) -- (0.4,0);
\draw (-0.4, -0.2) -- (0.4, -0.2);
\filldraw [fill=white] (0,0.1) circle [radius=2.5mm] ;
\node at (0,0.1) {$-$} ;
\end{tikzpicture}
-
\begin{tikzpicture}[baseline=-2]
\draw (-0.65,0.2) -- (-0.3,0.2);
\draw (-0.65,0) -- (-0.3,0);
\draw (0.65,0.2) -- (0.3,0.2);
\draw (0.65,0) -- (0.3,0);
\draw (-0.65,-0.2) -- (0.65,-0.2);
\draw[line width=5pt] (-0.3,0.1) -- (0.3,0.1);
\filldraw [fill=white] (-0.3,0.1) circle [radius=2.5mm] ;
\node at (-0.3,0.1) {$+$} ;
\filldraw [fill=white] (0.3,0.1) circle [radius=2.5mm] ;
\node at (0.3,0.1) {$-$} ;
\end{tikzpicture}
\right)
\nn
&+\Big(
\Athree{+}-\Athree{-}
-
\begin{tikzpicture}[baseline=-2]
\draw (-1,0.2) -- (-0.5,0.2);
\draw (-1,0) -- (-0.5,0);
\draw (-1,-0.2) -- (-0.5,-0.2);
\draw (1,0.2) -- (0.5,0.2);
\draw (1,0) -- (0.5,0);
\draw (1,-0.2) -- (0.5,-0.2);
\draw[line width=5pt] (-0.5,0) -- (0.5,0);
\filldraw [fill=white] (-0.5,0) circle [radius=4mm] ;
\node at (-0.5,0) {$+$} ;
\filldraw [fill=white] (0.5,0) circle [radius=4mm] ;
\node at (0.5,0) {$-$} ;
\end{tikzpicture}
\nn
&\qquad
-
\sum
\begin{tikzpicture}[baseline=-2]
\draw (-1,0.2) -- (-0.5,0.2);
\draw (-1,0) -- (-0.5,0);
\draw (-1,-0.2) -- (0.65,-0.2);
\draw (0.65,0.2) -- (0.3,0.2);
\draw (0.65,0) -- (0.3,0);
\draw[line width=5pt] (-0.5,0.1) -- (0.3,0.1);
\filldraw [fill=white] (-0.5,0) circle [radius=4mm] ;
\node at (-0.5,0) {$+$} ;
\filldraw [fill=white] (0.3,0.1) circle [radius=2.5mm] ;
\node at (0.3,0.1) {$-$} ;
\end{tikzpicture}
-
\sum
\begin{tikzpicture}[baseline=-2]
\draw (1,0.2) -- (0.5,0.2);
\draw (1,0) -- (0.5,0);
\draw (1,-0.2) -- (-0.65,-0.2);
\draw (-0.65,0.2) -- (-0.35,0.2);
\draw (-0.65,0) -- (-0.35,0);
\draw[line width=5pt] (0.5,0.1) -- (-0.3,0.1);
\filldraw [fill=white] (0.5,0) circle [radius=4mm] ;
\node at (0.5,0) {$-$} ;
\filldraw [fill=white] (-0.3,0.1) circle [radius=2.5mm] ;
\node at (-0.3,0.1) {$+$} ;
\end{tikzpicture}
-
\sum
\begin{tikzpicture}[baseline=-2]
\draw (-0.65,0.2) -- (0.65,0.2);
\draw (-0.65,0) -- (-0.3,0);
\draw (0.65,0) -- (0.3,0);
\draw (-0.65,-0.2) -- (0.65,-0.2);
\draw[line width=5pt] (-0.3,0.1) -- (0.3,-0.1);
\filldraw [fill=white] (-0.3,0.1) circle [radius=2.5mm] ;
\node at (-0.3,0.1) {$+$} ;
\filldraw [fill=white] (0.3,-0.1) circle [radius=2.5mm] ;
\node at (0.3,-0.1) {$-$} ;
\end{tikzpicture}
\Big)
\,.
\end{align}
The connected part yields the 3-to-3 unitarity equation \eqref{unitarity33}.

The 4-to-4 unitarity equation can be similarly derived and the complete expression in $(4\mu)^2<s<(5\mu)^2$ is given in~\cite{Eden:1966dnq}. However, for our purpose, we only need terms that have certain singular structures in the unitarity equation. As we have seen, the unitarity equations arise from the connected parts. Hence, we write
\begin{align}
0&=
\left(
\begin{tikzpicture}[baseline=-2]
\draw (-1,0.3) -- (-0.5,0.3);
\draw (-1,0.1) -- (-0.5,0.1);
\draw (-1,-0.1) -- (-0.5,-0.1);
\draw (-1,-0.3) -- (-0.5,-0.3);
\draw (1,0.3) -- (0.5,0.3);
\draw (1,0.1) -- (0.5,0.1);
\draw (1,-0.1) -- (0.5,-0.1);
\draw (1,-0.3) -- (0.5,-0.3);
\draw[line width=5pt] (-0.5,0) -- (0.5,0);
\filldraw [fill=white] (-0.5,0) circle [radius=4mm] ;
\node (O1) at (-0.5,0) {$S$} ;
\filldraw [fill=white] (0.5,0) circle [radius=4mm] ;
\node (O2) at (0.5,0) {$S^{\dagger}$} ;
\end{tikzpicture}
\right)_c
\nn
&=
\Big[
\Big(
\begin{tikzpicture}[baseline=-2]
\draw (-1,0.3) -- (-0.5,0.3);
\draw (-1,0.1) -- (-0.5,0.1);
\draw (-1,-0.1) -- (-0.5,-0.1);
\draw (-1,-0.3) -- (-0.5,-0.3);
\end{tikzpicture}
+
\sum
\begin{tikzpicture}[baseline=-2]
\draw (-1,0.3) -- (-0.5,0.3);
\draw (-1,0.1) -- (-0.5,0.1);
\draw (-1,-0.1) -- (-0.1,-0.1);
\draw (-1,-0.3) -- (-0.1,-0.3);
\draw[line width=5pt] (-0.5,0.2) -- (-0.1,0.2);
\filldraw [fill=white] (-0.5,0.2) circle [radius=2mm] ;
\node at (-0.5,0.2) {$+$} ;
\end{tikzpicture}
+
\sum
\begin{tikzpicture}[baseline=-2]
\draw (-1,0.3) -- (-0.5,0.3);
\draw (-1,0.1) -- (-0.5,0.1);
\draw (-1,-0.1) -- (-0.5,-0.1);
\draw (-1,-0.3) -- (-0.5,-0.3);
\draw[line width=5pt] (-0.5,0.2) -- (-0.1,0.2);
\filldraw [fill=white] (-0.5,0.2) circle [radius=1.8mm] ;
\node at (-0.5,0.2) {$+$} ;
\draw[line width=5pt] (-0.5,-0.2) -- (-0.1, -0.2);
\filldraw [fill=white] (-0.5, -0.2) circle [radius=1.8mm] ;
\node at (-0.5, -0.2) {$+$} ;
\end{tikzpicture}
\nn
&\qquad +
\sum
\begin{tikzpicture}[baseline=-2]
\draw (-1,0.3) -- (-0.5,0.3);
\draw (-1,0.1) -- (-0.5,0.1);
\draw (-1,-0.1) -- (-0.5,-0.1);
\draw (-1,-0.3) -- (-0,-0.3);
\draw[line width=5pt] (-0.5,0.1) -- (-0,0.1);
\filldraw [fill=white] (-0.5,0.1) circle [radius=3mm] ;
\node at (-0.5,0.1) {$+$} ;
\end{tikzpicture}
+
\begin{tikzpicture}[baseline=-2]
\draw (-1,0.3) -- (-0.5,0.3);
\draw (-1,0.1) -- (-0.5,0.1);
\draw (-1,-0.1) -- (-0.5,-0.1);
\draw (-1,-0.3) -- (-0.5,-0.3);
\draw[line width=5pt] (-0.5,0) -- (0.1,0);
\filldraw [fill=white] (-0.5,0) circle [radius=4mm] ;
\node at (-0.5,0) {$+$} ;
\end{tikzpicture}
\Big)
\nn
&\qquad
\times \Big(
\begin{tikzpicture}[baseline=-2]
\draw (-1,0.3) -- (-0.5,0.3);
\draw (-1,0.1) -- (-0.5,0.1);
\draw (-1,-0.1) -- (-0.5,-0.1);
\draw (-1,-0.3) -- (-0.5,-0.3);
\end{tikzpicture}
-
\sum
\begin{tikzpicture}[baseline=-2]
\draw (1,0.3) -- (0.5,0.3);
\draw (1,0.1) -- (0.5,0.1);
\draw (1,-0.1) -- (0.1,-0.1);
\draw (1,-0.3) -- (0.1,-0.3);
\draw[line width=5pt] (0.5,0.2) -- (0.1,0.2);
\filldraw [fill=white] (0.5,0.2) circle [radius=2mm] ;
\node at (0.5,0.2) {$-$} ;
\end{tikzpicture}
+
\sum
\begin{tikzpicture}[baseline=-2]
\draw (1,0.3) -- (0.5,0.3);
\draw (1,0.1) -- (0.5,0.1);
\draw (1,-0.1) -- (0.5,-0.1);
\draw (1,-0.3) -- (0.5,-0.3);
\draw[line width=5pt] (0.5,0.2) -- (0.1,0.2);
\filldraw [fill=white] (0.5,0.2) circle [radius=1.8mm] ;
\node at (0.5,0.2) {$-$} ;
\draw[line width=5pt] (0.5,-0.2) -- (0.1, -0.2);
\filldraw [fill=white] (0.5, -0.2) circle [radius=1.8mm] ;
\node at (0.5, -0.2) {$-$} ;
\end{tikzpicture}
\nn
&\qquad-
\sum
\begin{tikzpicture}[baseline=-2]
\draw (1,0.3) -- (0.5,0.3);
\draw (1,0.1) -- (0.5,0.1);
\draw (1,-0.1) -- (0.5,-0.1);
\draw (1,-0.3) -- (0,-0.3);
\draw[line width=5pt] (0.5,0.1) -- (0,0.1);
\filldraw [fill=white] (0.5,0.1) circle [radius=3mm] ;
\node at (0.5,0.1) {$-$} ;
\end{tikzpicture}
-
\begin{tikzpicture}[baseline=-2]
\draw (1,0.3) -- (0.5,0.3);
\draw (1,0.1) -- (0.5,0.1);
\draw (1,-0.1) -- (0.5,-0.1);
\draw (1,-0.3) -- (0.5,-0.3);
\draw[line width=5pt] (0.5,0) -- (-0.1,0);
\filldraw [fill=white] (0.5,0) circle [radius=4mm] ;
\node at (0.5,0) {$-$} ;
\end{tikzpicture}
\Big)
\Big]_c
\,.
\label{unitarity44c}
\end{align}
To obtain the expression \eqref{unitarity44}, we only need to keep the terms that the pairs of external lines can connect to a single bubble. Hence, we can ignore terms with
\scalebox{0.7}{
\begin{tikzpicture}[baseline=-2]
\draw (-1,0.3) -- (-0.5,0.3);
\draw (-1,0.1) -- (-0.5,0.1);
\draw (-1,-0.1) -- (-0.5,-0.1);
\draw (-1,-0.3) -- (-0,-0.3);
\draw[line width=5pt] (-0.5,0.1) -- (-0,0.1);
\filldraw [fill=white] (-0.5,0.1) circle [radius=3mm] ;
\node at (-0.5,0.1) {$+$} ;
\end{tikzpicture}}
and
\scalebox{0.7}{
\begin{tikzpicture}[baseline=-2]
\draw (1,0.3) -- (0.5,0.3);
\draw (1,0.1) -- (0.5,0.1);
\draw (1,-0.1) -- (0.5,-0.1);
\draw (1,-0.3) -- (0,-0.3);
\draw[line width=5pt] (0.5,0.1) -- (0,0.1);
\filldraw [fill=white] (0.5,0.1) circle [radius=3mm] ;
\node at (0.5,0.1) {$-$} ;
\end{tikzpicture}},
and find \eqref{unitarity44} by expanding \eqref{unitarity44c} with only keeping the connected terms that the pairs of external lines connect to a single bubble. The equation \eqref{unitarity44emb} is obtained by expanding
\begin{align}
\Big[ &\Big(
\sum
\begin{tikzpicture}[baseline=-2]
\draw (-1,0.3) -- (-0.5,0.3);
\draw (-1,0.1) -- (-0.5,0.1);
\draw (-1,-0.1) -- (-0.1,-0.1);
\draw (-1,-0.3) -- (-0.1,-0.3);
\draw[line width=5pt] (-0.5,0.2) -- (-0.1,0.2);
\filldraw [fill=white] (-0.5,0.2) circle [radius=2mm] ;
\node at (-0.5,0.2) {$+$} ;
\end{tikzpicture}
+
\sum
\begin{tikzpicture}[baseline=-2]
\draw (-1,0.3) -- (-0.5,0.3);
\draw (-1,0.1) -- (-0.5,0.1);
\draw (-1,-0.1) -- (-0.5,-0.1);
\draw (-1,-0.3) -- (-0,-0.3);
\draw[line width=5pt] (-0.5,0.1) -- (-0,0.1);
\filldraw [fill=white] (-0.5,0.1) circle [radius=3mm] ;
\node at (-0.5,0.1) {$+$} ;
\end{tikzpicture}
+
\begin{tikzpicture}[baseline=-2]
\draw (-1,0.3) -- (-0.5,0.3);
\draw (-1,0.1) -- (-0.5,0.1);
\draw (-1,-0.1) -- (-0.5,-0.1);
\draw (-1,-0.3) -- (-0.5,-0.3);
\draw[line width=5pt] (-0.5,0) -- (0.1,0);
\filldraw [fill=white] (-0.5,0) circle [radius=4mm] ;
\node at (-0.5,0) {$+$} ;
\end{tikzpicture}
\Big)
\nn
&\times \Big(
-
\sum
\begin{tikzpicture}[baseline=-2]
\draw (1,0.3) -- (0.5,0.3);
\draw (1,0.1) -- (0.5,0.1);
\draw (1,-0.1) -- (0.1,-0.1);
\draw (1,-0.3) -- (0.1,-0.3);
\draw[line width=5pt] (0.5,0.2) -- (0.1,0.2);
\filldraw [fill=white] (0.5,0.2) circle [radius=2mm] ;
\node at (0.5,0.2) {$-$} ;
\end{tikzpicture}
+
\sum
\begin{tikzpicture}[baseline=-2]
\draw (1,0.3) -- (0.5,0.3);
\draw (1,0.1) -- (0.5,0.1);
\draw (1,-0.1) -- (0.5,-0.1);
\draw (1,-0.3) -- (0.5,-0.3);
\draw[line width=5pt] (0.5,0.2) -- (0.1,0.2);
\filldraw [fill=white] (0.5,0.2) circle [radius=1.8mm] ;
\node at (0.5,0.2) {$-$} ;
\draw[line width=5pt] (0.5,-0.2) -- (0.1, -0.2);
\filldraw [fill=white] (0.5, -0.2) circle [radius=1.8mm] ;
\node at (0.5, -0.2) {$-$} ;
\end{tikzpicture}
-
\begin{tikzpicture}[baseline=-2]
\draw (1,0.3) -- (0.5,0.3);
\draw (1,0.1) -- (0.5,0.1);
\draw (1,-0.1) -- (0.5,-0.1);
\draw (1,-0.3) -- (0.5,-0.3);
\draw[line width=5pt] (0.5,0) -- (-0.1,0);
\filldraw [fill=white] (0.5,0) circle [radius=4mm] ;
\node at (0.5,0) {$-$} ;
\end{tikzpicture}
\Big)
\Big]_c \sim 0
\,,
\end{align}
and picking up terms satisfying (i), (ii), and (iii) mentioned in Sec.~\ref{sec:extended}.
Here, 
\scalebox{0.7}{\begin{tikzpicture}[baseline=-2]
\draw (-1,0.3) -- (-0.5,0.3);
\draw (-1,0.1) -- (-0.5,0.1);
\draw (-1,-0.1) -- (-0.5,-0.1);
\draw (-1,-0.3) -- (-0.5,-0.3);
\end{tikzpicture}},
\scalebox{0.7}{\begin{tikzpicture}[baseline=-2]
\draw (-1,0.3) -- (-0.5,0.3);
\draw (-1,0.1) -- (-0.5,0.1);
\draw (-1,-0.1) -- (-0.5,-0.1);
\draw (-1,-0.3) -- (-0.5,-0.3);
\draw[line width=5pt] (-0.5,0.2) -- (-0.1,0.2);
\filldraw [fill=white] (-0.5,0.2) circle [radius=1.8mm] ;
\node at (-0.5,0.2) {$+$} ;
\draw[line width=5pt] (-0.5,-0.2) -- (-0.1, -0.2);
\filldraw [fill=white] (-0.5, -0.2) circle [radius=1.8mm] ;
\node at (-0.5, -0.2) {$+$} ;
\end{tikzpicture}},
and
\scalebox{0.7}{\begin{tikzpicture}[baseline=-2]
\draw (1,0.3) -- (0.5,0.3);
\draw (1,0.1) -- (0.5,0.1);
\draw (1,-0.1) -- (0.5,-0.1);
\draw (1,-0.3) -- (0,-0.3);
\draw[line width=5pt] (0.5,0.1) -- (0,0.1);
\filldraw [fill=white] (0.5,0.1) circle [radius=3mm] ;
\node at (0.5,0.1) {$-$} ;
\end{tikzpicture}}
are omitted because they do not satisfy (i), (ii), and (iii). Note that in the main text the subenergy variables are assumed to satisfy $(2\mu)^2<s_{12},s_{34},s_{56},s_{78}<(3\mu)^2$ for simplicity. Hence, diagrams such as
\begin{align*}
\begin{tikzpicture}[baseline=-2]
\draw (-0.7,0.3) -- (0.7,0.3);
\draw (-0.7,0.1) -- (0.7,0.1);
\draw (-0.7,-0.3) -- (0.7,-0.3);
\draw (-0.7,-0.1) -- (0.7,-0.1);
\draw (-0.35,0.1) -- (0.35,-0.1);
\filldraw [fill=white] (-0.35,0.2) circle [radius=2.5mm] ;
\node at (-0.35,0.2) {$+$} ;
\filldraw [fill=white] (0.35,-0.2) circle [radius=2.5mm] ;
\node at (0.35,-0.2) {$-$} ;
\end{tikzpicture}
\,, \quad
\begin{tikzpicture}[baseline=-2]
\draw (-0.7,0.3) -- (0.7,0.3);
\draw (-0.7,0.1) -- (0.7,0.1);
\draw (-0.7,-0.3) -- (0.7,-0.3);
\draw (-0.7,-0.1) -- (0.7,-0.1);
\draw (-0.35,-0.1) -- (0.35,0.1);
\filldraw [fill=white] (-0.35,0.2) circle [radius=1.8mm] ;
\node at (-0.35,0.2) {$+$} ;
\filldraw [fill=white] (-0.35,-0.2) circle [radius=1.8mm] ;
\node at (-0.35,-0.2) {$+$} ;
\filldraw [fill=white] (0.35,0.2) circle [radius=1.8mm] ;
\node at (0.35, 0.2) {$-$} ;
\end{tikzpicture}
\,,
\end{align*}
do not appear in the unitarity equation.

%%%%%%%%%%%%%%%%%%%%%%%%%%%%%
%%%%%%%%%%%%%%%%%%%%%%%%%%%%%
%%%%%%%%%%%%%%%%%%%%%%%%%%%%%
%%%%%%%%%%%%%%%%%%%%%%%%%%%%%
%%%%%%%%%%%%%%%%%%%%%%%%%%%%%

\section{Triangle diagram}
\label{sec:triangle}
In this Appendix, we consider the one-loop Feynman diagram
\begin{align}
-i\Amp_{\rm tri}(t,s_1,s_3)=
\begin{tikzpicture}[baseline=-2]
\draw  [decorate, decoration={snake}] (-0.4,0.3) -- (-0.8, 0.3) ;
\draw  [decorate, decoration={snake}] (0.4,0.3) -- (0.8, 0.3) ;
\draw (-0.4,0.3) -- (0.4,0.3) ; 
\draw (-0.8,-0.3) -- (0.8,-0.3);
\draw (-0.4,0.3) -- (0,-0.3) -- (0.4,0.3);
\node at (1,0.3) {\scalebox{0.8}{$p_1$}};
\node at (1,-0.3) {\scalebox{0.8}{$p_2$}};
\node at (-1,0.3) {\scalebox{0.8}{$p_3$}};
\node at (-1,-0.3) {\scalebox{0.8}{$p_4$}};
\node at (0,0.5) {\scalebox{0.8}{$q_2$}};
\node at (0.4,-0.05) {\scalebox{0.8}{$q_1$}};
\node at (-0.4,-0.05) {\scalebox{0.8}{$q_3$}};
\end{tikzpicture}
\label{deftri}
\end{align}
with the assigned external and internal four-momenta where $q_i$ are fixed by $p_i$ and the loop momentum $k$. Here, we have introduced $t=-(p_1-p_3)^2=-(p_2-p_4)^2, s_1=-p_1^2$ and $s_3=-p_3^2$. The internal lines are supposed to have a real mass $\mu$ while the external lines are ``on the mass shell'' $m_{X_i}^2=-p_i^2~(i=1,3)$ where $m_{X_i}^2$ is either $M^2$ or $(M^2)^*$. The external states 2 and 4 can be either the stable particle or the unstable particle since the diagram \eqref{deftri} is independent of $p_2^2$ and $p_4^2$ but we assign the stable particles to them for simplicity of the discussion. Strictly speaking, complexifying $m_{X_i}^2$ requires embedding the diagram in a 3-to-3 amplitude and a resummation of loop corrections to the propagator of the unstable particle. The variables $s_i=-p_i^2$ should be understood as the subenergy variables of the 3-to-3 diagram and then \eqref{deftri} is extracted by considering integration contours encircling the complex poles in the complex $s_i$-planes. In practice, we may directly compute the Feynman diagram \eqref{deftri} for real $s_i$ by following the standard $i \varepsilon$ prescription and then impose the ``on-shell'' conditions $s_i = m_{X_i}^2$ by analytic continuation (see Fig.~\ref{fig:complexpole} for the paths of the continuation). The Feynman diagram, as is well-known, gives
\begin{align}
-i \Amp_{\rm tri}&= -\int \frac{\D^4 k}{ (2 \pi)^4 } \prod_{i=1}^3 \frac{1}{q_i^2+\mu^2-i \varepsilon} = \frac{i}{(4\pi)^2} \mathcal{I}_{\rm tri}
\,, \\
 \mathcal{I}_{\rm tri}&:= \int_0^1 \left[ \prod_{i=1}^3 \D \alpha_i \right] \delta(1-\sum_{i=1}^3 \alpha_i ) \frac{1}{D}
 \,, \label{triintegral}
\end{align}
with
\begin{align}
D &= \alpha_1 \alpha_2  s_1 + \alpha_2 \alpha_3 s_3 + \alpha_1  \alpha_3 t 
\nn
&- (\mu^2 -i \varepsilon)(\alpha_1+\alpha_2+\alpha_3)^2
\,,
\end{align}
where the coupling constants are omitted.

Unitarity predicts that the all-$(+)$ amplitude has the triangle singularity while the mixed-type amplitude does not. 
We study the analytic structure of the integral \eqref{triintegral} by studying the Landau equations~\cite{Landau:1959fi} and check this statement at the perturbative level. An extensive review on this subject is given in~\cite{Eden:1966dnq} and we follow their methods. Since the ``on-shell'' conditions are imposed after the analytic condition, we first regard $s_1$ and $s_3$ as complex variables. The integral \eqref{triintegral} may be singular when the integrand is singular, i.e.~$D=0$. However, in general, the integration hypercontour can be distorted to avoid the $D=0$ hypersurface. The singularity of \eqref{triintegral} can be found only if such a distortion ceases to exist. When the Landau equations
\begin{align}
\alpha_i \frac{\partial D}{\partial \alpha_i} = 0 \quad \text{for each $i$}
\,,
\label{Landau_tri}
\end{align}
have a non-trivial solution for $\alpha_i$ where $\varepsilon \to 0$ is understood, a part of the $\alpha_i$-space is trapped by the $D=0$ hypersurface (and the boundary of the integration hypercontour $\alpha_i=0$), leading to a necessary condition for the integral \eqref{triintegral} to be singular.
Note that since $D$ is a homogeneous function of $\alpha_i$, the delta function in \eqref{triintegral} can be ignored to analyse \eqref{Landau_tri} and $D=0$ is automatically satisfied when \eqref{Landau_tri} holds. The sufficient condition is that the integration hypercontour is actually trapped, which we will discuss later. The singularity with $\alpha_i \neq 0$ for all $i$ is called the leading singularity and the singularity having $\alpha_i=0$ is called the lower-order singularity, respectively. The leading singularity corresponds to the singularity that all the internal lines are on-shell which is the triangle singularity in the case of \eqref{deftri}. The necessary condition for the leading singularity is
\begin{align}
&{\rm det} \frac{\partial^2 D}{\partial \alpha_i \partial \alpha_j }
\nn
&= 2\mu^2 [t^2-t(2s_1+2s_3-s_1 s_3/\mu^2)+(s_1-s_3)^2]
=0
\,. \label{Landautri}
\end{align}
On the other hand, the lower-order singularities may be found when either one of the following is satisfied:
\begin{align}
t(t-4\mu^2) &= 0 
\,, \label{Landaut} \\
s_1 (s_1-4\mu^2) &=0
\,, \label{Landaus1} \\
s_3 (s_3-4\mu^2) &=0
\,. \label{Landaus3}
\end{align}
The equations \eqref{Landautri}-\eqref{Landaus3} describe the would-be singular hypersurfaces in the complex $(t,s_1,s_3)$-space which we denote by $\Sigma_{\rm tri}, \Sigma_t, \Sigma_1,$ and $\Sigma_3$, respectively.

The $i\varepsilon$ prescription guarantees that the integral \eqref{triintegral} is regular in the real $(t,s_1,s_3)$-space that defines the $(+)$ region on the Riemann surface. The integral \eqref{triintegral} is continued analytically into the region ${\rm Im}\, t>0, {\rm Im}\, s_1 >0, {\rm Im}\, s_3>0$ because $D$ does not vanish for positive $\alpha_i$ which is the undistorted integration hypercontour. On the other hand, singularities may be found when the real axis is passed. Having understood the $(+)$ region, we omit $i\varepsilon$ in the following. For real $(t,s_1,s_3)$, the singular hypersurface $D=0$ in the $\alpha_i$-space is symmetrical about the real $\alpha_i$-plane; if a part of $D=0$ intersects the real $\alpha_i$-plane, the complex conjugate part also intersects the real $\alpha_i$-plane, trapping the real $\alpha_i$-plane. Therefore, singularities in the vicinity of the $(+)$ region are found when the Landau equations have a non-trivial solution for positive $\alpha_i$.

\begin{figure*}[t]
\centering
 \includegraphics[width=0.8\linewidth]{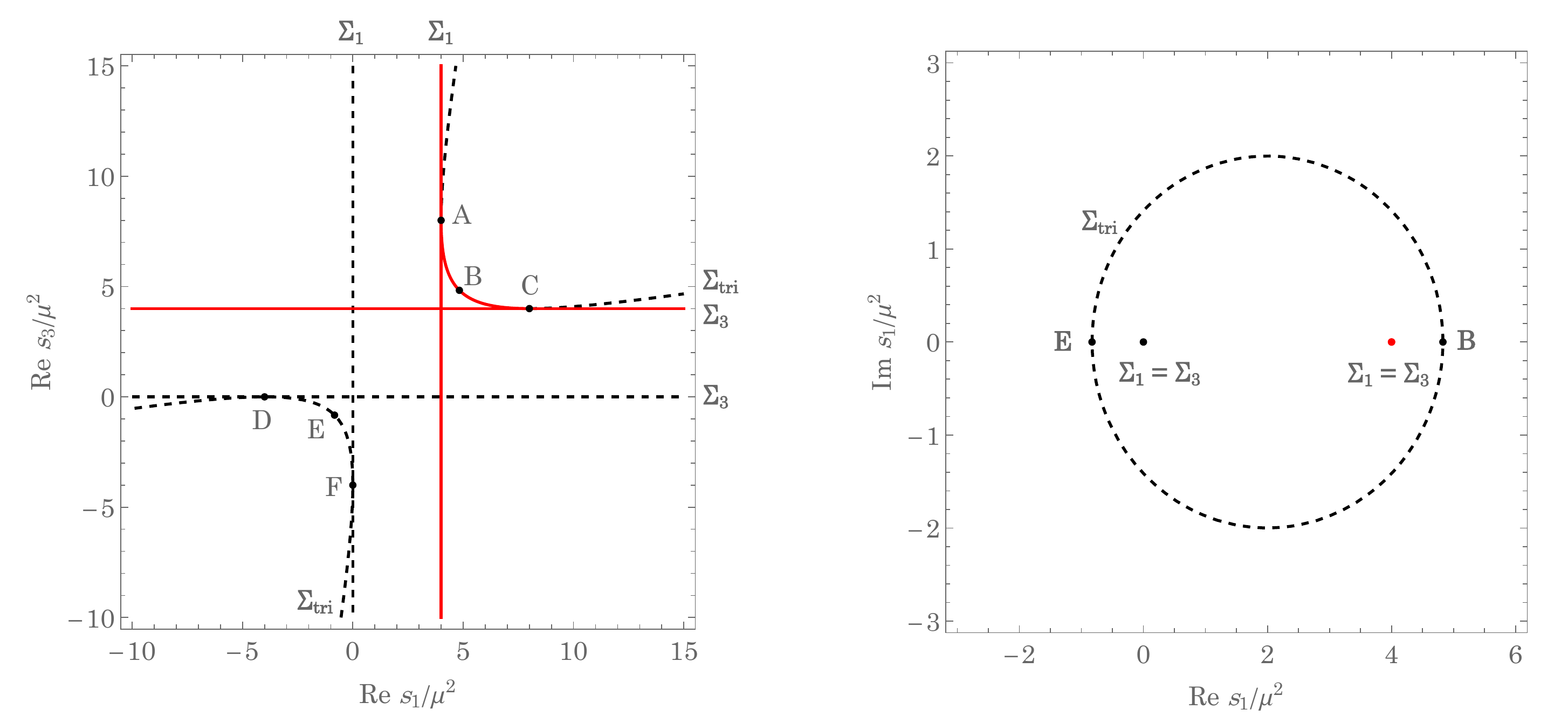}	
\caption{Two-dimensional sections of the complex $(s_1,s_3)$-space: ${\rm Im}\,s_1={\rm Im}\, s_3=0$ (left) and $s_1=s_3^*$ (right). The black dashed curves are regular parts of the surfaces $\Sigma_{\rm tri}, \Sigma_1, \Sigma_3$ while the red sold curves are singular when they are approached from ${\rm Im}\, s_1, {\rm Im}\, s_3>0$. The right figure shows that the surface $\Sigma_{\rm tri}$ spreads in the complex $(s_1,s_3)$-space and the point B and the point E approached from ${\rm Im}\, s_1>0, {\rm Im}\, s_3 < 0$ (or ${\rm Im}\, s_1<0, {\rm Im}\, s_3 > 0$) reside on the same surface $\Sigma_{\rm tri}$; on the other hand, the point B and the point E approached from ${\rm Im}\, s_1, {\rm Im}\, s_3>0$ are on different surfaces. Here, we set $t=-4\mu^2$.}
\label{fig:tri}
\end{figure*}

We are interested in the analytic structure away from the $(+)$ region; $s_3$ should follow the path $(a)$ of Fig.~\ref{fig:complexpole} while $s_1$ should follow the path $(b)$ to compute the amplitude \scalebox{0.7}{\AmixL}. Hence, we consider the analytic structure in the complex $(s_1,s_3)$-space by fixing $t$ at a physical value $(t<0)$. The would-be singular hypersurfaces $\Sigma_{\rm tri},\Sigma_1,\Sigma_3$ are described by two-dimensional surfaces in the four-dimensional space. They are generically curves in two-dimensional sections e.g., ${\rm Im}\,s_1={\rm Im}\, s_3=0$, as shown in Fig.~\ref{fig:tri}. $\Sigma_t$ does not appear because $t$ is fixed. In the $(+)$ region, that is the real $(s_1,s_3)$-plane approached from ${\rm Im}\,s_1, {\rm Im}\, s_3 > 0$, whether $\Sigma_{\rm tri},\Sigma_1,\Sigma_3$ are actually singular or not is checked by positivity of $\alpha_i$. The lines $s_1=4\mu^2$ and $s_3=4\mu^2$ are the lower-order singularities. There are branch cuts in the region $s_1, s_3 >4\mu^2$ and the way to approach the region $s_1, s_3 >4\mu^2$ determines whether the unstable-particle amplitude is \scalebox{0.7}{\Auns{+}} or \scalebox{0.7}{\AmixL}.
The curve ABC is singular if it is approached from the $(+)$ direction, showing that the diagram \eqref{deftri} of the type \scalebox{0.7}{\Auns{+}} has the triangle singularity (cf.~\cite{Eden:1966dnq,Hannesdottir:2022bmo}). On the other hand, we need to approach the curve ABC from the directions ${\rm Im}\, s_1 <0$ and ${\rm Im}\, s_3 >0$ to discuss the analytic structure of \scalebox{0.7}{\AmixL}. Let us consider the curve DEF which is the real section of the surface $\Sigma_{\rm tri}$ but is not singular in the $(+)$ region. The surface $\Sigma_{\rm tri}$ spreads in the complex $(s_1,s_3)$-space and the curve DEF is connected to the curve ABC through the surface $\Sigma_{\rm tri}$ as shown in Fig.~\ref{fig:tri} (right). Note that the curve BE exists on the surface $s_1=s_3^*$ so the point B in the $(+)$ region and the point E in the $(+)$ region do not exist on the same surface $\Sigma_{\rm tri}$. The point E in the $(+)$ region is connected to the point B approached from ${\rm Im}\, s_1 <0 $ and ${\rm Im}\, s_3 >0$ (or ${\rm Im}\, s_1 >0$ and ${\rm Im}\, s_3 <0$). Since the point E is regular and the curve BE does not intersect another singularity, the integral \eqref{triintegral} remains regular along the curve BE. Hence, the diagram \eqref{deftri} of the type \scalebox{0.7}{\AmixL} has no triangle singularity. All in all, the analytic structure of the one-loop diagram \eqref{deftri} is consistent with the predictions of unitarity.

Finally, let us briefly discuss so-called external-mass singularities~\cite{Hannesdottir:2022bmo,nakanishi1963external,Nakanishi:1974wm}. As $s_1$ or $s_3$ exceeds $4\mu^2$, the external line can decay to the pair of the internal lines which would lead to a non-vanishing imaginary part of the diagram. On the other hand, the unitarity equation \eqref{unitarityAphisu} implies that the imaginary part arises only from the $s$-channel and/or $u$-channel cut in the amplitude \scalebox{0.7}{\AmixL}.
The external-mass singularity should not provide an imaginary part of the amplitude \scalebox{0.7}{\AmixL},
while it can give an imaginary part of \scalebox{0.7}{\Auns{+}}.
To confirm the absence/existence of the imaginary part explicitly, we consider $t\to -0$ at which the simple analytic result is available~\cite{Patel:2015tea}:
\begin{align}
\mathcal{I}_{\rm tri}|_{t=0}=
\frac{1}{2(s_1-s_3)}\left[ L(s_1/\mu^2)- L(s_3/\mu^2) \right] 
\label{trit0}
\end{align}
where
\begin{align}
L(x):=\ln^2  \left( \frac{\sqrt{x(x-4)}-x}{\sqrt{x(x-4)}+x} \right)
\,,
\end{align}
is an analytic function with a branch cut in $x>4$ on the first sheet and ${\rm Im} L(x) =0$ in $x<4$. The $i\varepsilon$ prescription dictates that the physical region is approached from the upper-half plane. The Schwarz reflection principle concludes $L(x^*)=L^*(x)$. Therefore, when we set the masses of the external states in the complex-conjugate positions $s_1=(M^2)^*, s_3=M^2$, we find
\begin{align}
\mathcal{I}_{\rm tri}|_{t=0,s_3=s_1^*=M^2}=\frac{{\rm Im} L(M^2/\mu^2)}{2{\rm Im} M^2}
\end{align}
for ${\rm Im}M^2 \neq 0$. Hence, the diagram of the type \scalebox{0.7}{\AmixL} has no imaginary part, consistently with unitarity. On the other hand, when $s_1=s_3$, the integral is given by
\begin{align}
    \mathcal{I}_{\rm tri}|_{t=0,s_3=s_1}=-\frac{1}{\mu^2\sqrt{x(x-4)}}\ln  \left( \frac{\sqrt{x(x-4)}-x}{\sqrt{x(x-4)}+x} \right)
\end{align}
with $x=s_1/\mu^2=s_3/\mu^2$, which has a non-vanishing imaginary part at ${\rm Re}\,x>4$. The diagram of the type \scalebox{0.7}{\Auns{+}} has an imaginary part due to the external-mass singularity.

\bibliography{ref}
\bibliographystyle{JHEP}

\end{document}